\documentclass[12pt]{article}
\usepackage{natbib}
\usepackage{amssymb}
\usepackage[utf8]{inputenc}
\usepackage{mathtools}
\usepackage{amsmath,amsfonts,graphicx}
\usepackage{float}
\usepackage{bbm}
\usepackage{algorithm}
\usepackage{algpseudocode}
\usepackage{caption}
\usepackage{subcaption}
\usepackage{url}

\newtheorem{assumption}{Assumption}
\newtheorem{theorem}{Theorem}[section]
\newtheorem{lemma}[theorem]{Lemma}

\newcommand{\blind}{1}

\begin{document}
\def\spacingset#1{\renewcommand{\baselinestretch}%
{#1}\small\normalsize} \spacingset{1}

\if1\blind
{
  \title{\bf Flexible Regularized Estimation in High-Dimensional Mixed Membership Models}
  \author{Nicholas Marco\\
    Department of Biostatistics, University of California,
Los Angeles, USA.\\
    and \\
    Damla \c{S}ent\"{u}rk \\
    Department of Biostatistics, University of California,
Los Angeles, USA. \\
    and \\
    Shafali Jeste \\
    Division of Neurology and Neurological Institute,\\ Children’s Hospital Los Angeles, Los Angeles, USA.\\
    and \\
    Charlotte DiStefano \\
    Division of Psychiatry, Children’s Hospital Los Angeles, Los Angeles, USA.\\
    and \\
    Abigail Dickinson \\
    Department of Psychiatry and Biobehavioral Sciences,\\ University of California, Los Angeles, USA.\\
    and \\
    Donatello Telesca \thanks{
    The authors gratefully acknowledge \textit{funding from the NIH/NIMH R01MH122428-01 (DS,DT)}}\hspace{.2cm}\\
    Department of Biostatistics, University of California,
Los Angeles, USA.\\
    }
  \maketitle
} \fi

\newpage

\begin{abstract}
    Mixed membership models are an extension of finite mixture models, where each observation can partially belong to more than one mixture component. A probabilistic framework for mixed membership models of high-dimensional continuous data is proposed with a focus on scalability and interpretability. The novel probabilistic representation of mixed membership is based on convex combinations of dependent multivariate Gaussian random vectors. In this setting, scalability is ensured through approximations of a tensor covariance structure through multivariate eigen-approximations with adaptive regularization imposed through shrinkage priors. Conditional weak posterior consistency is established on an unconstrained model, allowing for a simple posterior sampling scheme while keeping many of the desired theoretical properties of our model. The model is motivated by two biomedical case studies: a case study on functional brain imaging of children with autism spectrum disorder (ASD) and a case study on gene expression data from breast cancer tissue. These applications highlight how the typical assumption made in cluster analysis, that each observation comes from one homogeneous subgroup, may often be restrictive in several applications, leading to unnatural interpretations of data features.
\end{abstract}


\section{Introduction}
Cluster analysis is generally thought of as an exploratory task aimed at assigning observations to homogeneous subgroups, so that the latent features characterizing the data distributions can better describe the heterogeneity observed in the data \citep{clusteringHandbook}. In cluster analysis, we often posit that there exists a finite number of these subgroups and that each observation belongs to only one distinct subgroup or cluster. Probabilistic cluster analysis aims to specify the probability that an observation belongs to one of the finitely many clusters, leading to the notion of \textit{uncertain membership}. In many applications, the assumption that each observation only belongs to one cluster is restrictive, and generalizations of the clustering framework have led to the formulation of \textit{mixed membership models} or \textit{partial membership models} \citep{heller2008statistical, Erosheva04}. In mixed membership models, each observation can partially belong to multiple clusters, leading to the idea of \textit{mixed membership}. This manuscript aims to derive a probabilistic modeling framework for mixed membership models of high-dimensional continuous data that is both flexible and easily interpretable while retaining features of scalable and regularized inference.

The idea that observations may be allowed to belong to more than one cluster is not novel. Starting from the seminal concept of \textit{fuzzy sets} by \citet{zadeth1965fuzzy}, several methodological contributions have been made under the nomenclature of \textit{fuzzy clustering} \citep{ruspini2019fuzzy}. Historically, fuzzy clustering has referred to cost-based algorithms, meaning probabilistic uncertainty on the memberships to each cluster is unobtainable. In this manuscript, we will use the term \textit{mixed membership models} to refer to probabilistic representations of the general concept of fuzzy clustering. 
In the context of biological applications, one of the first uses of mixed membership models was to model individuals into sub-populations using genotype data \citep{pritchard2000inference}. The added flexibility of a mixed membership model allowed for the study of admixed individuals, or individuals that have parents that belong to two separate sub-populations. The use of mixed membership models in the field of genetics, sometimes referred to as \textit{admixture models} in the genetics literature, became a popular model for reconstructing ancestries from genotype data \citep{tang2005estimation, alexander2009fast}. Mixed membership models also became popular in topic modeling with the advent of the latent Dirichlet allocation (LDA) model \citep{BleiJordan03}, specifically within topic modeling of text corpora. 

Although a mature body of work has focused on the probabilistic characterization of latent allocation structures \citep{Broderick2013}, the associated representation of mixed membership sampling models has been largely application or domain specific \citep{xu2016bayesian, Lee2016, Hu:Zhai:Williamson:Boyd-Graber-2012}.  In particular, the extension of mixed membership models to continuous and mixed data remains a challenge \citep{Galyardt2014}. Some important exceptions include the work of \citet{heller2008statistical} and \citet{Gahramani2014} who introduced a fully probabilistic mixed membership framework for exponential family distributions. Although the continuous case is covered in their framework through Gaussian mixed membership, the implied sampling model remains quite rigid, and the model parameters lead to a potentially unwieldy substantive interpretation \citep{Gruhl2014}. The same issue remains true in more recent generalizations of the same modeling framework \citep{hou2022chimeral}. 


This manuscript introduces the following elements of innovation in the definition of coherent probabilistic representations of mixed or partial membership: (i) a flexible and easily interpreted representation of mixed membership is introduced through convex combinations of \textit{dependent} Gaussian random vectors, (ii) regularized estimation in high-dimensional settings is achieved through a parsimonious eigen-representation of the latent features distribution, combined with multiplicative gamma shrinkage priors, and (iii) some theoretical guarantees are derived by proving weak conditional posterior consistency in generally regular settings.
The rest of the manuscript is organized as follows. Once the general from of our proposed mixed membership model is derived in Section \ref{sec: BPMM}, we propose a concise eigendecomposition of the latent Gaussian features to ensure scalability and regularization in high-dimensional estimation. Statistical inference is based on posterior simulation. Our proposal leverages the multiplicative gamma process shrinkage prior proposed by \citet{bhattacharya2011sparse}, allowing us to relax the orthogonality constraints on the eigenvectors and obtain efficient Markov chain exploration of the posterior target. In Section \ref{sec: posterior_consistency} we discuss potential identifiability issues and show how the proposed model maintains desired theoretical properties independent of the relaxations on the requisite of orthogonality. Section \ref{Simulations} explores the operating characteristics of the proposed modeling framework through two simulation studies and two case studies on real data. The first simulation study focuses on the recovery of our model parameters, while the second one focuses on the performance of various information criteria in choosing the number of features in our proposed model. We conclude this manuscript with a critical discussion in Section \ref{sec: Discussion}.

\section{Finite Mixture and Mixed Membership Models}
\label{sec: PMM}
Let $\mathbf{y}_1, \dots, \mathbf{y}_N$ be the observed data, where $\mathbf{y}_i \in \mathbb{R}^P$ is the $P$-dimensional outcome for the observational unit $i$, $(i=1,2,\ldots,N)$. We denote by $K$, the number of pure mixture components or {\it latent features} and defer to Section \ref{sec: info_criteria} for an in-depth discussion on the use of information criteria to select the number of features. 

In the framework of finite mixture models, a typical assumption is that each observation is drawn from one of $K$ sub-populations. When $\mathbf{y}_i$ is continuous, a popular sampling model assumes a mixture of multivariate Gaussian distributions. In this case, the sampling model for each mixture component $k$ is fully determined by a mean vector $\boldsymbol{\nu}_{k}\in\mathbb{R}^P$, and a covariance matrix $\mathbf{C}_{k}\in S^P_+$ (where $S^P_+$ denotes the set of all symmetric positive semi-definite $P \times P$ matrices). Letting $\rho_{k}\in (0,1)$, $(k=1,2,\ldots,K)$, be the marginal probability for any $\mathbf{y}_i$ to be drawn from the component $k$, the final sampling model assumes  
\begin{equation}
    \label{eq: fmm1}
p\left(\mathbf{y}_i\mid \rho_{(1:K)}, \boldsymbol{\nu}_{(1:K)}, \mathbf{C}_{(1:K)}\right) = \sum_{k=1}^K\rho_{k}\mathcal{N}\left(\mathbf{y}_i\mid\boldsymbol{\nu}_k, \mathbf{C}_k\right),
\end{equation}
s.t. $\sum_{k=1}^K \rho_{k} = 1$.  The mixing proportion, $\rho_{k}$, quantifies the uncertain membership of any observation $\mathbf{y}_i$ to the mixture component $k$. An equivalent representation of the finite mixture model in Equation (\ref{eq: fmm1}) relies on the introduction of latent membership indicator variables,  $\boldsymbol{\pi}_i = [\pi_{i1}, \dots, \pi_{iK}] \sim \mbox{Cat}(K, \boldsymbol{\rho} = (\rho_1, \dots, \rho_K))$, s.t.
\begin{equation}
    \label{eq: fmm}
    p\left(\mathbf{y}_i\mid \boldsymbol{\pi}_i, \boldsymbol{\nu}_{(1:K)}, \mathbf{C}_{(1:K)}\right) = \prod_{k=1}^K\mathcal{N}\left(\mathbf{y}_i\mid\boldsymbol{\nu}_{k}, \mathbf{C}_{k}\right)^{\pi_{ik}},
\end{equation}
where $\pi_{ik}\in \{0,1\}$ is interpreted as the $i^{th}$ observations membership indicator to the $k^{th}$ mixture component ($\sum_{k=1}^K \pi_{ik} = 1$). While, in this setting, probabilistic finite mixture models provide little room for ambiguity in interpretation, their generalizations to probability models that describe {\it mixed} or {\it partial membership} are open to alternative conceptualizations \citep{Galyardt2014, Gruhl2014}. A popular and direct generalization approach simply replaces the binary membership indicator variables $\pi_{ik}\in\{0,1\}$ with continuous membership scores $Z_{ik}\in(0,1)$. \citet{heller2008statistical} and, similarly,  \citet{Gahramani2014}, propose a direct application of the membership scores $\mathbf{z}_i = [Z_{i1}, \cdots, Z_{ik}]$ to the latent membership representation in Equation (\ref{eq: fmm}), obtaining:   
\begin{equation}
    \label{eq: pmm}
    p\left(\mathbf{y}_i\mid\mathbf{z}_i, \boldsymbol{\nu}_{(1:K)}, \mathbf{C}_{(1:K)}\right) \propto \prod_{k=1}^K\mathcal{N}\left(\mathbf{y}_i\mid\boldsymbol{\nu}_{k}, \mathbf{C}_{k}\right)^{Z_{ik}},
\end{equation}
s.t $\sum_{k=1}^K Z_{ik} =1$. From this, defining $\mathbf{h}_i = \sum_{k=1}^K Z_{ik}\mathbf{C}_k^{-1}\boldsymbol{\nu}_k$ and $\mathbf{H}_i = \left(\sum_{k=1}^K Z_{ik}\mathbf{C}_k^{-1}\right)^{-1}$ as convex combinations of the natural parameters, $\left(\mathbf{C}^{-1}_k\boldsymbol{\nu}_k,\, \mathbf{C}^{-1}_k\right)$ , of a multivariate Gaussian distribution, we obtain:
\begin{equation}
    \label{eq: Heller}
    \mathbf{y}_i\mid \mathbf{z}_{i}, \boldsymbol{\nu}_{(1:K)}, \mathbf{C}_{(1:K)} \sim \mathcal{N}\left(\mathbf{H}_i\mathbf{h}_i, \mathbf{H}_i\right).
\end{equation}
In the following section, we argue that while seemingly natural, this probabilistic conceptualization of mixed membership may prove too rigid for many applications. We propose an alternative representation based on convex combinations of dependent Gaussian random vectors.

\subsection{Mixed Membership through Convex Combinations of Dependent Gaussian Features}
\label{sec: BPMM}
The proposed probabilistic representation of mixed membership for continuous data starts with the introduction of $K$ {\it dependent} latent Gaussian feature vectors $\mathbf{f}_k \sim \mathcal{N}(\boldsymbol{\nu}_k, \mathbf{C}_k)$, with cross-covariance $\text{Cov}(\mathbf{f}_k, \mathbf{f}_{k'}) = \mathbf{C}^{(k,k')}$, for $k \neq k'=(1,2,\ldots, K)$.  

Like before, our representation maintains reliance on unit-specific mixed membership scores $\mathbf{z}_i = [Z_{i1},\cdots,Z_{iK}]$, defined on the $K$-dimensional standard simplex $\Delta^K$. However, in contrast to the representation found in Section \ref{sec: PMM}, the application of our continuous membership scores does not rely on the latent membership representation of Equation (\ref{eq: fmm}), but exploits the direct convex combination of our latent Gaussian features. Specifically, we note that conditioning on the membership indicator variables $\boldsymbol{\pi}_1, \dots, \boldsymbol{\pi}_N$, the model in Equation (\ref{eq: fmm}) may be restated through equivalence in distribution as follows:
\begin{equation}
    \mathbf{y}_i\mid\boldsymbol{\pi}_{i}=_d \sum_{k=1}^K \pi_{ik} \mathbf{f}_k,
    \label{eq: fmm_sum}
\end{equation}
leading to the idea that $\mathbf{y}_i \mid \{\pi_{ik}= 1\} =_d\mathbf{f}_k \sim \mathcal{N}\left(\boldsymbol{\nu}_{k}, \mathbf{C}_{k}\right)$. 

At this point, a direct application of the continuous membership scores to Equation (\ref{eq: fmm_sum}), rather than Equation (\ref{eq: fmm}), leads to the natural definition of a sampling model through distributional equivalence with a convex combination of dependent Gaussian random vectors s.t. 
\begin{equation}
    \mathbf{y}_i\mid \mathbf{z}_{i}=_d \sum_{k=1}^K Z_{ik} \mathbf{f}_k.
    \label{eq: pmm_sum}
\end{equation}
\begin{figure}
    \centering
    \begin{subfigure}[b]{0.3\textwidth}
         \centering
         \includegraphics[width=\textwidth]{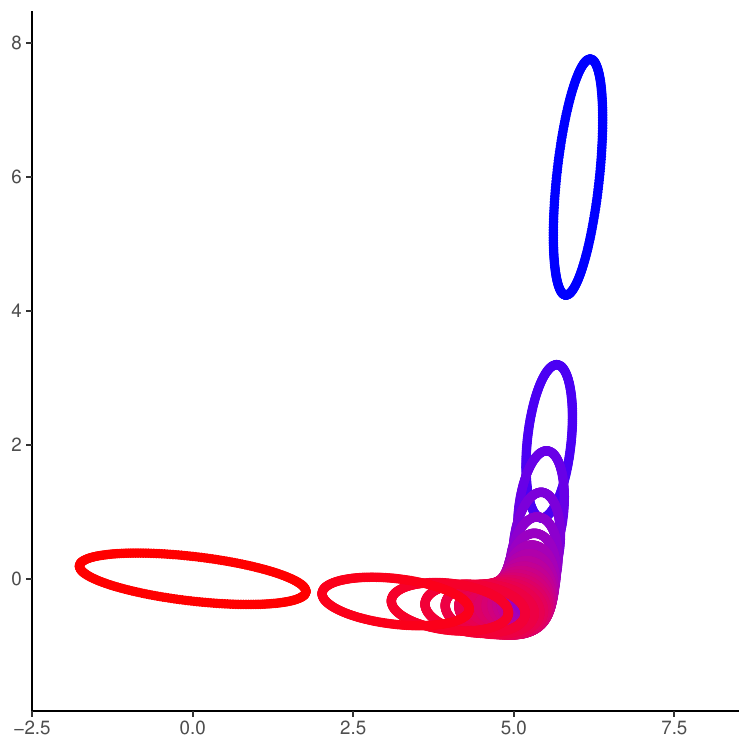}
         \caption{}
         \label{fig: sim1}
     \end{subfigure}
     \hfill
     \begin{subfigure}[b]{0.3\textwidth}
         \centering
         \includegraphics[width=\textwidth]{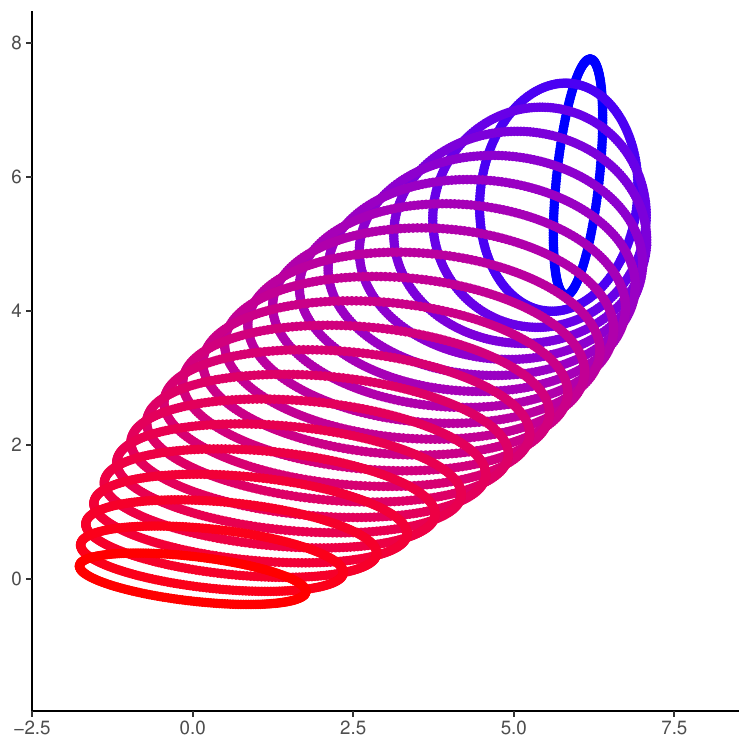}
         \caption{}
         \label{fig: sim2}
     \end{subfigure}
     \hfill
     \begin{subfigure}[b]{0.3\textwidth}
         \centering
         \includegraphics[width=\textwidth]{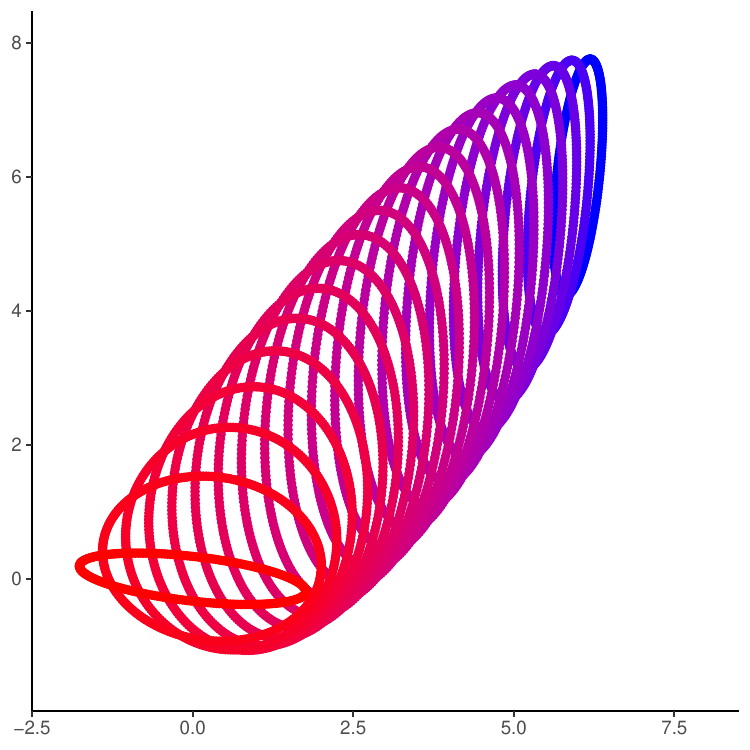}
         \caption{}
         \label{fig: sim3}
     \end{subfigure}
     \begin{subfigure}[b]{0.3\textwidth}
         \centering
         \includegraphics[width=\textwidth]{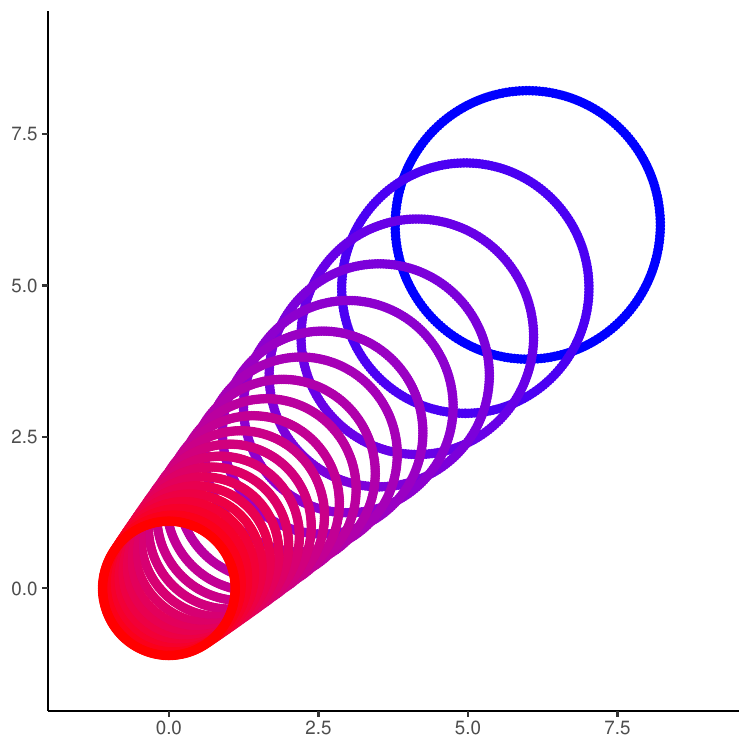}
         \caption{}
         \label{fig: sim4}
     \end{subfigure}
     \hfill
     \begin{subfigure}[b]{0.3\textwidth}
         \centering
         \includegraphics[width=\textwidth]{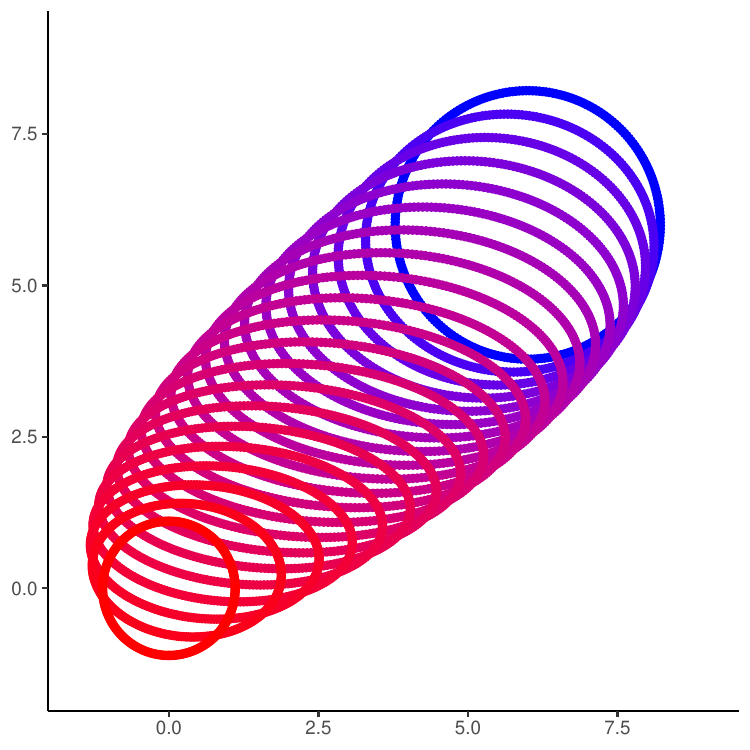}
         \caption{}
         \label{fig: sim5}
     \end{subfigure}
     \hfill
     \begin{subfigure}[b]{0.3\textwidth}
         \centering
         \includegraphics[width=\textwidth]{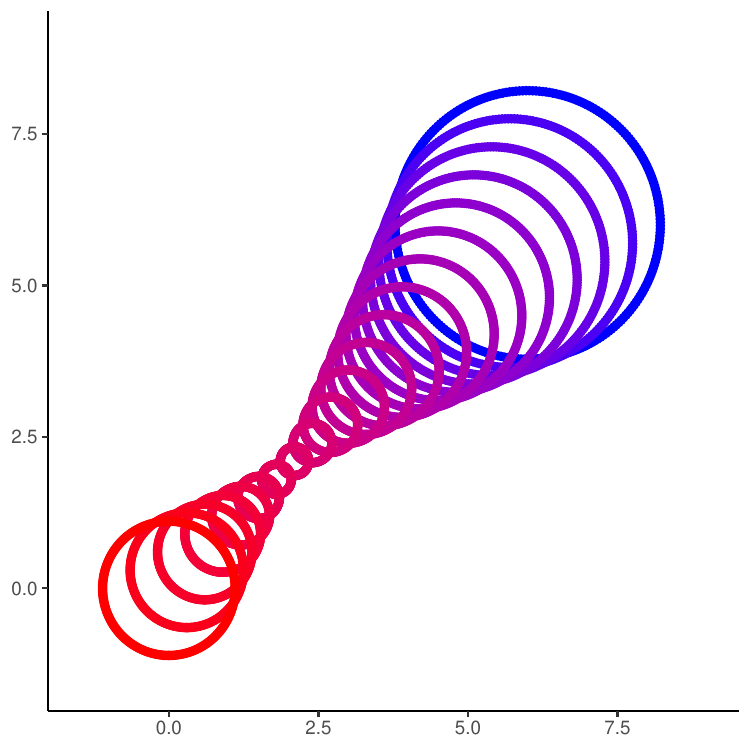}
         \caption{}
         \label{fig: sim6}
     \end{subfigure}
    \caption{Subfigures \ref{fig: sim1} and \ref{fig: sim4} depict two examples of the generative distribution of a two-feature Gaussian mixed membership model as specified in \citet{heller2008statistical}. Subfigures \ref{fig: sim2} and \ref{fig: sim3} show two examples of the generative distribution of our proposed mixed membership model with the same mean vectors and covariance matrices as the clusters in Subfigure \ref{fig: sim1}, but with varying cross-covariance matrices. Similarly, the generative distributions of Subfigures \ref{fig: sim5} and \ref{fig: sim6} were constructed with the same mean vectors and covariance matrices as the clusters in Subfigure \ref{fig: sim4}, but with varying cross-covariance matrices.}
 \label{fig:sim_data}
\end{figure}

In this context, the latent membership scores $Z_{ik}$ can be interpreted, arguably more naturally, as the $i^{th}$ observation's proportion of membership to the $k^{th}$ feature. Furthermore, denoting with $\boldsymbol{\mathcal{C}}$ the collection of covariance and cross-covariance matrices, we obtain a sampling model defined in terms of the original means and covariances, s.t.
\begin{equation}
    \label{eq: pmm_norm}
    \mathbf{y}_i\mid \mathbf{z}_i, \boldsymbol{\nu}_{(1:K)}, \boldsymbol{\mathcal{C}} \sim \mathcal{N}\left(\sum_{k=1}^K Z_{ik}\boldsymbol{\nu}_k, \sum_{k=1}^K Z_{ik}^2 \mathbf{C}_k + \sum_{k=1}^K\sum_{k \ne k'} Z_{ik}Z_{ik'}\mathbf{C}^{(k, k')}\right).
\end{equation}
While the random variables $\mathbf{f}_k$ could be assumed independent, the inclusion of cross-covariance components allows for greater expressivity and flexibility of the subsequent sampling model. A comparative visualization of the representation offered in Equation (\ref{eq: Heller}) versus the model proposed in Equation (\ref{eq: pmm_norm}), including the effects of cross-covariance components can be seen in Figure \ref{fig:sim_data}.  Subfigures \ref{fig: sim1} and \ref{fig: sim4} visualize the distribution of data generated from the Gaussian mixed membership model proposed by \citet{heller2008statistical}. Both scenarios have the same cluster-specific means; however, they have different covariances associated with each cluster. We note that when the major axes of concentration for each cluster are close to orthogonal (Subfigure \ref{fig: sim1}), the generated data appear to lie on a manifold.  Alternatively, Subfigures \ref{fig: sim2} and \ref{fig: sim3} show data generated from the proposed mixed membership model. In these two scenarios, each feature has the same mean and covariance as in Subfigure \ref{fig: sim1}, but with varying cross-covariance matrices. Even in less extreme settings, such as when component mixtures are spherical (Subfigure \ref{fig: sim4}), our model generalizes sampling expressivity through the introduction of cross-covariance components (Subfigures \ref{fig: sim5} and \ref{fig: sim6}).  

From this simple visualization, it is evident that the mixed membership model specified by Equation (\ref{eq: Heller}) may lead to mean structures that are hard to interpret, especially in higher dimensions. This interpretability challenge is largely due to the individual feature covariances appearing in the mean term in Equation (\ref{eq: Heller}). In contrast, the mean structure in our model is a simple convex combination of individual feature means, where the weighting is directly determined by the membership proportions. More details on our analysis of alternative mixed membership representations are offered in Section \ref{sec: Discussion}. 

However, increased representational flexibility comes at the cost of an increased dimensional parameter space. Therefore, a concise representation of the covariance structure is needed in order to ensure scalability and regularizability in estimation. In particular, if we were to implement a na\"{i}ve characterization of the covariance structure, we would need $\mathcal{O}(K^2P^2)$ parameters just to represent the covariance structure, which does not scale well as we increase the dimension of the data ($P$) or the number of clusters ($K$). Instead, we will use a joint decomposition of the $K$ features described in Section \ref{sec: JFD}, which will allow us to represent the covariance structure with $\mathcal{O}(KPM)$ parameters, where $M$ is a user-determined variable that controls the accuracy of our representation of the covariance structure. Once we have a concise representation of our $K$ features, we will fully specify a Bayesian version of our proposed mixed membership model in Section \ref{sec: model specification}.

\subsection{Joint Feature Decomposition}
\label{sec: JFD}
In this section, we construct a joint representation of the $K$ features based on a multivariate eigendecomposition. This joint representation allows for a scalable representation of the underlying high-dimensional covariance structure. 

Let $\mathbf{F} = \left[\mathbf{f}_1, \dots, \mathbf{f}_K\right]\in \mathbb{R}^{P\times K},$ be a random matrix that stacks the $K$ latent Gaussian features. We define the corresponding mean matrix,
$\boldsymbol{\mu} = \left[\boldsymbol{\nu}_1, \dots, \boldsymbol{\nu}_K \right]$,
where $\boldsymbol{\nu}_k$ is the mean corresponding to the $k^{th}$ feature. Let $\mathbf{C}^{(k,k')} = \text{Cov}\left(\mathbf{f}_k, \mathbf{f}_{k'}\right)$ for $1 \le k, k' \le K$. By vectorizing the matrix $\mathbf{F}$, we obtain
\begin{equation}
     \text{Cov}(\text{vec}(\mathbf{F})) = \boldsymbol{\Sigma} = \begin{bmatrix}
\mathbf{C}^{(1,1)} & \dots & \mathbf{C}^{(1,K)} \\
\vdots & \ddots & \vdots \\
\mathbf{C}^{(K,1)} & \dots & \mathbf{C}^{(K,K)}
\end{bmatrix}.
\label{eq: Cov}
\end{equation}
Since $\boldsymbol{\Sigma}$ is a positive semi-definite matrix, we know that there exists a set of eigenvalues, $\lambda_1 \ge \lambda_2, \ge \dots \ge \lambda_{KP} \ge 0$, and a set of eigenvectors, $\boldsymbol{\Psi}_1, \dots, \boldsymbol{\Psi}_{KP}$, such that
$$\boldsymbol{\Sigma} \boldsymbol{\Psi}_m = \lambda_m \boldsymbol{\Psi}_m.$$
Since the $\boldsymbol{\Psi}_m$ are eigenvectors, we know that they are orthonormal. We will define a set of parameters $\boldsymbol{\Phi}_m$ such that $\boldsymbol{\Phi}_m = \sqrt{\lambda_m} \boldsymbol{\Psi}_m$. Thus, we can see that $\boldsymbol{\Phi}_m$ are still mutually orthogonal, but are scaled by the square root of the corresponding eigenvalue. Consider partitioning $\boldsymbol{\Phi}_m$ so that
$$\boldsymbol{\Phi}_m = \begin{bmatrix} 
\boldsymbol{\phi}_{1m} \\
\vdots \\
\boldsymbol{\phi}_{Km}
\end{bmatrix}.$$
From Equation (\ref{eq: Cov}), by using a spectral decomposition on $\boldsymbol{\Sigma}$, we can see that
\begin{equation}
    \mathbf{C}^{(k,k')} = \sum_{m = 1}^{KP} \boldsymbol{\phi}_{km}\boldsymbol{\phi}_{k'm}'.
    \label{eq: cross-covariance}
\end{equation}
Since $\boldsymbol{\Phi}_m$ form a basis for $\mathbb{R}^{KP}$, we have that
\begin{align}
    \text{vec}(\mathbf{F}) - \text{vec}(\boldsymbol{\mu}) & = \mathbf{P} \circ \left( \text{vec}(\mathbf{F}) - \text{vec}(\boldsymbol{\mu})\right) \nonumber \\
     & = \sum_{m=1}^{KP}\lambda_m^{-1} \left\langle \text{vec}(\mathbf{F}) - \text{vec}(\boldsymbol{\mu}),\boldsymbol{\Phi}_m \right\rangle \boldsymbol{\Phi}_m, 
     \label{eq: projection}
\end{align}
where $\mathbf{P}$ is the projection operator on $\mathbb{R}^{KP}$. Letting $\chi_m = \lambda_m^{-1} \left\langle \text{vec}(\mathbf{F}) - \text{vec}(\boldsymbol{\mu}),\boldsymbol{\Phi}_m \right\rangle$, we can see that
\begin{align}
    \nonumber \mathbb{E}(\chi_m) & = \lambda_m^{-1} \boldsymbol{\Phi}_m' \mathbb{E}\left(\text{vec}(\mathbf{F}) - \text{vec}(\boldsymbol{\mu}) \right) = 0, \\
    \nonumber \text{Var}\left(\chi_m \right) &= \text{Cov}\left(\lambda_m^{-1} \langle \text{vec}(\mathbf{F}), \boldsymbol{\Phi}_m \rangle\right) \\
    \nonumber & = \lambda_m^{-2} \boldsymbol{\Phi}_m'\text{Cov}(\mathbf{F}) \boldsymbol{\Phi}_m \\
    \nonumber & = \lambda_m^{-2} \boldsymbol{\Phi}_m'\left(\sum_{j=1}^{KP}\boldsymbol{\Phi}_{j}\boldsymbol{\Phi}_{j}'\right) \boldsymbol{\Phi}_m = 1.
\end{align}
Therefore, using the decomposition in Equation (\ref{eq: projection}), we have that
$$\text{vec}(\mathbf{F}) = \text{vec}(\boldsymbol{\mu}) + \sum_{m=1}^{KP} \chi_m \boldsymbol{\Phi}_m,$$
where $\chi_m \sim \mathcal{N}(0,1)$.
 To reduce the dimension of the model, we often approximate $\mathbf{F}$ by only using the first $M$ scaled eigenvectors, where $M \le KP$. Using the first $M$ eigencomponents to approximate $\boldsymbol{\Sigma}$ will give us the best rank $M$  approximation of $\boldsymbol{\Sigma}$ as measured by the Frobenius norm (Eckart-Young-Mirsky Theorem \citep{eckart1936approximation}). Thus, for sufficiently large $M$, we have
 $$\text{vec}(\mathbf{F}) \approx \text{vec}(\boldsymbol{\mu}) + \sum_{m=1}^{M} \chi_m \boldsymbol{\Phi}_m.$$
 Equivalently, we can express this in terms of the $K$ partitioned vectors, such that
\begin{equation} \mathbf{f}_k \approx \boldsymbol{\nu}_k + \sum_{m=1}^{M} \chi_m \boldsymbol{\phi}_{km}.
\label{eq: eigen_decomp}\end{equation}
From Equation (\ref{eq: eigen_decomp}), we can see that each feature can be represented by a mean component, $\boldsymbol{\nu}_k$, and deviations from the mean controlled by the eigenstructure of the covariance structure of our model, $\boldsymbol{\phi}_{km}$. In this context, the hyperparameter $M$ is defined by the user and controls the approximation accuracy of the covariance structure. If $M = KP$, we are able to recover the true covariance structure. However, in most applications, a relatively small $M$, allows for a good fit to large models, while still ensuring relatively good approximations to the true covariance structure. In the following section, we discuss Bayesian estimation and adopt a regularization approach to select the effective dimension of $M$.

\subsection{Sampling Model and Prior Distributions}
\label{sec: model specification}
In this section we combine the convolutional representation of mixed membership introduced in Section \ref{sec: BPMM}, with the multivariate eigen-approximation of Section \ref{sec: JFD} 
to define a probability model of mixed membership amenable to formal Bayesian analysis. 
Using our approximation in Equation (\ref{eq: eigen_decomp}), we let $\boldsymbol{\Theta}$ be the full collection of model parameters and define the following sampling model:
\begin{equation}
    \mathbf{y}_i\mid \boldsymbol{\Theta} \sim \mathcal{N} \left\{\sum_{k=1}^K Z_{ik}\left(\boldsymbol{\nu}_k + \sum_{m=1}^M \chi_{im} \boldsymbol{\phi}_{km} \right), \;\sigma^2\mathbf{I}_P \right\},
    \label{eq: likelihood}
\end{equation}
where we assume $\chi_{im} \sim_{iid} \mathcal{N}(0,1)$, $(i=1,2,\ldots,N;\;m=1,2,\ldots,M)$.
Marginally, integrating out $\chi_{im}$, we recover an approximation to the model in (\ref{eq: pmm_norm}):
\begin{equation}
    \mathbf{y}_i\mid \boldsymbol{\Theta}_{-\chi} \sim \mathcal{N} \left\{\sum_{k=1}^K Z_{ik}\boldsymbol{\nu}_k,\;\sum_{k=1}^K\sum_{k'=1}^K Z_{ik}Z_{ik'}\left(\sum_{m=1}^M \boldsymbol{\phi}_{km}\boldsymbol{\phi}_{k'm}'\right) +  \sigma^2\mathbf{I}_P \right\},
    \label{eq: likelihood_chi_int}
\end{equation}
where $\boldsymbol{\Theta}_{-\chi}$ denotes the full collection of model parameters excluding the $\chi_{im}$ parameters. Using Equation (\ref{eq: cross-covariance}), we can see that the covariance in Equation (\ref{eq: likelihood_chi_int}) is a weighted sum of the individual feature covariances and the cross-covariances between the features, with an additional diagonal matrix to capture any residual noise. Similarly, the mean is a weighted sum of the individual feature means. Thus, the model in Equation (\ref{eq: likelihood_chi_int}) is what we would expect from a traditional additive model; however, we only need $\mathcal{O}(KPM)$ parameters to estimate the covariance structure. Selecting the number of eigencomponents, $M$, allows one to balance computational cost and model flexibility. 

From our derivations in Section \ref{sec: JFD}, we know that $\boldsymbol{\Phi}_{k}$ parameters are often assumed to be orthogonal. While aiding likelihood identifiability of the $\boldsymbol{\Phi}$ parameters, this constraint would require posterior simulation on a non-compact Stiefel manifold, which would be computationally challenging, and negatively affect mixing of algorithms using Markov chain Monte Carlo (MCMC) simulations. Section \ref{sec: posterior_consistency} shows that we can sample in an unconstrained space while still maintaining many of the desired theoretical properties of our model. Although the $\boldsymbol{\Phi}$ parameters can no longer be interpreted as scaled eigenvectors, posterior samples of the eigenpairs can still be obtained through the posterior samples of the covariance matrix $\boldsymbol{\Sigma}$ in Equation (\ref{eq: Cov}).

Furthermore, the eigen-representation of our model in Equation (\ref{eq: likelihood_chi_int}), allows for adaptive regularization through shrinkage priors.  Specifically, we know that the $\boldsymbol{\phi}_{km}$ should shrink in magnitude as they are scaled by the corresponding eigenvalues. Thus, we leverage the multiplicative gamma process shrinkage prior proposed by \citet{bhattacharya2011sparse}. Letting $\phi_{kpm}$ be the $p^{th}$ element of $\boldsymbol{\phi}_{km}$, we can specify our prior as:
$$\phi_{kpm}|\gamma_{kpm}, \tilde{\tau}_{mk} \sim \mathcal{N}\left(0, \gamma_{kpm}^{-1}\tilde{\tau}_{mk}^{-1}\right), \;\;\; \gamma_{kpm} \sim \Gamma\left(\nu_\gamma /2 , \nu_\gamma /2\right), \;\;\; \tilde{\tau}_{mk} = \prod_{n=1}^m \delta_{nk},$$
$$ \delta_{1k}| a_{1k} \sim \Gamma(a_{1k}, 1), \;\;\; \delta_{jk}|a_{2k} \sim \Gamma(a_{2k}, 1), \;\;\; a_{1k} \sim \Gamma(\alpha_1, \beta_1), \;\;\; a_{2k} \sim \Gamma(\alpha_2, \beta_2),$$
where $1 \le k \le K$, $1 \le p \le P$, $1 \le m \le M$, and $2 \le j \le M$ and the gamma distribution is specified by the shape and rate parameters. Since the $\boldsymbol{\Phi}$ parameters can be thought of as eigenvectors scaled by the square root of their corresponding eigenvalues, we would like a prior that promotes shrinkage as $m$ increases. By setting $\alpha_2 > \beta_2$, we have $\mathbb{E}(\delta_{jk}) > 1$, which will promote shrinkage in $\boldsymbol{\Phi}_i$ as $m$ increases. Although we utilize the multiplicative gamma process shrinkage prior \citep{bhattacharya2011sparse}, alternative methods for regularized inference in factor models \citep{schiavon2022generalized, matuk2022bayesian} can be used to obtain the desired regularized inference.

The model is completed through a conjugate prior for the mean parameters, s.t. 
$$\boldsymbol{\nu}_k|\tau_k \sim \mathcal{N}\left(\mathbf{0}_P, \tau_k\mathbf{I}_P \right) \;\; \text{and} \;\; \tau_k \sim IG(\alpha, \beta),$$
and, following \citet{heller2008statistical} and \citet{Gahramani2014}, we assume that
$$\mathbf{z}_i|\boldsymbol{\pi}, \alpha_3 \sim Dir(\alpha_3\boldsymbol{\pi}), \;\;\; \boldsymbol{\pi} \sim Dir(\mathbf{c}), \;\;\;
\alpha_3 \sim exp(b),$$
for $i = 1,\dots, N$. The parameter $\boldsymbol{\pi}$ mainly controls where the prior probability is centered for the allocation parameters. In practice, we often assume that all features are equally likely to be present in the data, and set $\mathbf{c} = \mathbf{1}_K$, corresponding to a uniform prior. The concentration of the prior probability on the allocation parameters is controlled by $\alpha_3$. In practice, we set $b = 10$, which puts slightly higher prior probability that the allocation parameters are close to the vertices of the simplex. In practice, we found that our model was not sensitive to the choice of $\mathbf{c}$ and $b$. Lastly, we assume that $\sigma^2 \sim IG(\alpha_0, \beta_0)$. 

\subsection{Identifiability and Posterior Consistency}
\label{sec: posterior_consistency}
Mixed membership models are subject to non-identifiability problems similar to the ones affecting finite mixture models. As in finite mixture models, a common source of non-identifiability in  mixed membership models is what is commonly known as the \textit{label switching} problem.  To solve this problem, we leverage the work of \citet{stephens2000dealing} and implement relabelling algorithms to post-process the posterior samples after running MCMC simulations. 

A second form of non-identifiability stems from the additional flexibility of the allocation parameters, namely that $Z_{ik} \in (0,1)$ rather than just being a binary variable like in a finite mixture model. Consider a two-feature model, where $\boldsymbol{\Theta}_0$ denotes the set of ``true'' parameters (i.e. $(Z_{ik})_0$ denotes the true value of $Z_{ik}$). Let $Z_{i1}^* = \frac{1}{3}(Z_{i 1})_0$ and $Z_{i2}^* =  (Z_{i2})_0 + \frac{2}{3}(Z_{i1})_0$ (transformation preserves the constraint that $Z_{i1}^* + Z_{i2}^* = 1$). If we let $\boldsymbol{\nu}_1^* = 3(\boldsymbol{\nu}_1)_0 - 2(\boldsymbol{\nu}_2)_0$, $\boldsymbol{\nu}_2^* = (\boldsymbol{\nu}_2)_0$, $\boldsymbol{\phi}_{1m}^* = 3(\boldsymbol{\phi}_{1m})_0 - 2(\boldsymbol{\phi}_{2m})_0$,  $\boldsymbol{\phi}_{2m}^* = (\boldsymbol{\phi}_{2m})_0 $, $\chi_{im}^* = (\chi_{im})_0$, and $(\sigma^2)^* = \sigma^2_0$, then from Equation (\ref{eq: likelihood}), we have $P(\mathbf{y}_i|\boldsymbol{\Theta}_0) = P(\mathbf{y}_i|\boldsymbol{\Theta}^*)$. We will refer to this type of non-identifiability as the \textit{rescaling problem}. To mitigate the effects of the rescaling problem, we derived the membership rescale algorithm (Algorithm 2 in Section 2.4 of the Supplementary Materials). In a two-feature mixed membership model, the membership rescale algorithm ensures that at least one observation completely lies in each of the features. This assumption that at least one observation completely belongs to each of the features is known as the \textit{separability condition} \citep{papadimitriou1998latent, mcsherry2001spectral, azar2001spectral, chen2022learning}. In the case of the two-feature mixed membership model, this assumption is relatively minor; however, when considering mixed membership models with three or more features, the separability assumption makes relatively strong geometric assumptions. The \textit{sufficiently scattered condition} \citep{huang2016anchor, jang2019minimum, chen2022learning} makes weaker geometric assumptions, however, it is often more complicated to implement into sampling schemes. In the case when we have more than two features, we leverage the work of \citet{chen2022learning}, which essentially rescales the allocation parameters so that they cover as much of the unit simplex as possible. 

A final source of non-identifiability arises due to the additional flexibility of the allocation parameters. Specifically, when $(\boldsymbol{\nu}_{k'})_0 \propto (\boldsymbol{\phi}_{km})_0$ in Equation (\ref{eq: likelihood}), we can see that the mean vector and covariance matrices are unidentifiable. By assuming the separability condition, we no longer have this non-identifiability issue; however, as stated above, this can be a relatively strong assumption when $K \ge 2$. When assuming weaker assumptions and when $(\boldsymbol{\nu}_{k'})_0 \propto (\boldsymbol{\phi}_{km})_0$, then we may not be able to identify the relative magnitude of the parameters. An underestimate of the magnitude of $(\boldsymbol{\phi}_{km})_0$ will lead to additional variability in the allocation parameters, leading to wider credible intervals of the $Z_{ik}$ parameters. On the contrary, an overestimate of $(\boldsymbol{\phi}_{km})_0$ will lead to smaller credible intervals of the allocation parameters. Since the model is still capturing the variability of the data in this direction, this form of non-identifiability may not be of great concern in practice.

In Section \ref{sec: model specification}, we formulated a model where the $\boldsymbol{\Phi}_m$ parameters are no longer mutually orthogonal. Therefore, the $\boldsymbol{\Phi}_m$ parameters can no longer be interpreted as scaled eigenvectors. However, the relaxation of the orthogonality constraint facilitates easier sampling schemes and the use of ``black-box'' samplers to obtain samples from the posterior distribution. Relaxation of the orthogonality constraint also tends to lead to better mixing of the Markov chain. 
Assuming that we can still recover the mean and covariance structure, we can still obtain posterior samples of the eigenvectors by reconstructing posterior draws of the covariance matrix and then calculating the eigenvectors of the sampled covariance matrices. The remaining part of this section will focus on proving that we can recover the mean and covariance structure. However, due to the identifiability issues described earlier in this section, we will be proving weak posterior consistency conditional on knowing the mixing allocation parameters.

Since our main goal is to prove that we can recover the mean and covariance structure, we will be proving weak posterior consistency using the likelihood in Equation (\ref{eq: likelihood_chi_int}) (integrating out the $\chi$ parameters). Since $\boldsymbol{\phi}_{km}$ are not identifiable, we will prove posterior consistency with respect to $\boldsymbol{\Sigma}_{kk'} := \sum_{p=1}^{KP}\left(\boldsymbol{\phi}_{kp}\boldsymbol{\phi}'_{k'p}\right)$. Let $\boldsymbol{\Pi}$ be the prior distribution on $\boldsymbol{\omega} := \left\{\boldsymbol{\nu}_1, \dots, \boldsymbol{\nu}_K, \boldsymbol{\Sigma}_{11}, \dots, \boldsymbol{\Sigma}_{1K},\dots, \boldsymbol{\Sigma}_{KK}, \sigma^2 \right\}$. We will denote the set of true parameters as
$$\boldsymbol{\omega}_0 =\left\{(\boldsymbol{\nu}_1)_0, \dots, (\boldsymbol{\nu}_K)_0, (\boldsymbol{\Sigma}_{11})_0, \dots, (\boldsymbol{\Sigma}_{1K})_0,\dots, (\boldsymbol{\Sigma}_{KK})_0, \sigma^2_0 \right\}.$$
In order to prove weak posterior consistency, we will have to make the following two assumptions: 
\begin{assumption}
The variables $Z_{ik}$ are known a-priori for $i =1, \dots, N$ and $k = 1,\dots, K$.
\label{known_Z}
\end{assumption}
\begin{assumption}
The true parameter that models random noise is positive ($\sigma^2_0 > 0$).
\label{positive_var}
\end{assumption}
Under assumptions \ref{known_Z} and \ref{positive_var}, we would like to prove that the posterior distribution $\boldsymbol{\Pi}_N(. | \mathbf{y}_1, \dots, \mathbf{y}_N)$, is weakly consistent at $\boldsymbol{\omega}_0 \in \boldsymbol{\Omega}$. To do that, we will have to show that there is positive prior probability around the set of true parameters. To do this, we will first define some quantities related to the Kullback–Leibler (KL) divergence. Following the notation of \citet{choi2005posterior}, we will define the following quantities:
$$ \Lambda_i(\boldsymbol{\omega}_0, \boldsymbol{\omega}) = \log \left(\frac{f_i(\mathbf{y}_i;\boldsymbol{\omega}_0)}{f_i(\mathbf{y}_i;\boldsymbol{\omega})} \right), \;\; K_i(\boldsymbol{\omega}_0, \boldsymbol{\omega}) = \mathbb{E}_{\boldsymbol{\omega}_0}(\Lambda_i(\boldsymbol{\omega}_0, \boldsymbol{\omega})),\;\; V_i(\boldsymbol{\omega}_0, \boldsymbol{\omega}) = \text{Var}_{\boldsymbol{\omega}_0}(\Lambda_i(\boldsymbol{\omega}_0, \boldsymbol{\omega})),$$

 where $f_i(\mathbf{y}_i;\boldsymbol{\omega}_0)$ is the likelihood when we have the parameters $\boldsymbol{\omega}_0$. To simplify the notation, we will let
 \begin{align}
     \nonumber
     \boldsymbol{\mu}_i &= \sum_{k=1}^KZ_{ik} \boldsymbol{\nu}_k,\\
     \nonumber 
     \boldsymbol{\Sigma}_i &= \sum_{k=1}^K\sum_{k'=1}^K Z_{ik}Z_{ik'}\left(\sum_{p=1}^{KP}\left(\boldsymbol{\phi}_{kp}\boldsymbol{\phi}'_{k'p}\right)\right) + \sigma^2\mathbf{I}_P = \mathbf{U}_i'\mathbf{D}_i\mathbf{U}_i + \sigma^2\mathbf{I}_P,
 \end{align}
where $\mathbf{U}_i'\mathbf{D}_i\mathbf{U}_i$ is the spectral decomposition of $\sum_{k=1}^K\sum_{k'=1}^K Z_{ik}Z_{ik'}\left(\sum_{p=1}^{KP}\left(\boldsymbol{\phi}_{kp}\boldsymbol{\phi}'_{k'p}\right)\right)$. Let $\boldsymbol{\Omega}_\epsilon(\boldsymbol{\omega}_0)$ be the set of parameters such that $\boldsymbol{\Omega}_\epsilon(\boldsymbol{\omega}_0):= \{\boldsymbol{\omega}: K_i(\boldsymbol{\omega}_0, \boldsymbol{\omega}) < \epsilon \text{ for all i} \}$. Thus, we have that $\boldsymbol{\Omega}_\epsilon(\boldsymbol{\omega}_0)$ is the set of parameters such that the KL divergence is less than $\epsilon$. We will let $\mathcal{B}(\boldsymbol{\omega}_0)$ be the set of parameters such that 
$$\mathcal{B}(\boldsymbol{\omega}_0) := \left\{\boldsymbol{\omega}: \frac{1}{a}\left((d_{il})_0 + \sigma^2_0\right) \le d_{il} + \sigma^2 \le a\left((d_{il})_0 + \sigma^2_0\right), \| \left(\boldsymbol{\mu}_i\right)_0 - \boldsymbol{\mu}_i \| \le b \right\},$$
for some $a,b \in \mathbb{R}$, where $d_{il}$ is the $l^{th}$ diagonal element of $\mathbf{D}_i$.
\begin{lemma}
 Let $\mathcal{C}(\boldsymbol{\omega}_0, \epsilon) :=\mathcal{B}(\boldsymbol{\omega}_0) \cap \boldsymbol{\Omega}_\epsilon(\boldsymbol{\omega}_0)$. Thus, for $\boldsymbol{\omega}_0 \in \boldsymbol{\Omega}$ and $\epsilon > 0$, there exist $a > 1$ and $b > 0$ such that 
 \begin{enumerate}
     \item $\sum_{i=1}^\infty \frac{V_i(\boldsymbol{\omega}_0, \boldsymbol{\omega})}{i^2} < \infty$, for any $\boldsymbol{\omega} \in \mathcal{C}(\boldsymbol{\omega}_0, \epsilon)$,
     \item $\boldsymbol{\Pi}\left(\boldsymbol{\omega} \in  \mathcal{C}(\boldsymbol{\omega}_0, \epsilon)\right) > 0$.
 \end{enumerate}
 \label{pos_prior_prob}
 \end{lemma}
Lemma \ref{pos_prior_prob} shows us that there are neighborhoods around $\boldsymbol{\omega}_0$ that have positive prior probability. Since $\mathbf{y}_1, \dots, \mathbf{y}_N$ are not identically distributed, the first condition of Lemma \ref{pos_prior_prob} is needed to prove weak posterior consistency.
\begin{lemma}
Under assumptions \ref{known_Z} and \ref{positive_var}, the posterior distribution, $\boldsymbol{\Pi}_N(. | \mathbf{y}_1, \dots, \mathbf{y}_N)$, is weakly consistent at $\boldsymbol{\omega}_0 \in \boldsymbol{\Omega}$.
\label{weak_consistency}
\end{lemma}
Lemma \ref{weak_consistency} states that given the known allocation structure, we are able to recover the mean and covariance structure. Thus, while we cannot directly make inference on the eigenvectors of the covariance structure, we can still recover the covariance structure without enforcing the orthogonality constraint on the $\boldsymbol{\Phi}$ parameters. Although the allocation parameters are usually not known, Lemma \ref{pos_prior_prob} shows that we have large KL support of the prior in our model specified in Section \ref{sec: model specification}. Thus, without enforcing orthogonality constraints on the $\boldsymbol{\Phi}$ and using shrinkage priors on the $\boldsymbol{\Phi}$ parameters to give us the desired regularized estimation, we still have large KL support of the prior; supporting that the modeling decisions of the mean and covariance structures are not restrictive. Due to the complex identifiability problems discussed earlier in this section, it remains elusive to establish weak posterior consistency when $\mathbf{z}_i$ are unknown. Although the allocation parameters are usually not known, the empirical evidence in Section \ref{sec: SS1} shows that the mean and covariance structures converge as we obtain more information.

\section{Simulations and Case Studies}
\label{Simulations}

We conducted two simulation studies to explore the empirical performance of our model and two case studies to illustrate the application of mixed membership models to substantive scientific applications in functional brain imaging and cancer genomics. The first simulation study in Section \ref{sec: SS1} explores the empirical convergence of our model with simulated data. We then look at how information criteria can aid in choosing the number of features in Section \ref{sec: info_criteria}. In Section \ref{sec: EEG}, we look at using a mixed membership model to model electroencephalography (EEG) signals in children with Autism spectrum disorder (ASD) and typically developing (TD) children. Lastly, in Section \ref{sec: BreastCancer}, we look at how a mixed membership model can be used to classify different breast cancer subtypes using targeted gene expression data.

\begin{figure}
    \centering
    \includegraphics[width=.95\textwidth]{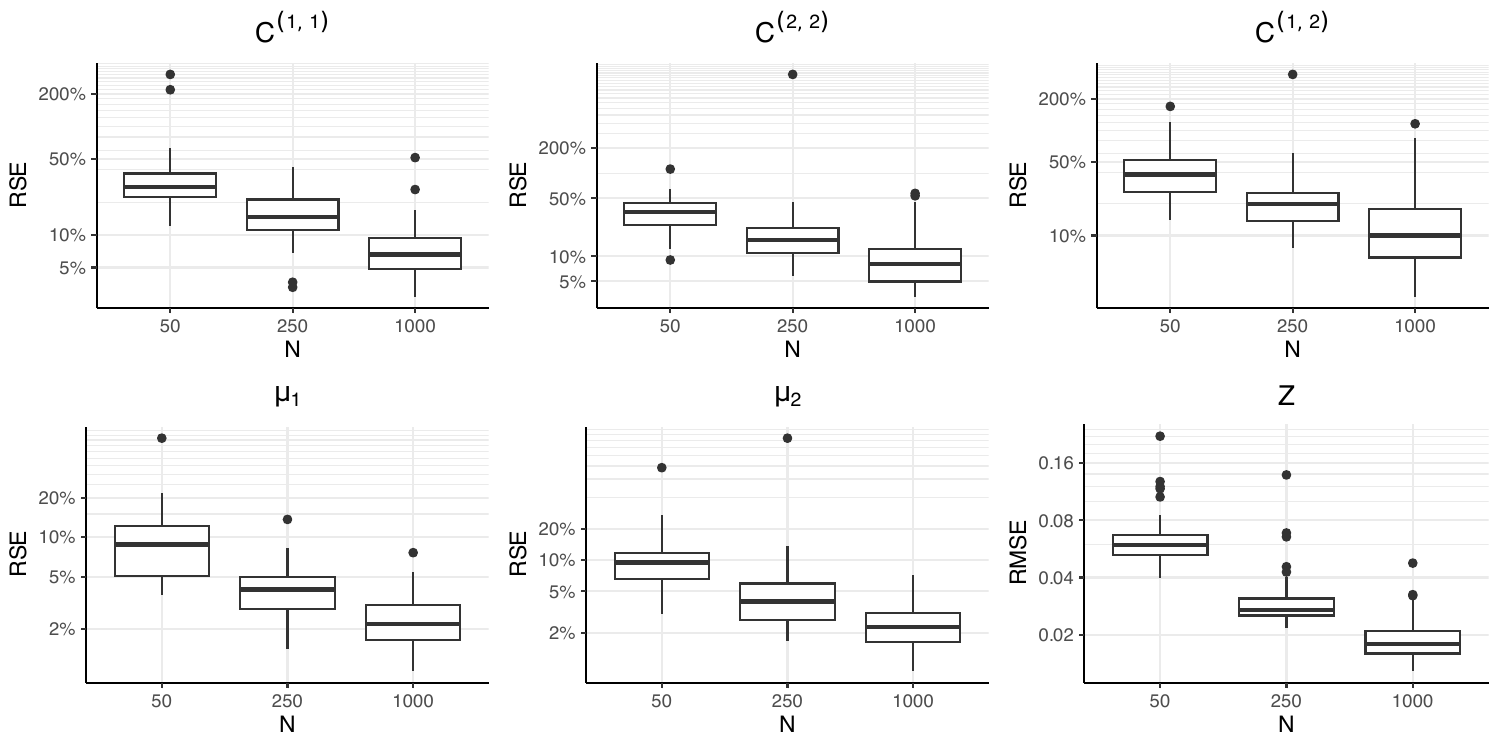}
    \caption{The relative squared error (RSE) for the mean and covariance structure evaluated under various sample sizes. To evaluate the recovery of the allocation parameters, we used the root mean squared error (RMSE).}
    \label{fig:sim1}
\end{figure}

\subsection{Simulation Study 1: Operating Characteristics under Increasing Sample Size}
\label{sec: SS1}

This simulation study focuses on how well we can recover the mean, covariance, and allocation structures under different sample sizes. In this simulation study, we will specifically look at the case when we have 50, 250, and 1000 observations. For each of the three different sample sizes, we generated 50 different datasets with observations $\mathbf{y}_i \in \mathbb{R}^{10}$. The data were generated from a two-feature model with $M =4$. More information on how the simulation study was conducted can be found in Section 3.1 of the Supplementary Materials.

To measure how well we can recover the mean vector, covariance matrices, and cross-covariance matrices, we will use the relative squared error (RSE). The RSE is defined as
$$\text{RSE} = \frac{\|f - \hat{f}\|_2}{\|f\|_2} \times 100\%,$$
where $\|\cdot\|_2$ is the Euclidean norm if $f$ is a vector and the Frobenius norm if $f$ is a matrix. In this simulation study, we will use the posterior median as our estimate, $\hat{f}$. To measure how well we can recover the allocation parameters, we will use the root mean squared error (RMSE).

From Figure \ref{fig:sim1}, we can see that our recovery of the mean and covariance structure improves as we obtain more data. While the RSE was relatively large in the simulations when $N = 50$, the credible intervals were also significantly wider to account for the uncertainty in the estimate. Specifically, the average credible interval width for the mean vectors when 50 observations were observed was roughly 4.8 times wider than the average credible interval width for the mean vectors when 1000 observations were observed. Overall, this simulation study provides empirical support that the mean and covariance structure converges to the true parameters, even when the allocation parameters are unknown.

\subsection{Simulation Study 2: Information Criteria for the Number of Latent Features}
\label{sec: info_criteria}

In the finite mixed membership setting, inference and interpretation depend on the chosen number of features. To aid in this choice, practitioners often use information criteria (IC) to employ a data-driven approach for the selection of the number of clusters or features for their model. In this section, we study how IC such as AIC \citep{akaike1974new}, BIC \citep{schwarz1978estimating}, and DIC \citep{celeux2006deviance, spiegelhalter2002bayesian} perform in choosing the optimal number of mixtures. In this section, we will also observe how heuristics such as the ``elbow-method'' perform in choosing the optimal number of clusters.

To evaluate the performance of the IC, we simulated 50 different data sets, where the true number of features was 3. For each data set, four mixed membership models were fit using a varying number of features ($K = 2,3,4,$ and $5$). Once the models were fit, we calculated the BIC, AIC, and DIC for each of the models and evaluated the performance of these IC. The definitions of the IC used in this section, as well as detailed information on how the simulation study was conducted, can be found in Section 3.2 of the Supplementary Materials.
  \begin{figure}
    \centering
    \includegraphics[width=.99\textwidth]{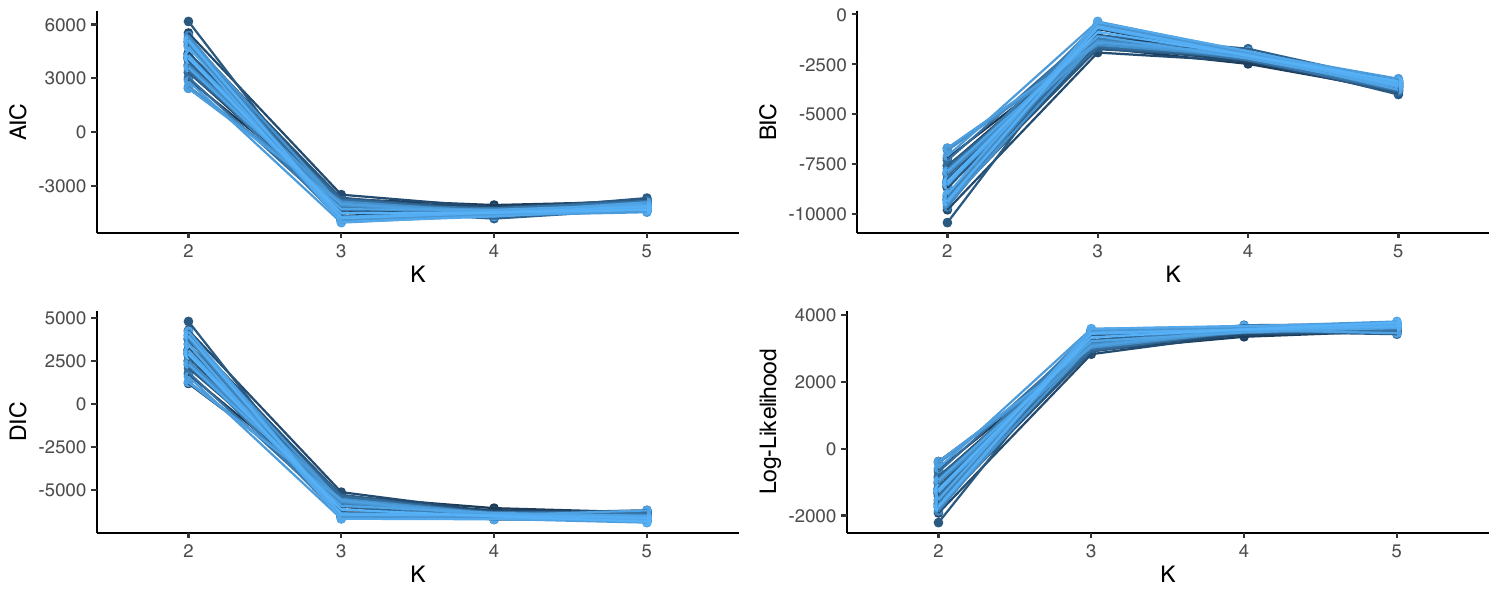}
    \caption{Information Criteria evaluated on our proposed Bayesian mixed membership model for $K = 2, 3, 4,$ and $5$. Information criteria were evaluated on 50 different data sets simulated from a three-feature model.}
    \label{fig: Multiple_K_plot}
\end{figure}

 From Figure \ref{fig: Multiple_K_plot}, we note that the BIC exhibited the most reliable performance, selecting the true number of features all 50 times. Since AIC does not penalize excess parameters as much as BIC, we can see that AIC sometimes selected a model with 4 features over a model with 3 features. Overall, the AIC selected the true number of features only 23 times out of the 50 different datasets and selected a model with four features in 27 of the 50 datasets. In this setting, the DIC was shown to be the least reliable IC, selecting a model with four or more features all 50 times. These results seem to support previous findings that AIC and DIC tend to overestimate the number of features in a model \citep{fruhwirth2006finite,meyer2014deviance}. On the contrary, BIC has been shown to be asymptotically consistent for i.i.d. observations from exponential families \citep{kass1995bayes} and asymptotically consistent for estimating the number of clusters in Gaussian finite mixture models \citep{keribin2000consistent}. From Figure \ref{fig: Multiple_K_plot}, we can see that all the IC considered in this section were unlikely to underestimate the number of features in our model, supporting the results of \citet{leroux1992consistent} who showed that AIC and BIC asymptotically do not underestimate the number of components in mixture models.
 
 As discussed in \citet{marco2022functional}, the average log-likelihood can aid in the selection of the optimal number of features in mixed membership models. From Figure \ref{fig: Multiple_K_plot}, we can see that there is a distinct ``elbow'' at $K = 3$. Using the elbow method, we would correctly identify the correct number of features all 50 times. Although BIC had the same performance as the ``elbow-method'', we can see from Figure \ref{fig: Multiple_K_plot} that BIC sometimes returned very similar values for three-feature and four-feature models, suggesting that there may be a slight tendency to choose an overfit model. However, the results from the ``elbow-method'' were more definitive, as there is a distinct elbow at $K =3$ for each dataset. However, the ``elbow-method'' cannot be used when deciding between two-feature and three-feature models. Therefore, we recommend using BIC in conjunction with the ``elbow-method'' to select the optimal number of features for our proposed model.

\subsection{A Case Study on Functional Brain Imaging through EEG}
\label{sec: EEG}
\begin{figure}
    \centering
    \includegraphics[width =  0.95\textwidth]{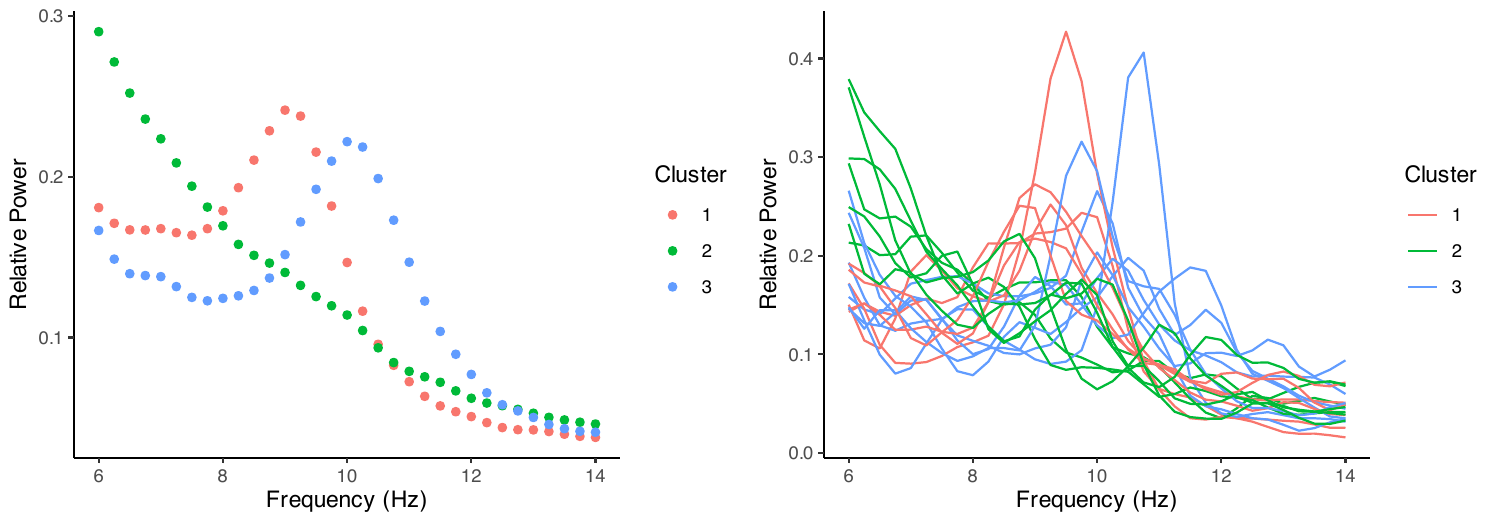}
    \caption{(Left Panel) Recovered means from fitting a $k$-means model with 3 clusters. (Right Panel) Data from the T8 electrode of 20 individuals with varying clinical diagnosis (TD vs. ASD), colored by the estimated cluster membership.}
    \label{fig: clustering_asd}
\end{figure}

Autism spectrum disorder (ASD) is a disorder characterized by social communication deficits and/or unusual sensory-motor behaviors \citep{lord2018autism, edition2013diagnostic}. Although once a more narrowly defined disorder, autism is now seen as a wide spectrum of symptoms, ranging from very mild symptoms to severe symptoms that can require lifelong care. In this case study, we analyze electroencephalogram (EEG) data collected in 39 typically developing (TD) children and 58 children diagnosed with ASD, between the ages of 2 and 12 years old \citep{dickinson2018peak}. EEG data in the present analysis are considered spontaneous because they reflect intrinsic functional brain dynamics under task-free conditions. EEG data were recorded for two minutes using a 128-channel HydroCel Geodesic Sensory net. During the recording, the children were seated in front of a computer monitor that displayed floating bubbles, a commonly used approach to collect resting EEG data in developmental populations \citep{dawson1995subgroups, stroganova1999eeg, tierney2012developmental, mcevoy2015physiologic}. The signals from the sensors were filtered to remove signals outside of the 0.1 to 100 Hz band range and then interpolated to match the international 10-20 system 25 channel montage. A fast Fourier transform was then applied to the data to transform the data into the frequency domain. In this analysis, we consider the relative power in the frequency domain, which means that each function is scaled to integrate to 1. Visualizations of the data, as well as the results of a $k$-means analysis \citep{lloyd1982least}, can be seen in Figure \ref{fig: clustering_asd}.

When neuroscientists evaluate resting-state EEG data, one area of interest is the location of a single prominent peak in the spectral density located in the alpha band of frequencies (6-14 Hz), called the \textit{peak alpha frequency} (PAF). The emergence of this peak has been shown to be a biomarker of neural development in typically developing children \citep{Rodriguez2017}. A clear alpha peak gradually emerges during the first year of life in typically developing children and increases in frequency over childhood before reaching stability in adolescence/early adulthood \citep{Rodriguez2017, scheffler2019covariate}. On the other hand, both the emergence of peak alpha frequency and developmental frequency shifts have been shown to be atypical in children diagnosed with ASD. More specifically, children are less likely to display a clear alpha peak than age-matched peers, and do not show the same age-related increase that is well documented in typical children \citep{scheffler2019covariate}.

 In this case study, we performed a multivariate analysis of the EEG analysis using the average power of the 25 electrodes at 33 frequencies of interest in the alpha frequency band (from 6 Hz to 14 Hz with step sizes of 0.25 Hz). Using the information criteria discussed in Section \ref{sec: info_criteria}, we found that a mixed membership model with 2 features seemed to be optimal. Thus, we fit a two-feature mixed membership model ($K = 2$) with five eigenvectors ($M = 5$), and ran our chain for 500,000 iterations. To save computational resources, the Markov chain was thinned, saving every $10^{th}$ iteration. 

\begin{figure}
    \centering
    \includegraphics[width=.99\textwidth]{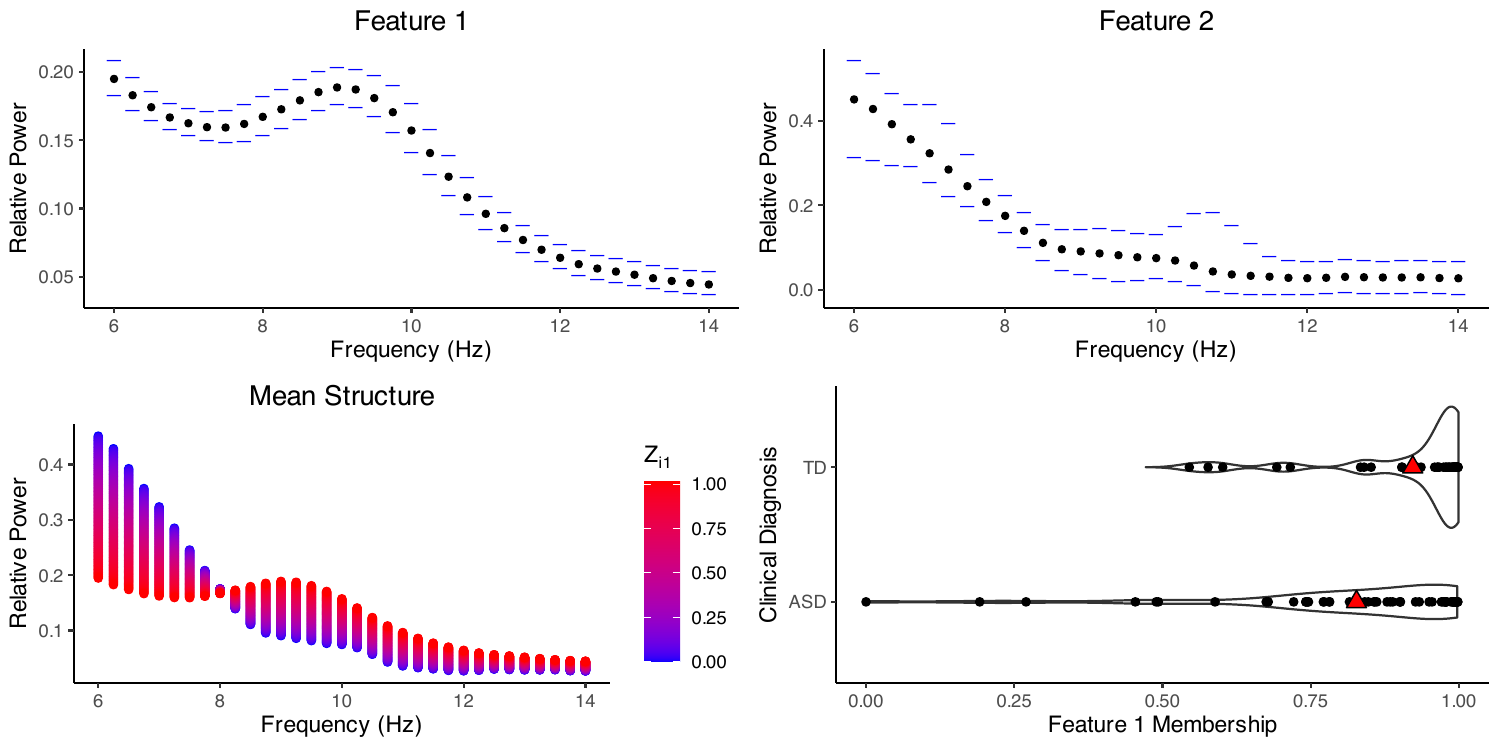}
    \caption{(Top Panels) Posterior median estimates of the mean vectors of the features, with corresponding 95\% credible intervals. (Bottom Left Panel) Posterior median estimates of the mean structure conditional on feature membership. (Bottom Right Panel) Posterior median estimates of the membership to the first feature, stratified by clinical cohort. The red triangles represent the mean membership to the first feature.}
    \label{fig:ASD_mean_Z}
\end{figure}

From Figure \ref{fig:ASD_mean_Z}, we can see that the first functional feature can be interpreted as a distinct alpha peak. On the other hand, the second feature can be interpreted as a $1/f$ trend, or \textit{pink noise}. These two features help to differentiate between periodic (alpha waves) and aperiodic ($1/f$ trend) neural activity patterns, which coexist in the EEG spectra. Loading heavily on feature 2 suggests that the 1/f trend is the most prominent in an individual's EEG recording, suggesting that a clear alpha peak has not yet emerged. From Figure \ref{fig:ASD_mean_Z}, we can see that typically developing children seem to heavily load on the first feature, representing individuals with a distinct alpha peak. On the other hand, children with ASD seem to have relatively high heterogeneity in their loadings. We can see that on average children with ASD tend to have a more attenuated alpha peak (more loading on feature 2), with some individuals having no discernible alpha peak at all. These results closely match the results obtained by \citet{marco2022functional}, who used a functional mixed membership model to model the spectral density as a random function.

Given that the presence and/or emergence of an alpha peak is a developmental biomarker, objectively quantifying the presence of a peak is important. By allowing individuals to partially belong to both features, we are able to objectively quantify the presence of a clear alpha peak over and above $1/f$ aperiodic patterns. This approach offers a novel way to quantify the progression of alpha peak emergence in individuals where the peak has not yet reached maturity. Visualizations of the recovered covariance structure can be found in Section 3.3 of the Supplementary Materials.

Figures \ref{fig: clustering_asd} and \ref{fig:ASD_mean_Z} clearly depict the differences between traditional clustering models, like a $k$-means model \citep{lloyd1982least}, and our proposed mixed membership model. Information criteria determined that 3 clusters were the optimal number of clusters for a $k$-means analysis, illustrating how the increased flexibility of mixed membership models can potentially represent data using fewer clusters than traditional clustering. In the $k$-means analysis, the first and third clusters can be interpreted as distinct alpha peaks, while the second cluster can be interpreted as primarily $1/f$ aperiodic patterns. Although the cluster means found in the $k$-means analysis are relatively interpretable, we lose the ability to quantify the ratio of periodic signals to aperiodic signals, which is of significant scientific interest. From a modeling standpoint, traditional clustering models assume that each observation comes from one of the three clusters, meaning that children who have a developing alpha peak are not represented by any of the three clusters. Alternatively, the mixed membership model assumes that each observation can be represented by a continuous mixture of periodic and aperiodic neural activity patterns, leading to a more natural sampling model that is supported by previous scientific literature.

\subsection{A Case Study on Molecular Subtypes in Breast Cancer}
\label{sec: BreastCancer}

\begin{figure}
    \centering
    \includegraphics[width=.85\textwidth]{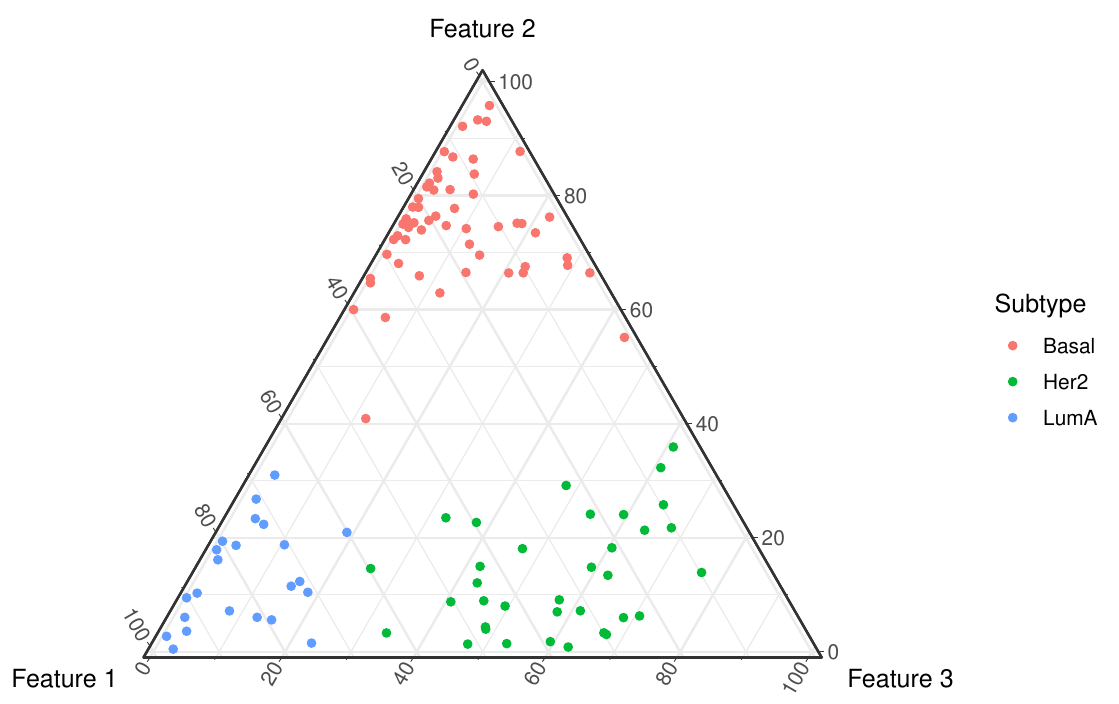}
    \caption{Estimated feature membership stratified by cancer subtype.}
    \label{fig:BC_group_membership}
\end{figure}

As of 2015, breast cancer is the $29^{th}$ leading cause of death in the world, with an estimated 534,000 deaths per year \citep{wang2016global}. Although there has been a significant increase in the number of deaths between 2005 and 2015 (21.3\%), the age-adjusted death rate has decreased by 6.8\% mainly due to our increased understanding of how to treat breast cancer. In the last two decades, five molecular subtypes of breast cancer have been discovered; each with a different prognosis, risk factors, and treatment sensitivity \citep{prat2015clinical}. \citet{parker2009supervised} discovered that the cancer subtype can be accurately classified using centroid-based prediction methods using gene expression data from 50 genes (known as PAM50). Although there are five commonly accepted subtypes, newer studies have found up to ten molecular subtypes \citep{curtis2012genomic}, while others theorize that multiple cancer subtypes can coexist within a tumor \citep{yeo2017breast}. Although the number of subgroups and even the definition of the subgroups have evolved over the years \citep{orrantia2022subtypes}, the main purpose of creating these subgroups is to explain the inter- and intratumoral heterogeneity leading to differential treatments to elicit the best response in patients \citep{yeo2017breast}. In this case study, we propose modeling heterogeneity through a mixed membership model, which allows for a more expressive model while building upon the well-accepted idea of five molecular subtypes.

Following \citet{xu2016bayesian}, we will use the PAM50 dataset to fit a mixed membership model on patients with LumA, Basal, and Her2 cancer subtypes. When restricting the PAM50 dataset to these three cancer subtypes, we have 115 patients with breast cancer ($N = 115$), and 50 genes of interest ($P = 50$). To build on the idea of the three well-accepted cancer subtypes, we fit a three-feature mixed membership model ($K = 3$) with four eigenvectors to approximate the covariance structure ($M = 4$).

From Figure \ref{fig:BC_group_membership}, we can see that each feature corresponds to a specific breast cancer subtype. While the data are still separable by cancer subtype, we can see that there is considerable overlap between features 1 and 3 for some of the patients who were classified with the Her2 cancer subtype. The added flexibility of a mixed membership model is crucial as physicians can offer more personalized treatment plans for each patient. Since each cancer subtype has particular risk factors and treatment sensitivity, physicians who have patients that load heavily on two or more features will be aware that they should be monitoring all risk factors in the corresponding features. Treatment plans can also be customized to account for the treatment sensitivity of two or more cancer subtypes. 

\begin{figure}[H]
    \centering
    \includegraphics[width=.49\textwidth]{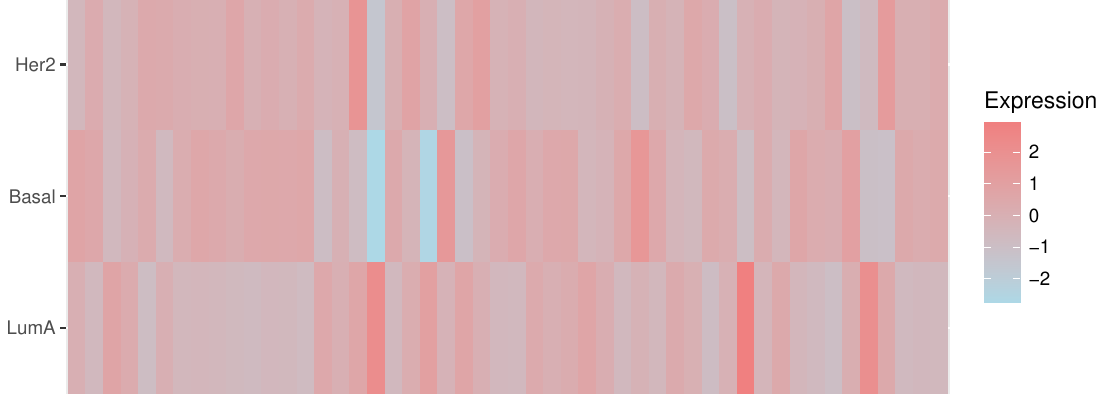}
    \includegraphics[width=.49\textwidth]{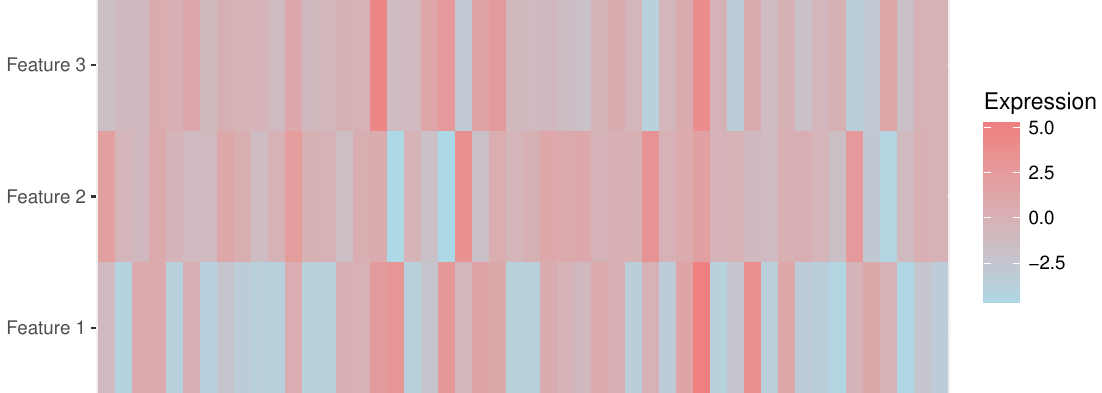}
    \caption{Cluster centroids for the model constructed by \citet{parker2009supervised}(left) and the feature means for our mixed membership model (right).}
    \label{fig:BC_means}
\end{figure}

From Figure \ref{fig:BC_means}, we can see that there is more heterogeneity in the mean structure of our model than in the cluster centroids found in \citet{parker2009supervised}. Conceptually, these mean structures represent two different ideas. In centroid-based clustering, the data is assumed to belong to only one of the cancer subtypes. Therefore, the mean structure in the centroid-based clustering model represents the average gene expression values for an individual belonging to that cluster. On the other hand, our model assumes that each individual can simultaneously belong to multiple cancer subtype groups. Therefore, the mean for a particular feature represents the average gene expression values for an individual that only belongs to that feature. Thus, the mean structure for the mixed membership model is more heterogeneous because they represent the ``extreme'' cases, or cases that are the most dissimilar from the other cancer subtypes. Overall, the added flexibility offered by our proposed mixed membership model allows us to maintain the well-accepted subgroups of LumA, Basal, and Her2, while having a more expressive model that explains more heterogeneity than comparable centroid-based models. The ability to explain more heterogeneity is advantageous from a clinical perspective as more personalized treatment plans can be constructed. Moreover, the use of mixed membership model over adding more clusters to a centroid-based model is advantageous as it will be more parsimonious. Lastly, the idea that multiple cancer subtypes can exist within a tumor \citep{yeo2017breast} is closely related to our model; since the mixed membership model in this section suggests that each individual can be represented as a continuous mixture of the three cancer subtypes modeled in this section.
Visualizations of the correlation structure of features can be found in Section 3.4 of the Supplementary Materials.

\section{Discussion}
\label{sec: Discussion}
This manuscript introduces a flexible, yet scalable mixed membership model for continuous multivariate data. Mixed membership models, sometimes referred to as partial membership models or admixture models, can be thought of as a generalization of clustering, where each observation can partially belong to multiple clusters. In Section \ref{sec: PMM}, we derive our mixed membership model by extending the framework of finite mixture models. To have a scalable framework, we use a spectral decomposition of the $K$ features, leveraging the multiplicative gamma shrinkage prior to ensure that the scaled eigenvectors are stochastically shrunk. To facilitate the use of simple sampling methods, we removed the constraint that the scaled eigenvectors, $\boldsymbol{\Phi}_m$, had to be mutually orthogonal. Within this context, we proved that the model had conditional weak posterior consistency of our mean and covariance structures, allowing us to facilitate relatively simple sampling schemes. Compared to previous work on mixed membership models, our proposed model has an easily interpretable mean and covariance structure, allowing practitioners to easily and effectively communicate their findings from the model.

For the case of continuous data, the mixed membership model representation proposed by \citet{heller2008statistical} and discussed by \citep{Gahramani2014, Gruhl2014} is seemingly natural and generally applicable to  
data assumed to be sampled from a multivariate exponential family of distributions. We note that for the case of Normally distributed mixture components, this framework may prove to be too rigid to model more complex datasets. The Subfigures \ref{fig: sim1} and \ref{fig: sim4} contain visualizations of the distribution of data generated from the Gaussian mixed membership model proposed by \citet{heller2008statistical}. 
Both scenarios have the same cluster-specific means; however, they have different covariances associated with each cluster. 
In this setting, the generated data follow the direction associated with the eigenvectors of the pure mixture covariance matrices, which can lead to unwieldy implied trajectories for plausible data realizations, e.g. data lying on a manifold (Subfigure \ref{fig: sim1}). Alternatively, Subfigures \ref{fig: sim2} and \ref{fig: sim3} show the data generated from our proposed mixed membership model. In these two scenarios, each feature has the same mean and covariance as in Subfigure \ref{fig: sim1}, but they have different cross-covariance matrices. In these cases, we can see a more natural trajectory of the data, where an assumption of a Euclidean metric seems appropriate. In less extreme settings, such as assuming spherical contours for the pure mixture components (Subfigure \ref{fig: sim4}), the representation in Equation (\ref{eq: Heller}) still leads to natural trajectories for the generated data, where the variance monotonically grows as the observation moves from the red to the blue cluster.  Crucially, the proposed mixed membership model in Equation (\ref{eq: pmm_norm}) recovers the model in Equation (\ref{eq: Heller}) as a special case, by setting the cross-covariance matrix equal to the zero matrix.  However, the inclusion of cross-covariance elements can express more flexible sampling scenarios, as shown in Subfigures \ref{fig: sim5} and \ref{fig: sim6}. Specifically, in Subfigure \ref{fig: sim5} we show the distribution of data where the symmetric part of the cross-covariance matrix has positive eigenvalues, while \ref{fig: sim6} shows the distribution of data where the symmetric part of the cross-covariance matrix has negative eigenvalues. Overall, the inclusion of explicit cross-covariance elements and the additive nature of our model leads to more natural trajectories of the implied model and an increased expressivity of the implied data generating process.

\begin{figure}
    \centering
    \includegraphics[width = 0.9\textwidth]{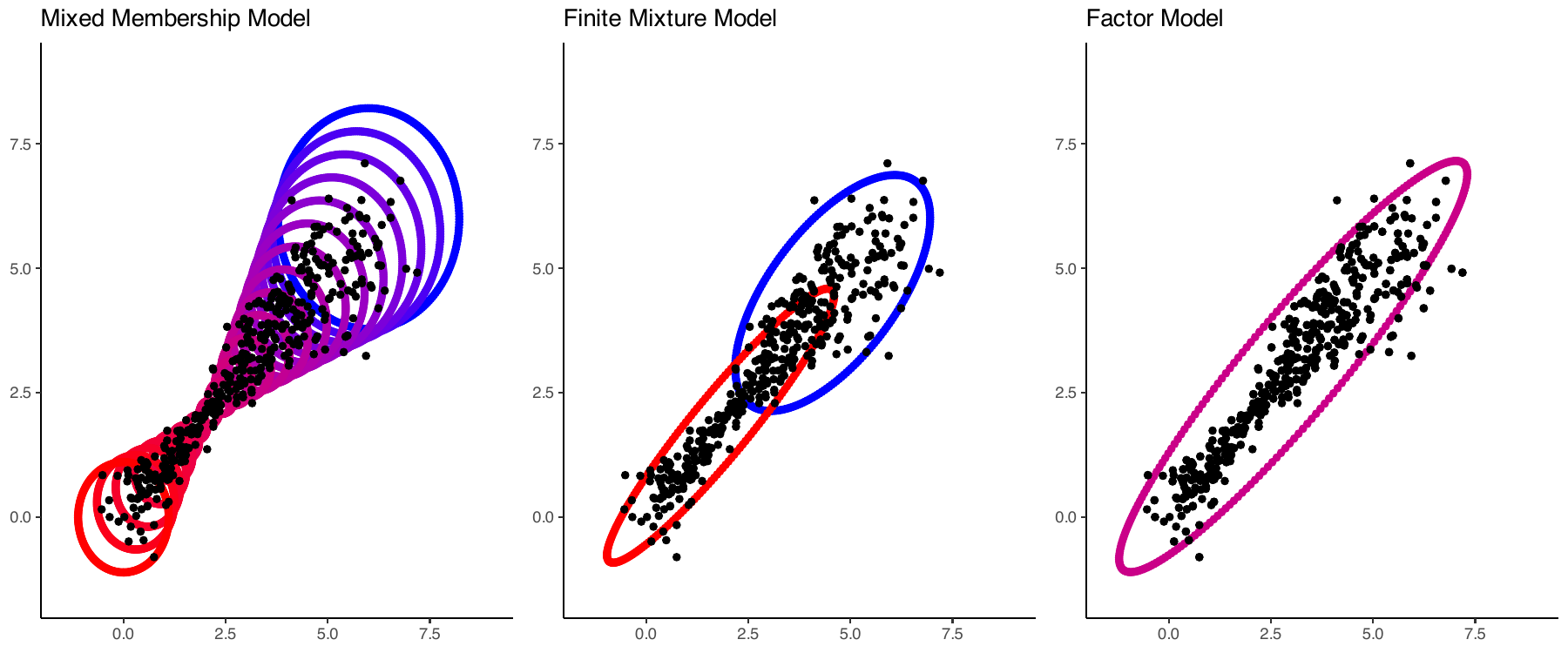}
    \caption{Comparative visualization of the differences between mixed membership models, finite mixture models, and factor models. Each of the models were fit on the same set of data, illustrated by the black dots.}
    \label{fig: MM_FMM_FM}
\end{figure}

Mixed membership models for continuous data, as encoded in our representation in Equation (\ref{eq: fmm_sum}), are related to latent factor models, as they rely on similar additive structures. Nevertheless, mixed membership models obtain an alternative decomposition of data variation, leading to a different interpretation of the model parameters. An illustration of the differences between Gaussian finite mixture models, factor models, and our proposed mixed membership model can be seen in Figure \ref{fig: MM_FMM_FM}. The principal difference between factor models and mixed membership models being that mixed membership models restrict mixing between latent features to be defined on the standard simplex. As a consequence, the ensuing model defines margins which are not constrained to be elliptically symmetric, as in the case of factor models.  An in-depth discussion of how factor analysis compares to mixed membership modeling is provided in Section 4 of the Supplementary Materials.

Although flexibility is crucial to ensure that the model fits the data adequately, interpretability of a model is paramount when communicating scientific findings. Under the Gaussian mixed membership model framework described by \citet{heller2008statistical}, each observation is distributed normally, where the natural parameters of the distribution are a convex combination of the natural parameters associated with each cluster. The natural parameter representation can be particularly difficult to work with because the mean structure is influenced by the covariance matrices of each individual cluster, leading to unnatural trajectories as seen in Subfigure \ref{fig: sim1}. Although the mean structure can be easily visualized in low dimensions, visualization of the mean structure can be particularly challenging in a high-dimensional model, leading to challenges in communicating the overall scientific findings. In our proposed model, each observation is also normally distributed; however, the mean parameter is simply a convex combination of the individual features' mean parameters. Similarly, the covariance of each observation can be written as a weighted sum of the covariance and cross-covariance functions of the features. Overall, this simple structure allows practitioners to easily describe the distribution of observations that belong to multiple features while maintaining a relatively flexible model.

From a practitioner's perspective, the main challenge in fitting a mixed membership model is choosing the number of features ($K$). Similarly to finite mixture models \citep{fruhwirth2006finite}, mixed membership models also suffer from non-identifiability problems due to overfitting. Therefore, choosing the correct number of features in a mixed membership model is crucial, as underspecifying the number of features will lead to a model that lacks the required flexibility needed to model the variability of the data, while overspecifying the number of features can lead to a lack of interpretability.  As shown in Section \ref{sec: info_criteria}, information criteria and simple heuristics, such as the ``elbow'' method, can be very informative in choosing the number of features in a mixed membership model. In traditional finite mixture models, \citet{rousseau2011asymptotic} and \citet{nguyen2013convergence} have shown that under certain conditions, an overfit mixture model would have a posterior distribution where only the ``true'' parameters would have positive weight, and the rest of the parameters would converge to zero. In both manuscripts, they assumed that the parameters were identifiable if we disregard the non-identifiability caused by the label-switching problem. However, as discussed in Section \ref{sec: posterior_consistency}, the continuous nature of the allocation parameters create multiple non-identifiability problems. These types of non-identifiability problems make applying the results from \citet{rousseau2011asymptotic} and \citet{nguyen2013convergence} non-trivial. An alternative non-parametric approach can be used by leveraging the Indian Buffet Process \citep{Griffiths2011}, allowing for an infinite number of potential clusters. However, implementing and conducting inference across changing dimensions is a non-trivial task and still an active area of research. In addition to the number of features, users are also tasked with choosing the number of eigenvectors ($M$) to use in the mixed membership model. The number of eigenvectors controls how accurately we can approximate the covariance structure of the model. If $M = KP$, then we have a fully saturated model, and we can have an exact representation of the covariance structure. In large models, setting $M = KP$ may be computationally intractable, which means that we will have to approximate the covariance structure by using a low-rank approximation of the covariance structure. Thus, the choice of $M$ is a user-defined choice that depends primarily on the computational budget afforded to fit the model.

Due to the added flexibility of mixed membership models in general, we are often faced with multiple modal posterior distributions. Due to the potential multiple-modal nature of the posterior distribution, we implemented tempered transitions (Section 2.3 of the Supplementary Materials) to ensure adequate exploration of the posterior distribution by allowing the chain to traverse areas of low posterior probability. Although tempered transitions theoretically allow you to move between modes of the posterior distribution, they can be computationally expensive and sometimes tricky to tune. To limit the computational burden, we suggest a mixture of tempered transitions and untempered transitions when performing MCMC. To speed up convergence of the Markov chain, we pick an informative starting position for our parameters using the multiple start algorithm (Algorithm 1) described in Section 2.2 of the Supplementary Materials. The added flexibility of the allocation parameters also causes the \textit{rescaling problem} described in Section \ref{sec: posterior_consistency}. To make interpretation easier for practitioners, we recommend using the membership rescale algorithm (Algorithm 2) described in Section Section 2.4 of the Supplementary Materials. In the two-feature case, this algorithm ensures that at least one observation belongs entirely to one feature. In the case where we have more than two features, the algorithm can be reformulated as an optimization problem. An R package for our proposed Gaussian mixed membership model is available for download at \url{https://github.com/ndmarco/BayesFMMM}.

\vskip.2in
\noindent{\bf Acknowledgements}\\
The authors gratefully acknowledge \textit{funding from the NIH/NIMH R01MH122428-01 (DS,DT)}.

\bibliographystyle{elsarticle-harv}
\bibliography{BPMM}

\end{document}


\def\spacingset#1{\renewcommand{\baselinestretch}%
{#1}\small\normalsize} \spacingset{1}

\if1\blind
{
  \title{\bf Supplementary Material: Flexible Regularized Estimation in High-Dimensional Mixed Membership Models}
  \author{Nicholas Marco\\
    Department of Biostatistics, University of California,
Los Angeles, USA.\\
    and \\
    Damla \c{S}ent\"{u}rk \\
    Department of Biostatistics, University of California,
Los Angeles, USA. \\
    and \\
    Shafali Jeste \\
    Division of Neurology and Neurological Institute,\\ Children’s Hospital Los Angeles, Los Angeles, USA.\\
    and \\
    Charlotte DiStefano \\
    Division of Psychiatry, Children’s Hospital Los Angeles, Los Angeles, USA.\\
    and \\
    Abigail Dickinson \\
    Department of Psychiatry and Biobehavioral Sciences,\\ University of California, Los Angeles, USA.\\
    and \\
    Donatello Telesca \thanks{
    The authors gratefully acknowledge \textit{funding from the NIH/NIMH R01MH122428-01 (DS,DT)}}\hspace{.2cm}\\
    Department of Biostatistics, University of California,
Los Angeles, USA.\\
    }
  \maketitle
} \fi

\newpage
\begin{abstract}
    Section \ref{sec: Proofs} contains all of the proofs for the lemmas found in the main manuscript. Section \ref{sec: Computation} specifies the posterior distributions for our model, a tempered transition sampling scheme, as well as two of the main algorithms (Multiple Start Algorithm and Membership Rescaling Algorithm) used to conduct inference. Section \ref{sec: sim_case_studies} contains detailed information on how the case studies and simulation studies were conducted. Section \ref{sec: sim_case_studies} also contains definitions of the information criteria used in the second simulation study. Section \ref{sec: factor} includes an in-depth discussion of latent factor models as they relate to mixed membership models.
\end{abstract}

\section{Proofs}
\label{sec: Proofs}
\subsection{Proof of Lemma 2.1}

We will start by explicitly defining the functions $\Lambda_i(\boldsymbol{\omega}_0 ,\boldsymbol{\omega})$, $K_i(\boldsymbol{\omega}_0, \boldsymbol{\omega})$, and $V_i(\boldsymbol{\omega}_0, \boldsymbol{\omega})$. Thus we have
 \begin{align}
     \nonumber
    \Lambda_i(\boldsymbol{\omega}_0 ,\boldsymbol{\omega}) = & \log \left(\frac{\left|\left(\boldsymbol{\Sigma}_i\right)_0\right|^{-1/2} \text{exp}\left\{-\frac{1}{2}\left(\mathbf{y}_i - \left(\boldsymbol{\mu}_i\right)_0\right)'\left(\boldsymbol{\Sigma}_i\right)^{-1}_0\left(\mathbf{y}_i - \left(\boldsymbol{\mu}_i\right)_0\right)\right\}}{\left|\boldsymbol{\Sigma}_i\right|^{-1/2} \text{exp}\left\{-\frac{1}{2}\left(\mathbf{y}_i - \boldsymbol{\mu}_i\right)'\left(\boldsymbol{\Sigma}_i\right)^{-1}\left(\mathbf{y}_i - \boldsymbol{\mu}_i\right)\right\}} \right)\\
    \nonumber
    =& -\frac{1}{2}\left[\log \left(\left|\left(\boldsymbol{\Sigma}_i\right)_0 \right|\right) - \log \left(\left|\boldsymbol{\Sigma}_i \right|\right) \right]  \\
    \nonumber
    & -\frac{1}{2} \left[\left(\mathbf{y}_i - \left(\boldsymbol{\mu}_i\right)_0\right)'\left(\boldsymbol{\Sigma}_i\right)^{-1}_0\left(\mathbf{y}_i - \left(\boldsymbol{\mu}_i\right)_0\right) - \left(\mathbf{y}_i - \boldsymbol{\mu}_i\right)'\left(\boldsymbol{\Sigma}_i\right)^{-1}\left(\mathbf{y}_i - \boldsymbol{\mu}_i\right) \right]\\
    \nonumber
    =& -\frac{1}{2}\left[\sum_{l=1}^{P}\log \left((d_{il})_0 + \sigma^2_0\right) - \log \left(d_{il} + \sigma^2\right) \right]  \\
    & -\frac{1}{2} \left[\left(\mathbf{y}_i - \left(\boldsymbol{\mu}_i\right)_0\right)'\left(\boldsymbol{\Sigma}_i\right)^{-1}_0\left(\mathbf{y}_i - \left(\boldsymbol{\mu}_i\right)_0\right) - \left(\mathbf{y}_i - \boldsymbol{\mu}_i\right)'\left(\boldsymbol{\Sigma}_i\right)^{-1}\left(\mathbf{y}_i - \boldsymbol{\mu}_i\right) \right]
 \end{align}
 \begin{align}
     \nonumber
     K_i(\boldsymbol{\omega}_0, \boldsymbol{\omega}) =& -\frac{1}{2}\left[\sum_{l=1}^P\log \left((d_{il})_0 + \sigma^2_0\right) - \log \left(d_{il} + \sigma^2\right) \right] \\
     \nonumber
     & -\frac{1}{2} \mathbb{E}_{\boldsymbol{\omega}_0}\left[\left(\mathbf{y}_i - \left(\boldsymbol{\mu}_i\right)_0\right)'\left(\boldsymbol{\Sigma}_i\right)^{-1}_0\left(\mathbf{y}_i - \left(\boldsymbol{\mu}_i\right)_0\right) -\left(\mathbf{y}_i - \boldsymbol{\mu}_i\right)'\left(\boldsymbol{\Sigma}_i\right)^{-1}\left(\mathbf{y}_i - \boldsymbol{\mu}_i\right) \right]\\
     \nonumber
      =& -\frac{1}{2}\left[\sum_{l=1}^{P}\log \left((d_{il})_0 + \sigma^2_0\right) - \log \left(d_{il} + \sigma^2\right) \right] \\
     & -\frac{1}{2} \left[P - \left(\text{tr}\left(\boldsymbol{\Sigma}_i^{-1}\left(\boldsymbol{\Sigma}_i\right)_0\right) + \left(\left(\boldsymbol{\mu}_i\right)_0- \boldsymbol{\mu}_i\right)'\left(\boldsymbol{\Sigma}_i\right)^{-1}\left(\left(\boldsymbol{\mu}_i\right)_0 - \boldsymbol{\mu}_i\right) \right) \right]
 \end{align}
 \begin{align}
     \nonumber
     V_i(\boldsymbol{\omega}_0, \boldsymbol{\omega}) = & \frac{1}{4}\text{Var}_{\boldsymbol{\omega}_0}\left[\left(\mathbf{y}_i - \left(\boldsymbol{\mu}_i\right)_0\right)'\left(\boldsymbol{\Sigma}_i\right)^{-1}_0\left(\mathbf{y}_i - \left(\boldsymbol{\mu}_i\right)_0\right) - \left(\mathbf{y}_i - \boldsymbol{\mu}_i\right)'\left(\boldsymbol{\Sigma}_i\right)^{-1}\left(\mathbf{y}_i - \boldsymbol{\mu}_i\right) \right] \\
     \nonumber
     = & \frac{1}{4}\text{Var}_{\boldsymbol{\omega}_0}\left[ \mathbf{y}_i'\left(\left(\boldsymbol{\Sigma}_i \right)_0^{-1} + \boldsymbol{\Sigma}_i^{-1} \right)\mathbf{y}_i - 2\mathbf{y}_i'\left(\left( \boldsymbol{\Sigma}_i \right)_0^{-1}\left(\boldsymbol{\mu}_i\right)_0 + \boldsymbol{\Sigma}_i^{-1}\boldsymbol{\mu}_i\right)\right]
 \end{align}
 Letting $\mathbf{M}_v = \left(\boldsymbol{\Sigma}_i \right)_0^{-1} + \boldsymbol{\Sigma}_i^{-1}$, and $\mathbf{m}_v = \left( \boldsymbol{\Sigma}_i \right)_0^{-1}\left(\boldsymbol{\mu}_i\right)_0 + \boldsymbol{\Sigma}_i^{-1}\boldsymbol{\mu}_i$, we have
 \begin{align}
     \nonumber
     V_i(\boldsymbol{\omega}_0, \boldsymbol{\omega}) = & \frac{1}{4}\text{Var}_{\boldsymbol{\omega}_0}\left[(\mathbf{y}_i - \mathbf{M}^{-1}_v\mathbf{m}_v)'\mathbf{M}_v(\mathbf{y}_i - \mathbf{M}^{-1}_v\mathbf{m}_v) \right]\\
     \nonumber
     = & \frac{1}{4}\left[2\text{tr}\left(\mathbf{M}_v\left(\boldsymbol{\Sigma}_i \right)_0\mathbf{M}_v\left(\boldsymbol{\Sigma}_i \right)_0\right) + 4\left(\left(\boldsymbol{\mu}_i\right)_0 - \mathbf{M}^{-1}_v\mathbf{m}_v\right)' \left(\boldsymbol{\Sigma}_i \right)_0\left(\left(\boldsymbol{\mu}_i\right)_0 - \mathbf{M}^{-1}_v\mathbf{m}_v\right) \right]\\
     \nonumber
     = & \frac{1}{2}\left[P + 2 \text{tr}\left( \boldsymbol{\Sigma}_i^{-1}\left(\boldsymbol{\Sigma}_i\right)_0\right) + \text{tr}\left( \boldsymbol{\Sigma}_i^{-1}\left(\boldsymbol{\Sigma}_i\right)_0\boldsymbol{\Sigma}_i^{-1}\left(\boldsymbol{\Sigma}_i\right)_0\right)\right]\\
     & + \left(\left(\boldsymbol{\mu}_i\right)_0- \boldsymbol{\mu}_i\right)'\left(\boldsymbol{\Sigma}_i^{-1}\left(\boldsymbol{\Sigma}_i\right)_0\boldsymbol{\Sigma}_i^{-1}\right)\left(\left(\boldsymbol{\mu}_i\right)_0 - \boldsymbol{\mu}_i\right).
     \label{Var_theta}
 \end{align}
 Let $\boldsymbol{\Omega}_\epsilon(\boldsymbol{\omega}_0) = \{\boldsymbol{\omega}: K_i(\boldsymbol{\omega}_0, \boldsymbol{\omega}) < \epsilon \; \text{for all i}\}$ for some $\epsilon > 0$. We will assume that $\sigma^2_0 > 0$. Consider the set $\mathcal{B}(\boldsymbol{\omega}_0) = \{\boldsymbol{\omega}: \frac{1}{a}((d_{il})_0 + \sigma^2_0) \le d_{il} + \sigma^2 \le a((d_{il})_0 + \sigma^2_0 ), \|\left(\boldsymbol{\mu}_i\right)_0 - \boldsymbol{\mu}_i\| \le b\}$ for some $a,b \in \mathbb{R}$ such that $a > 1$ and $b > 0$. Thus for a fixed $\boldsymbol{\omega}_0 \in \boldsymbol{\Omega}$ and any $\boldsymbol{\omega} \in \mathcal{C}(\boldsymbol{\omega}_0, \epsilon) :=\mathcal{B}(\boldsymbol{\omega}_0) \cap \boldsymbol{\Omega}_\epsilon(\boldsymbol{\omega}_0)$, we can bound $V_i(\boldsymbol{\omega}_0, \boldsymbol{\omega})$. We will let $\lambda_r(\mathbf{A})$ denote the $r^{th}$ eigenvalue of the matrix $\mathbf{A}$, and $\lambda_{max}(\mathbf{A})$ denote the largest eigenvalue of $\mathbf{A}$. Thus we have 
$$\text{tr}\left( \boldsymbol{\Sigma}_i^{-1}\left(\boldsymbol{\Sigma}_i\right)_0\right) \le P \lambda_{max}\left( \boldsymbol{\Sigma}_i^{-1}\left(\boldsymbol{\Sigma}_i\right)_0\right)\le \frac{Pa}{\sigma_0^2}\left(\max_l(d_{il} + \sigma_0^2)\right)$$
     $$\text{tr}\left( \boldsymbol{\Sigma}_i^{-1}\left(\boldsymbol{\Sigma}_i\right)_0\boldsymbol{\Sigma}_i^{-1}\left(\boldsymbol{\Sigma}_i\right)_0\right) \le \text{tr}\left( \boldsymbol{\Sigma}_i^{-1}\left(\boldsymbol{\Sigma}_i\right)_0\right)^2 \le \left(\frac{Pa}{\sigma_0^2}\left(\max_l(d_{il} + \sigma_0^2)\right)\right)^2$$
     \begin{align}
         \nonumber \left(\left(\boldsymbol{\mu}_i\right)_0- \boldsymbol{\mu}_i\right)'\left(\boldsymbol{\Sigma}_i^{-1}\left(\boldsymbol{\Sigma}_i\right)_0\boldsymbol{\Sigma}_i^{-1}\right)\left(\left(\boldsymbol{\mu}_i\right)_0 - \boldsymbol{\mu}_i\right) & \le b^2\lambda_{max}\left(\boldsymbol{\Sigma}_i^{-1}\left(\boldsymbol{\Sigma}_i\right)_0\boldsymbol{\Sigma}_i^{-1} \right)\\
         \nonumber & \le \frac{a^2b^2}{\sigma_0^4}\max_l(d_{il} + \sigma_0^2)
         \end{align}
Thus we can see that for any $\boldsymbol{\omega} \in \mathcal{C}(\boldsymbol{\omega}_0, \epsilon)$, 
 \begin{align}
    \nonumber
     V_i(\boldsymbol{\omega}_0, \boldsymbol{\omega}) \le & \frac{1}{2}\left[ P + 2 \left(\frac{Pa}{\sigma_0^2}\left(\max_l(d_{il} + \sigma_0^2)\right)\right) + \left(\frac{Pa}{\sigma_0^2}\left(\max_l(d_{il} + \sigma_0^2)\right)\right)^2\right]\\
     \nonumber
     & +\frac{a^2b^2}{\sigma_0^4}\max_l(d_{il} + \sigma_0^2)\\ 
     \nonumber = &  M_{V}.
 \end{align}
If we can bound $\lambda_{max}\left((d_{il})_0 + \sigma_0\right)$, then we have that $V_i(\boldsymbol{\omega}_0, \boldsymbol{\omega})$ is bounded. Let $\|\cdot\|_F$ be the Frobenius norm. Using the triangle inequality, we have
\begin{align}
    \nonumber \|(\boldsymbol{\Sigma}_i)_0\|_{F} & \le \sum_{k=1}^K\sum_{j=1}^K\sum_{p=1}^{KP} Z_{ij}Z_{ik}\|(\boldsymbol{\phi}_{kp})_0(\boldsymbol{\phi}_{jp})_0'\|_{F} + \sigma^2_0\|\mathbf{I}_P\|_F\\
    \nonumber & \le \sum_{k=1}^K\sum_{j=1}^K\sum_{p=1}^{KP} \|(\boldsymbol{\phi}_{kp})_0(\boldsymbol{\phi}_{jp})_0'\|_{F}  + \sigma^2_0\|\mathbf{I}_P\|_F\\
    \nonumber &  = \sum_{k=1}^K\sum_{j=1}^K\sum_{p=1}^{KP}  \sqrt{\text{tr}\left((\boldsymbol{\phi}_{jp})_0(\boldsymbol{\phi}_{kp})_0'(\boldsymbol{\phi}_{kp})_0(\boldsymbol{\phi}_{jp})_0'\right)} + \sqrt{P}\sigma^2_0\\
    \nonumber &  = \sum_{k=1}^K\sum_{j=1}^K\sum_{p=1}^{KP}  \sqrt{\text{tr}\left(\boldsymbol{\phi}_{kp}'(\boldsymbol{\phi}_{kp})_0(\boldsymbol{\phi}_{jp})_0'(\boldsymbol{\phi}_{jp})_0\right)} + \sqrt{P}\sigma^2_0 \\
    \nonumber &  = \sum_{k=1}^K\sum_{j=1}^K\sum_{p=1}^{KP} \|(\boldsymbol{\phi}_{jp})_0\|_2\|(\boldsymbol{\phi}_{kp})_0\|_2 + \sqrt{P}\sigma^2_0\\
    \nonumber & = M_{\boldsymbol{\Sigma}_0} < \infty,
\end{align}
for all $i \in \mathbb{N}$. Therefore, we know that $\lambda_{max}\left((d_{il})_0 + \sigma_0\right) \le M_{\boldsymbol{\Sigma}_0}$, as the Frobenius is the squareroot of the sum of the squared eigenvalues for a square matrix. Therefore, we have for all $i \in \mathbb{N}$ and $\boldsymbol{\omega} \in \mathcal{C}(\boldsymbol{\omega}_0, \epsilon)$, we have that
 $$\frac{V_i(\boldsymbol{\omega}_0, \boldsymbol{\omega})}{i^2} \le \frac{M_{V}}{i^2}.$$
Since $\sum_{i=1}^\infty \frac{1}{i^2} = \frac{\pi^2}{6}$, we have that $\sum_{i=1}^\infty \frac{M_{V}}{i^2} = \frac{M_{V}\pi^2}{6} < \infty$. Thus we have 
 \begin{equation}
 \sum_{i = 1}^\infty \frac{V_i(\boldsymbol{\omega}_0, \boldsymbol{\omega})}{i^2}  < \infty.
     \label{finite_variance}
 \end{equation}
 We will next show that for $\boldsymbol{\omega}_0 \in \boldsymbol{\Omega}$ and $\epsilon > 0$, $\boldsymbol{\Pi}(\mathcal{C}(\boldsymbol{\omega}_0),\epsilon) > 0$. Fix $\boldsymbol{\omega}_0 \in \boldsymbol{\Omega}$. While the $(\boldsymbol{\phi}_{jp})_0$ may not be identifiable (for any orthogonal matrix $\mathbf{H}$, $(\boldsymbol{\phi}_{jp})_0 \mathbf{H}\mathbf{H}' (\boldsymbol{\phi}_{kp})_0 = (\boldsymbol{\phi}_{jp})_0 '  (\boldsymbol{\phi}_{kp})_0$), let $(\boldsymbol{\phi}_{jp})_0$ be such that $\sum_{p=1}^{KP} (\boldsymbol{\phi}_{jp})_0 ' (\boldsymbol{\phi}_{kp})_0 = (\boldsymbol{\Sigma}_{jk})_0$. Thus we can define the following sets:
 \begin{align}
    \nonumber
     \boldsymbol{\Omega}_{\boldsymbol{\phi}_{jp}}  &= \left\{\boldsymbol{\phi}_{jp} :(\boldsymbol{\phi}_{jp})_0 \le \boldsymbol{\phi}_{jp} \le (\boldsymbol{\phi}_{jp})_0 + \epsilon_1\mathbf{1} \right\}\\
     \nonumber
     \boldsymbol{\Omega}_{\boldsymbol{\nu}_k}  &= \left\{\boldsymbol{\nu}_k : (\boldsymbol{\nu}_k)_0 \le \boldsymbol{\nu}_k \le (\boldsymbol{\nu}_k)_0 + \epsilon_2\mathbf{1}\right\} \\
     \nonumber
     \boldsymbol{\Omega}_{\sigma^2} &= \left\{\sigma^2:  \sigma^2_0 \le \sigma^2 \le (1 + \epsilon_1)\sigma^2_0 \right\}.
 \end{align}
We define $\boldsymbol{\epsilon}_{1jp}$ and $\boldsymbol{\epsilon}_{2k}$ such that each element of $\boldsymbol{\epsilon}_{1jp}$ is between 0 and $\epsilon_1$, and each element of $\boldsymbol{\epsilon}_{2k}$ is between 0 and $\epsilon_2$. Therefore $(\boldsymbol{\phi}_{jp})_0 + \boldsymbol{\epsilon}_{1jp} \in \boldsymbol{\Omega}_{\boldsymbol{\phi}_{jp}}$ and $(\boldsymbol{\nu}_k)_0 + \boldsymbol{\epsilon}_{2k} \in \boldsymbol{\Omega}_{\boldsymbol{\nu}_k}$. 
We will define $$\boldsymbol{\Omega}_{\boldsymbol{\Sigma}_{jk}}:= \left\{\left.\sum_{p=1}^{KP} \boldsymbol{\phi}_{jp}'\boldsymbol{\phi}_{kp}\right| \boldsymbol{\phi}_{jp} \in \boldsymbol{\Omega}_{\boldsymbol{\phi}_{jp}}, \boldsymbol{\phi}_{kp} \in \boldsymbol{\Omega}_{\boldsymbol{\phi}_{kp}} \right\}.$$
Thus for $\boldsymbol{\Sigma}_i$ such that $\boldsymbol{\phi}_{jp} \in \boldsymbol{\Omega}_{\boldsymbol{\phi}_{jp}}$ and $\sigma^2 \in \boldsymbol{\Omega}_{\sigma^2}$, we have that 
\begin{align}
    \nonumber
    \boldsymbol{\Sigma}_i = & \sum_{k=1}^K \sum_{j=1}^K Z_{ik}Z_{ij} \left(\sum_{p=1}^{KP}\left(\left((\boldsymbol{\phi}_{kp})_0 + \boldsymbol{\epsilon}_{1kp}\right)\left((\boldsymbol{\phi}_{jp})_0+ \boldsymbol{\epsilon}_{1jp}\right)'\right)\right) + (1 + \epsilon_\sigma)\sigma_0^2\mathbf{I}_P\\
    \nonumber
    = & (\boldsymbol{\Sigma}_i)_0 + \sum_{k=1}^K \sum_{j=1}^K\sum_{p=1}^{KP} Z_{ik}Z_{ij} \left(\left(\boldsymbol{\epsilon}_{1kp}\right)(\boldsymbol{\phi}_{jp})_0'\right) \\
    \nonumber
    & + \sum_{k=1}^K \sum_{j=1}^K\sum_{p=1}^{KP} Z_{ik}Z_{ij} \left((\boldsymbol{\phi}_{kp})_0 \left( \boldsymbol{\epsilon}_{1jp}\right)'\right) \\
    \nonumber
    & +\sum_{k=1}^K \sum_{j=1}^K\sum_{p=1}^{KP} Z_{ik}Z_{ij}\left(\left(\boldsymbol{\epsilon}_{1kp} \right)\left( \boldsymbol{\epsilon}_{1jp}\right)'\right) + \epsilon_{\sigma}\sigma^2_0\mathbf{I}_P\\
    \nonumber
    = & (\boldsymbol{\Sigma}_i)_0 + \tilde{\boldsymbol{\Sigma}_i},
\end{align}
for some $\epsilon_{kp}$ and $\epsilon_\sigma$ such that $0 < \epsilon_\sigma \le \epsilon_1$. Thus, letting $\boldsymbol{\zeta}_{jkp} = \left(\left(\boldsymbol{\epsilon}_{1kp}\right)(\boldsymbol{\phi}_{jp})_0'\right) +\left((\boldsymbol{\phi}_{kp})_0 \left( \boldsymbol{\epsilon}_{1jp}\right)'\right)$, we have
\begin{align}
    \nonumber
    \left|\left|Z_{ik}Z_{ij} \boldsymbol{\zeta}_{jkp}\right|\right|_F^2  \le & \left|\left| \boldsymbol{\zeta}_{jkp}\right|\right|_F^2\\
    \nonumber
    = & \text{tr}\left(\left(\left(\boldsymbol{\epsilon}_{1kp}\right)(\boldsymbol{\phi}_{jp})_0'\right)\left((\boldsymbol{\phi}_{jp})_0\left(\boldsymbol{\epsilon}_{1kp}\right)'\right)\right) \\
    \nonumber
    & + \text{tr}\left(\left(\left(\boldsymbol{\epsilon}_{1kp}\right)(\boldsymbol{\phi}_{jp})_0'\right)\left(\left( \boldsymbol{\epsilon}_{1jp}\right)(\boldsymbol{\phi}_{kp})_0 '\right)\right)\\
    \nonumber
    & + \text{tr}\left(\left((\boldsymbol{\phi}_{kp})_0 \left( \boldsymbol{\epsilon}_{1jp}\right)'\right)\left((\boldsymbol{\phi}_{jp})_0\left(\boldsymbol{\epsilon}_{1kp}\right)'\right)\right) \\
    \nonumber
    & + \text{tr}\left(\left((\boldsymbol{\phi}_{kp})_0 \left( \boldsymbol{\epsilon}_{1jp}\right)'\right)\left(\left( \boldsymbol{\epsilon}_{1jp}\right)(\boldsymbol{\phi}_{kp})_0 '\right)\right)\\
    \nonumber
     \le &  \epsilon_1^2\text{tr}\left((\boldsymbol{\phi}_{jp})_0'(\boldsymbol{\phi}_{jp})_0\mathbf{1}'\mathbf{1}\right) \\
    \label{simp_1} & + 2\text{tr}\left((\boldsymbol{\phi}_{jp})_0'\left( \boldsymbol{\epsilon}_{1jp}\right)(\boldsymbol{\phi}_{kp})_0 '\left(\boldsymbol{\epsilon}_{1kp}\right)\right)\\
    \nonumber
    & + \epsilon_1^2\text{tr}\left(\mathbf{1}'\mathbf{1}(\boldsymbol{\phi}_{kp})_0 '(\boldsymbol{\phi}_{kp})_0\right).
\end{align}
Using the Cauchy-Schwarz inequality, we can simplify equation (\ref{simp_1}), such that
\begin{align}
    \nonumber (\ref{simp_1}) & = 2\langle(\boldsymbol{\phi}_{jp})_0, \boldsymbol{\epsilon}_{1jp} \rangle \langle (\boldsymbol{\phi}_{kp})_0, \boldsymbol{\epsilon}_{1kp}  \rangle \\
    \nonumber & \le 2\|(\boldsymbol{\phi}_{jp})_0\|_2 \|\boldsymbol{\epsilon}_{1jp}\|_2 \|(\boldsymbol{\phi}_{kp})_0\|_2 \|\boldsymbol{\epsilon}_{1kp}\|_2\\
    \nonumber& \le 2\epsilon_1^2P \|(\boldsymbol{\phi}_{jp})_0\|_2  \|(\boldsymbol{\phi}_{kp})_0\|_2 .
\end{align}
Letting 
\begin{align}
    \nonumber \tilde{M}_{jkp} & = P\left[\|(\boldsymbol{\phi}_{jp})_0\|_2^2 + \|(\boldsymbol{\phi}_{kp})_0\|_2^2 + 2 \|(\boldsymbol{\phi}_{jp})_0\|_2  \|(\boldsymbol{\phi}_{kp})_0\|_2\right] ,
\end{align} we have
$$\left|\left|Z_{ik}Z_{ij} \boldsymbol{\zeta}_{jkp}\right|\right|_F^2  \le \epsilon_1^2 \tilde{M}_{jkp}.$$

In a similar fashion, we can show that 
 $$\|Z_{ik}Z_{ij} \left(\left(\left(\boldsymbol{\epsilon}_{1kp} \right)\left(\boldsymbol{\epsilon}_{1jp}\right)'\right)\right)\|_F^2 \le \epsilon_1^2P$$
 and
 $$\|\epsilon_{\sigma}\sigma^2_0\mathbf{I}_P\|^2_F \le \epsilon_1^2\sigma_0^4 P.$$
 By using the triangle inequality we have
 \begin{equation}
     \|\tilde{\boldsymbol{\Sigma}_i}\|_F \le \epsilon_1\left(\sum_{j=1}^K\sum_{k=1}^K\sum_{p=1}^{KP}\left(\sqrt{\tilde{M}_{jkp}}\right) + JK^2P^{3/2} + \sigma_0^2 \sqrt{P} \right) := \epsilon_1M_{\boldsymbol{\Sigma}}
     \label{Frobenius_Bound}
 \end{equation}
 for all $i \in \mathbb{N}$. By the Wielandt-Hoffman Theorem (\citet{golub2013matrix} Theorem 8.1.4), we have that
 $$\sum_{p=1}^P \left(\lambda_p\left((\boldsymbol{\Sigma}_i)_0 + \tilde{\boldsymbol{\Sigma}}_i\right) -\lambda_p\left((\boldsymbol{\Sigma}_i)_0\right)  \right)^2 \le \|\tilde{\boldsymbol{\Sigma}}_i\|^2_F,$$
 which implies that
 \begin{equation}
 \label{eq: bound_eigen}
     \max_{p}\left|\lambda_p\left((\boldsymbol{\Sigma}_i)_0 + \tilde{\boldsymbol{\Sigma}}_i\right) -\lambda_p\left((\boldsymbol{\Sigma}_i)_0\right) \right| \le \|\tilde{\boldsymbol{\Sigma}}_i\|_F
 \end{equation}
 where $\lambda_p(\mathbf{A})$ are the eigenvalues of the matrix $\mathbf{A}$.
 By using equation (\ref{Frobenius_Bound}), we can bound the log-determinant of the ratio of the two covariance matrices as follows
\begin{align}
     \nonumber
     \log \left(\frac{|\boldsymbol{\Sigma}_i|}{|(\boldsymbol{\Sigma}_i)_0|} \right) & = \log \left(\frac{\prod_{p=1}^P\lambda_p\left((\boldsymbol{\Sigma}_i)_0 + \tilde{\boldsymbol{\Sigma}}_i\right)}{\prod_{p=1}^P\lambda_p\left((\boldsymbol{\Sigma}_i)_0\right)} \right)\\
     \nonumber
     & \le \log \left(\prod_{p=1}^P\frac{\left((d_{ip})_0 + \sigma_0^2\right) + \epsilon_1M_{\boldsymbol{\Sigma}}}{(d_{ip})_0 + \sigma_0^2}\right) \\
     & \le P\log \left( 1 + \frac{\epsilon_1M_{\boldsymbol{\Sigma}}}{\sigma_0^2}\right).
     \label{Bound_log_det}
\end{align}
We can also bound $\text{tr}\left(\boldsymbol{\Sigma}_i^{-1}\left(\boldsymbol{\Sigma}_i\right)_0\right)$. To do this, we will first consider the spectral norm, defined as $\|\mathbf{A}\|_2 = \sqrt{\mathbf{A}^*\mathbf{A}}$ for some matrix $\mathbf{A}$. In the case where $\mathbf{A}$ is symmetric, we have that $\|\mathbf{A}\|_2 = \max_{r}|\lambda_r(\mathbf{A})|$. By the submultiplicative property of induced norms, we have that
\begin{equation}
\label{eq: submult}
    \max_{p}|\lambda_p(\mathbf{A}\mathbf{B})| = \|\mathbf{A}\mathbf{B}\|_2 \le \|\mathbf{A}\|_2 \|\mathbf{B}\|_2 = \max_{p}|\lambda_p(\mathbf{A})| \max_{p}|\lambda_p(\mathbf{B})|,
\end{equation}
for two symmetric matrices $\mathbf{A}$ and $\mathbf{B}$. By using the Sherman–Morrison–Woodbury formula, we can see that
\begin{align}
    \nonumber \boldsymbol{\Sigma}_i^{-1} &= \left((\boldsymbol{\Sigma}_i)_0 + \tilde{\boldsymbol{\Sigma}}_i \right)^{-1}\\
    \nonumber &= (\boldsymbol{\Sigma}_i)_0^{-1} - (\boldsymbol{\Sigma}_i)_0^{-1}\tilde{\boldsymbol{\Sigma}}_i\left((\boldsymbol{\Sigma}_i)_0 + \tilde{\boldsymbol{\Sigma}}_i \right)^{-1}.
\end{align}
Thus, we have that 
\begin{equation}
    \label{eq: SMW_formula}
    \boldsymbol{\Sigma}_i^{-1}\left(\boldsymbol{\Sigma}_i\right)_0 =  \mathbf{I}_R - (\boldsymbol{\Sigma}_i)_0^{-1}\tilde{\boldsymbol{\Sigma}}_i\left((\boldsymbol{\Sigma}_i)_0 + \tilde{\boldsymbol{\Sigma}}_i \right)^{-1}\left(\boldsymbol{\Sigma}_i\right)_0.
\end{equation}
Using equation (\ref{eq: submult}), we would like to bound the magnitude of the eigenvalues of \\
$(\boldsymbol{\Sigma}_i)_0^{-1}\tilde{\boldsymbol{\Sigma}}_i\left((\boldsymbol{\Sigma}_i)_0 + \tilde{\boldsymbol{\Sigma}}_i \right)^{-1}\left(\boldsymbol{\Sigma}_i\right)_0$. We know that $$\max_p\left|\lambda_p\left( \left((\boldsymbol{\Sigma}_i)_0 + \tilde{\boldsymbol{\Sigma}}_i \right)^{-1}\right)\right| \le \frac{1}{\sigma_0^2}$$
and
$$\max_p\left|\lambda_p(\tilde{\boldsymbol{\Sigma}}_i )\right| \le \epsilon_1 M_{\boldsymbol{\Sigma}},$$
with the second inequality coming from equation (\ref{Frobenius_Bound}). From equation (\ref{eq: SMW_formula}) and basic properties of the trace, we have that
\begin{align}
    \nonumber \text{tr}\left(\boldsymbol{\Sigma}_i^{-1}\left(\boldsymbol{\Sigma}_i\right)_0 \right) & = \text{tr}\left(\mathbf{I}_P - (\boldsymbol{\Sigma}_i)_0^{-1}\tilde{\boldsymbol{\Sigma}}_i\left((\boldsymbol{\Sigma}_i)_0 + \tilde{\boldsymbol{\Sigma}}_i \right)^{-1}\left(\boldsymbol{\Sigma}_i\right)_0 \right) \\
    \nonumber & = \text{tr}\left(\mathbf{I}_P \right) - \text{tr}\left(  \tilde{\boldsymbol{\Sigma}}_i\left((\boldsymbol{\Sigma}_i)_0 + \tilde{\boldsymbol{\Sigma}}_i \right)^{-1}\left(\boldsymbol{\Sigma}_i\right)_0(\boldsymbol{\Sigma}_i)_0^{-1}\right) \\
    \nonumber & = \text{tr}\left(\mathbf{I}_P \right) - \text{tr}\left(  \tilde{\boldsymbol{\Sigma}}_i\left((\boldsymbol{\Sigma}_i)_0 + \tilde{\boldsymbol{\Sigma}}_i \right)^{-1}\right)
\end{align}
Thus, using the fact that the trace of a matrix is the sum of its eigenvalues, we have that 
    $$\text{tr}\left(\boldsymbol{\Sigma}_i^{-1}\left(\boldsymbol{\Sigma}_i\right)_0 \right) \le P + P\max_p\left|\lambda_p\left(\tilde{\boldsymbol{\Sigma}}_i\left((\boldsymbol{\Sigma}_i)_0 + \tilde{\boldsymbol{\Sigma}}_i \right)^{-1} \right)\right|.$$
Using the submultiplicative property stated in equation (\ref{eq: submult}), we have
\begin{equation}
\label{Bound_trace}
    \text{tr}\left(\boldsymbol{\Sigma}_i^{-1}\left(\boldsymbol{\Sigma}_i\right)_0 \right) \le P + \frac{P\epsilon_1M_{\boldsymbol{\Sigma}}}{\sigma_0^2}.
\end{equation}

Lastly, we can bound the quadratic term in $K_i(\boldsymbol{\omega}_0, \boldsymbol{\omega})$ in the following way:
\begin{align}
    \nonumber
    \left(\left(\boldsymbol{\mu}_i\right)_0- \boldsymbol{\mu}_i\right)'\left(\boldsymbol{\Sigma}_i\right)^{-1}\left(\left(\boldsymbol{\mu}_i\right)_0 - \boldsymbol{\mu}_i\right) &\le \left|\left|\left(\boldsymbol{\mu}_i\right)_0- \boldsymbol{\mu}_i \right| \right|_2^2 \max_{p} \lambda_p(\left(\boldsymbol{\Sigma}_i)^{-1}\right)\\
    \nonumber & \le \frac{1}{\sigma^2}\sum_{k=1}^K\|(\boldsymbol{\nu}_k)_0 - \boldsymbol{\nu}_k\|^2_2 \\
    \nonumber & = \frac{1}{\sigma^2}\sum_{k=1}^K\boldsymbol{\epsilon}_{2k}'\boldsymbol{\epsilon}_{2k}  \\
    & \le \frac{KP\epsilon_2^2}{\sigma^2_0}.
     \label{quadratic_bound}
\end{align}
Thus letting 
\begin{equation}
\epsilon_1 < \min\left\{\frac{\sigma^2_0}{M_{\boldsymbol{\Sigma}}}\left(\text{exp}\left( \frac{2\epsilon}{3P} \right) - 1\right),\frac{2\epsilon\sigma^2_0}{3PM_{\boldsymbol{\Sigma}}}\right\}
\label{epsilon_1}
\end{equation}
and 
\begin{equation}
    \epsilon_2 < \sqrt{\frac{2\sigma_0^2\epsilon}{3KP}},
\end{equation}
we have from equations \ref{Bound_log_det}, \ref{Bound_trace}, and \ref{quadratic_bound} that
$$K_i(\boldsymbol{\omega}_0,\boldsymbol{\omega}) < \epsilon \text{ for all } \boldsymbol{\omega} \in \boldsymbol{\Omega}_1$$
where $\boldsymbol{\Omega}_1 := \left(\bigtimes_{j=1}^K \bigtimes_{k = 1}^{K} \boldsymbol{\Omega}_{\boldsymbol{\Sigma}_{jk}} \right) \times \left(\bigtimes_{k=1}^K \boldsymbol{\Omega}_{\boldsymbol{\nu}_k} \right) \times \boldsymbol{\Omega}_{\sigma^2}$.
Letting $a > \max \left\{1 + \frac{\epsilon_1M_{\boldsymbol{\Sigma}}}{\sigma_0^2}, \left(1 - \frac{\epsilon_1M_{\boldsymbol{\Sigma}}}{\sigma_0^2} \right)^{-1} \right\}$ and $b > \sqrt{KP\epsilon_2^2
}$ in the definition of $\mathcal{C}(\boldsymbol{\omega}_0, \epsilon)$, we have that $ \boldsymbol{\Omega}_1 \subset \mathcal{C}(\boldsymbol{\omega}_0, \epsilon)$. Let $H_{\boldsymbol{\phi}}$ be the set of hyper-parameters corresponding to the $\boldsymbol{\phi}$ parameters, and let $\boldsymbol{\Pi}(\boldsymbol{\eta}_{\boldsymbol{\phi}})$ be the prior distribution on $\boldsymbol{\eta}_{\boldsymbol{\phi}}\in H_{\boldsymbol{\phi}}$. Thus we have that
\begin{align}
    \nonumber
    \boldsymbol{\Pi}\left(\boldsymbol{\omega} \in  \mathcal{C}(\boldsymbol{\omega}_0, \epsilon)\right) & \ge \int_{H_{\boldsymbol{\phi}}} \prod_{j=1}^K \prod_{p=1}^{KP} \prod_{r=1}^P \int_{(\phi_{jrp})_0}^{(\phi_{jrp})_0 + \epsilon_1} \sqrt{\frac{\gamma_{jrp}\tilde{\tau}_{pj}}{2\pi}} \text{exp}\left\{-\frac{\gamma_{jrp}\tilde{\tau}_{pj}}{2}\phi_{jrp}^2\right\} \text{d}\phi_{jrp} \text{d} \boldsymbol{\Pi}(\boldsymbol{\eta}_{\boldsymbol{\phi}})\\
    \nonumber & \times \prod_{k=1}^K \int_0^\infty \int_{(\boldsymbol{\nu}_k)_0}^{(\boldsymbol{\nu}_k)_0 + \epsilon_2\mathbf{1}}\left(\frac{\tau_k}{2\pi}\right)^{P/2} \text{exp}\left\{\frac{\tau_k}{2}\boldsymbol{\nu}_k' \boldsymbol{\nu}_k \right\} \text{d}\boldsymbol{\nu}_k \text{d}\boldsymbol{\Pi}(\tau_k)\\
    \nonumber
    & \times \int_{\sigma_0^2}^{(1 + \epsilon_1) \sigma_0^2}\frac{\beta_0^{\alpha_0}}{\Gamma(\alpha_0)}(\sigma^2)^{-\alpha_0 - 1} \text{exp} \left\{- \frac{\beta_0}{\sigma^2} \right\} \text{d} \sigma^2.
\end{align}
Restricting the hyper-parameters of $\boldsymbol{\phi}$ to only a subset of the support, say $\tilde{H}_{\boldsymbol{\phi}}$, where  $$\tilde{H}_{\boldsymbol{\phi}} = \left\{\boldsymbol{\eta}_{\boldsymbol{\phi}}: \frac{1}{10} \le \gamma_{jrp} \le 10, 1 \le \delta_{pj} \le 2, 1 \le a_{1j} \le 10, 1 \le a_{2j} \le 10\right\},$$ we can see that there exists a $M_{\phi_{jrp}} > 0$ such that $$\sqrt{\frac{\gamma_{jrp}\tilde{\tau}_{pj}}{2\pi}} \text{exp}\left\{-\frac{\gamma_{jrp}\tilde{\tau}_{pj}}{2}\phi_{jrp}^2\right\} \ge M_{\phi_{jrp}},$$ for all $\phi_{jrp} \in [(\phi_{jrp})_0,(\phi_{jrp})_0 + \epsilon_1]$. Similarly, we can find a lower bound $M_{\tilde{H}_{\boldsymbol{\phi}}} > 0$, such that $$\int_{\tilde{H}_{\boldsymbol{\phi}}} \text{d}(\boldsymbol{\eta}_{\boldsymbol{\phi}}) \ge M_{\tilde{H}_{\boldsymbol{\phi}}}.$$ Similarly, if we bound $\tau_k$ such that $\frac{1}{10} \le \tau_k \le 10$, it is easy to see that there exists constants $M_{\boldsymbol{\nu}_k}, M_{\tau_k}, M_{\sigma^2} > 0$ such that
$$\left(\frac{\tau_k}{2\pi}\right)^{P/2} \text{exp}\left\{\frac{\tau_k}{2}\boldsymbol{\nu}_k' \boldsymbol{\nu}_k \right\} \ge  M_{\boldsymbol{\nu}_k},$$
for all $\boldsymbol{\nu}_k \in [(\boldsymbol{\nu}_k)_0, (\boldsymbol{\nu}_k)_0 + \epsilon_2\mathbf{1}]$,
$$\int_{\frac{1}{10}}^{10} \boldsymbol{\Pi}(\tau_k) \ge M_{\tau_k},$$
and 
$$ \frac{\beta_0^{\alpha_0}}{\Gamma(\alpha_0)}(\sigma^2)^{-\alpha_0 - 1} \text{exp} \left\{- \frac{\beta_0}{\sigma^2} \right\} \ge M_{\sigma^2}$$
for all $\sigma^2 \in [\sigma^2_0, (1 + \epsilon_1)\sigma^2_0]$. Therefore we have that 
\begin{align}
    \nonumber
    \boldsymbol{\Pi}\left(\boldsymbol{\omega} \in  \mathcal{C}(\boldsymbol{\omega}_0, \epsilon)\right) & \ge M_{\tilde{H}_{\boldsymbol{\phi}}}  \prod_{j=1}^K \prod_{p=1}^{KP} \prod_{r=1}^{P} \epsilon_1 M_{\phi_{jrp}} \\
    \nonumber
    & \times \prod_{k=1}^K M_{\tau_k} \epsilon_2^PM_{\boldsymbol{\nu}_k} \\
    \nonumber
    & \times \epsilon_1\sigma^2_0 M_{\sigma^2_0}\\
    \nonumber
    & > 0.
\end{align}
Therefore, for $\epsilon > 0$, there exists $a$ and $b$ such that $\sum_{i=1}^\infty \frac{V_i(\boldsymbol{\omega}_0, \boldsymbol{\omega})}{i^2} < \infty$ for any $\boldsymbol{\omega} \in \mathcal{C}(\boldsymbol{\omega}_0, \epsilon)$ and $\boldsymbol{\Pi}\left(\boldsymbol{\omega} \in  \mathcal{C}(\boldsymbol{\omega}_0, \epsilon)\right) > 0$.

\subsection{Proof of Lemma 2.2}
 Following the notation of \citet{ghosal2017fundamentals}, we will let $P_{\boldsymbol{\omega}_0}^{(N)}$ denote the joint distribution of $\mathbf{y}_1, \dots, \mathbf{y}_N$ at $\boldsymbol{\omega}_0\in\boldsymbol{\Omega}$. In order to show that the posterior distribution, $\boldsymbol{\Pi}_N(. | \mathbf{y}_1, \dots, \mathbf{y}_N)$, is weakly consistent at $\boldsymbol{\omega}_0 \in \boldsymbol{\Omega}$, we need to show that $\boldsymbol{\Pi}_N(\mathcal{U}^c| \mathbf{y}_1, \dots, \mathbf{y}_N) \rightarrow 0$  a.s. $[P_{\boldsymbol{\omega}_0}]$ for every weak neighborhood, $\mathcal{U}$ of $\boldsymbol{\omega}_0$. Following a similar notation to \citet{ghosal2017fundamentals}, let $\psi_N$ be measurable mappings, $\psi_N: \boldsymbol{\mathcal{S}}^N \times \boldsymbol{\mathcal{Z}}^N \rightarrow [0,1]$, where $\boldsymbol{\mathcal{Z}}$ is the sample space of $\{Z_{i1}, \dots, Z_{iK}\}$. Let $\psi_N(\mathbf{y}_1, \dots, \mathbf{y}_N, \mathbf{z}_1, \dots, \mathbf{z}_N)$ be the corresponding test function, and \linebreak $P_{\boldsymbol{\omega}}^N\psi_N = \mathbb{E}_{P_{\boldsymbol{\omega}}^N}\psi_N(\mathbf{y}_1, \dots, \mathbf{y}_N, \mathbf{z}_1,\dots, \mathbf{z}_N) = \int \psi_N \text{d}P_{\boldsymbol{\omega}}^N$, where $P_{\boldsymbol{\omega}}^N$ denotes the joint distribution on $\mathbf{y}_1, \dots, \mathbf{y}_N$ with parameters $\boldsymbol{\omega}$. Suppose there exists tests $\psi_N$ such that $P_{\boldsymbol{\omega}_0}^{N}\psi_N \rightarrow 0$, and $\text{sup}_{\boldsymbol{\omega} \in \mathcal{U}^c} P_{\boldsymbol{\omega}}^{N}(1-\psi_N) \rightarrow 0$. Since $\psi_N(\mathbf{y}_1, \dots, \mathbf{y}_N, \mathbf{z}_1,\dots, \mathbf{z}_N) \in [0,1]$, we have 
 \begin{align}
     \nonumber
     \boldsymbol{\Pi}_n(U^c| \mathbf{y}_1, \dots, \mathbf{y}_N) & \le \boldsymbol{\Pi}_n(U^c| \mathbf{y}_1, \dots, \mathbf{y}_N) + \psi_N(\mathbf{y}_1, \dots, \mathbf{y}_N)\left( 1 - \boldsymbol{\Pi}_n(U^c| \mathbf{y}_1, \dots, \mathbf{y}_N)\right)\\
     & = \psi_N(\mathbf{y}_1, \dots, \mathbf{y}_N) + \frac{\left( 1- \psi_N(\mathbf{y}_1, \dots, \mathbf{y}_N)\right)\int_{U^c}\prod_{i=1}^N \frac{f_i(\mathbf{y}_i; \boldsymbol{\omega})}{f_i(\mathbf{y}_i; \boldsymbol{\omega}_0)}\text{d}\boldsymbol{\Pi}(\boldsymbol{\omega})}{\int_{\boldsymbol{\Omega}}\prod_{i=1}^N \frac{f_i(\mathbf{y}_i; \boldsymbol{\omega})}{f_i(\mathbf{y}_i; \boldsymbol{\omega}_0)}\text{d}\boldsymbol{\Pi}(\boldsymbol{\omega})}.
     \label{posterior}
 \end{align}
 To show that $\boldsymbol{\Pi}_n(U^c| \mathbf{y}_1, \dots, \mathbf{y}_N) \rightarrow 0$, it is sufficient to show the following three conditions:
 \begin{enumerate}
     \item $\psi_N(\mathbf{y}_1, \dots, \mathbf{y}_N,\mathbf{z}_1,\dots, \mathbf{z}_N) \rightarrow 0$ a.s. $[P_{\boldsymbol{\omega}_0}]$,
     \item $e^{\beta_1 N}\left( 1- \psi_N(\mathbf{y}_1, \dots, \mathbf{y}_N,\mathbf{z}_1,\dots, \mathbf{z}_N)\right)\int_{\mathcal{U}^c}\prod_{i=1}^N \frac{f_i(\mathbf{y}_i; \boldsymbol{\omega})}{f_i(\mathbf{y}_i; \boldsymbol{\omega}_0)}\text{d}\boldsymbol{\Pi}(\boldsymbol{\omega}) \rightarrow 0$ a.s. $[P_{\boldsymbol{\omega}_0}]$ for some $\beta_1 > 0$,
     \item $e^{\beta N}\left(\int_{\boldsymbol{\Omega}}\prod_{i=1}^N \frac{f_i(\mathbf{y}_i; \boldsymbol{\omega})}{f_i(\mathbf{y}_i; \boldsymbol{\omega}_0)}\text{d}\boldsymbol{\Pi}(\boldsymbol{\omega}) \right) \rightarrow \infty$ a.s. $[P_{\boldsymbol{\omega}_0}]$ for all $\beta > 0$.
 \end{enumerate}
 
 We will start by proving (c). Fix $\beta > 0$. Thus we have
 $$e^{\beta N}\left(\int_{\boldsymbol{\Omega}}\prod_{i=1}^N \frac{f_i(\mathbf{y}_i; \boldsymbol{\omega})}{f_i(\mathbf{y}_i; \boldsymbol{\omega}_0)}\text{d}\boldsymbol{\Pi}(\boldsymbol{\omega}) \right)= e^{\beta N}\left(\int_{\boldsymbol{\Omega}}\text{exp}\left[-\sum_{i=1}^N \log \left(\frac{f_i(\mathbf{y}_i; \boldsymbol{\omega}_0)}{f_i(\mathbf{y}_i; \boldsymbol{\omega})}\right)\right]\text{d}\boldsymbol{\Pi}(\boldsymbol{\omega})\right).$$
 By Fatou's Lemma, we have
 \begin{align}
    \nonumber & 
     \liminf_{N \rightarrow \infty} \int_{\boldsymbol{\Omega}}\text{exp}\left[\beta N -\sum_{i=1}^N \log \left(\frac{f_i(\mathbf{y}_i; \boldsymbol{\omega}_0)}{f_i(\mathbf{y}_i; \boldsymbol{\omega})}\right)\right]\text{d}\boldsymbol{\Pi}(\boldsymbol{\omega}) \\
     \nonumber \ge & \int_{\boldsymbol{\Omega}}\liminf_{N \rightarrow \infty}\text{exp}\left[\beta N -\sum_{i=1}^N \log \left(\frac{f_i(\mathbf{y}_i; \boldsymbol{\omega}_0)}{f_i(\mathbf{y}_i; \boldsymbol{\omega})}\right)\right]\text{d}\boldsymbol{\Pi}(\boldsymbol{\omega})
 \end{align} 
 Let $\beta >\epsilon > 0$ and $a,b > 0$ be defined such that lemma 3.1 holds. Since $\mathcal{C}(\boldsymbol{\omega}_0, \epsilon) \subset \boldsymbol{\Omega}$, we have that
 \begin{align}
     \nonumber & 
     \int_{\boldsymbol{\Omega}}\liminf_{N \rightarrow \infty}\text{exp}\left[\beta N -\sum_{i=1}^N \log \left(\frac{f_i(\mathbf{y}_i; \boldsymbol{\omega}_0)}{f_i(\mathbf{y}_i; \boldsymbol{\omega})}\right)\right]\text{d}\boldsymbol{\Pi}(\boldsymbol{\omega})\\
     \nonumber \ge & \int_{\mathcal{C}(\boldsymbol{\omega}_0, \epsilon)}\liminf_{N \rightarrow \infty}\text{exp}\left[\beta N -\sum_{i=1}^N \log \left(\frac{f_i(\mathbf{y}_i; \boldsymbol{\omega}_0)}{f_i(\mathbf{y}_i; \boldsymbol{\omega})}\right)\right]\text{d}\boldsymbol{\Pi}(\boldsymbol{\omega})
 \end{align}
 By Kolmogorov's strong law of large numbers for non-identically distributed  random variables, we have that 
 $$\frac{1}{N}\sum_{i=1}^N\left(\Lambda_i(\boldsymbol{\omega}_0, \boldsymbol{\omega}) - K_i(\boldsymbol{\omega}_0, \boldsymbol{\omega}) \right) \rightarrow 0$$
 a.s. $[P_{\boldsymbol{\omega}_0}]$. Thus for each $\boldsymbol{\omega} \in \mathcal{C}(\boldsymbol{\omega}_0, \epsilon)$, with $P_{\boldsymbol{\omega}_0}$-probability  1, 
 $$\frac{1}{N}\sum_{i=1}^N\Lambda_i(\boldsymbol{\omega}_0, \boldsymbol{\omega}) \rightarrow \mathbb{E}(\overline{K_i(\boldsymbol{\omega}_0, \boldsymbol{\omega})}) < \epsilon < B,$$
 since $\boldsymbol{\omega} \in \mathcal{C}(\boldsymbol{\omega}_0, \epsilon)$. Therefore, we have that
 $$\int_{\mathcal{C}(\boldsymbol{\omega}_0, \epsilon)}\liminf_{N \rightarrow \infty}\text{exp}\left[\beta N -\sum_{i=1}^N \log \left(\frac{f_i(\mathbf{y}_i; \boldsymbol{\omega}_0)}{f_i(\mathbf{y}_i; \boldsymbol{\omega})}\right)\right]\text{d}\boldsymbol{\Pi}(\boldsymbol{\omega}) \ge \int_{\mathcal{C}(\boldsymbol{\omega}_0, \epsilon)} \inf_{N \rightarrow \infty} \text{exp}\left\{N(\beta- \epsilon) \right\}\text{d}\boldsymbol{\Pi}(\boldsymbol{\omega}).$$
Since $\beta - \epsilon > 0$, and $\boldsymbol{\Pi}\left(\theta \in \mathcal{C}(\boldsymbol{\omega}_0, \epsilon)\right) > 0$ (lemma 3.1), we have that 
\begin{equation}
    e^{\beta N}\left(\int_{\boldsymbol{\Omega}}\prod_{i=1}^N \frac{f_i(\mathbf{y}_i; \boldsymbol{\omega})}{f_i(\mathbf{y}_i; \boldsymbol{\omega}_0)}\text{d}\boldsymbol{\Pi}(\boldsymbol{\omega}) \right) \rightarrow \infty
    \label{part_c}
\end{equation}
a.s. $[P_{\boldsymbol{\omega}_0}]$ for all $\beta > 0$. We will now show that exists measurable mappings such that $P_{\boldsymbol{\omega}_0}^{N}\psi_N \rightarrow 0$ and $\text{sup}_{\boldsymbol{\omega} \in \mathcal{U}^c} P_{\boldsymbol{\omega}}^{N}(1-\psi_N) \rightarrow 0$. Consider weak neighborhoods $\mathcal{U}$ of $\boldsymbol{\omega}_0$ of the form
\begin{equation}
    \mathcal{U} = \left\{\boldsymbol{\omega}: \left|\int f_i \text{d}P_{\boldsymbol{\omega}} -\int f_i \text{d}P_{\boldsymbol{\omega}_0}  \right| < \epsilon_i, \;\; i = 1,2, \dots, r \right\},
    \label{weak_neighborhood}
\end{equation}
where $r \in \mathbb{N}$, $\epsilon_i > 0$, and $f_i$ are continuous functions such that $f_i:\boldsymbol{\mathcal{S}} \times \boldsymbol{\mathcal{Z}} \rightarrow [0,1]$. As shown in \citet{GhoshNonParametrics}, for any particular $f_i$ and $\epsilon_i > 0$,  $\left|\int f_i \text{d}P_{\boldsymbol{\omega}} -\int f_i \text{d}P_{\boldsymbol{\omega}_0}  \right| < \epsilon_i$ iff $\int f_i \text{d}P_{\boldsymbol{\omega}} -\int f_i \text{d}P_{\boldsymbol{\omega}_0} < \epsilon_i$ and $\int (1 -f_i) \text{d}P_{\boldsymbol{\omega}} -\int (1 -f_i) \text{d}P_{\boldsymbol{\omega}_0} < \epsilon$. Since $\tilde{f}_i:=(1 - f_i)$ is still a continuous function such that $\tilde{f}_i:\boldsymbol{\mathcal{S}} \times \boldsymbol{\mathcal{Z}} \rightarrow [0,1]$, we can rewrite equation (\ref{weak_neighborhood}) as
\begin{equation}
    \mathcal{U} = \cap_{i=1}^{2r} \left\{\boldsymbol{\omega}: \int g_i \text{d}P_{\boldsymbol{\omega}} -\int g_i \text{d}P_{\boldsymbol{\omega}_0} < \epsilon_i \right\},
\end{equation}
where $g_i$ are continuous functions such that $g_i:\boldsymbol{\mathcal{S}} \times \boldsymbol{\mathcal{Z}} \rightarrow [0,1]$ and $\epsilon_i > 0$. Following \citet{ghosal2017fundamentals}, it can be shown by Hoeffding's inequality that using the test function $\tilde{\psi}$, defined as
\begin{equation}
    \tilde{\psi}_{iN}(\mathbf{y}_1, \dots, \mathbf{y}_N, \mathbf{z}_1, \dots, \mathbf{z}_N) := \mathbbm{1}\left\{\frac{1}{N}\sum_{j=1}^N g_i(\mathbf{y}_j, \mathbf{z}_j) > \int g_i \text{d}P_{\boldsymbol{\omega}_0} + \frac{\epsilon_i}{2}\right\},
\end{equation}
leads to 
$$\int \tilde{\psi}_{iN}(\mathbf{y}_1, \dots, \mathbf{y}_N, \mathbf{z}_1, \dots, \mathbf{z}_N) \text{d}P_{\boldsymbol{\omega}_0} \le e^{-N\epsilon_i^2/2}$$ 
and 
$$\int \left(1 - \tilde{\psi}_{iN}(\mathbf{y}_1, \dots, \mathbf{y}_N, \mathbf{z}_1, \dots, \mathbf{z}_N)\right) \text{d}P_{\boldsymbol{\omega}} \le e^{-N\epsilon_i^2/2}$$
for any $\boldsymbol{\omega} \in \mathcal{U}^c$. Let $\boldsymbol{\psi}_n = \max_{i}\tilde{\psi}_{iN}$ be our test function and $\epsilon = \min_{i}\epsilon_i$. Using the fact that $\mathbb{E}(\max_{i}\tilde{\psi}_{iN}) \le \sum_{i}\mathbb{E}(\tilde{\psi}_{iN})$ and $\mathbb{E}(1 - \max_{i}\tilde{\psi}_{iN}) \le \mathbb{E}(1 - \tilde{\psi}_{iN})$, we have
\begin{equation}
    \int \psi_N(\mathbf{y}_1, \dots, \mathbf{y}_N, \mathbf{z}_1, \dots, \mathbf{z}_N) \text{d}P_{\boldsymbol{\omega}_0} \le (2r)e^{-N\epsilon^2/2}
    \label{test_bound1}
\end{equation}
and
\begin{equation}
    \int \left(1 - \psi_N(\mathbf{y}_1, \dots, \mathbf{y}_N, \mathbf{z}_1, \dots, \mathbf{z}_N)\right) \text{d}P_{\boldsymbol{\omega}} \le e^{-N\epsilon^2/2},
    \label{test_bound2}
\end{equation}
for any $\boldsymbol{\omega} \in \mathcal{U}^c$.
Using Markov's inequality on equation (\ref{test_bound1}), we have that 
\begin{align}
\nonumber
    P\left(\psi_N(\mathbf{y}_1, \dots, \mathbf{y}_N, \mathbf{z}_1, \dots, \mathbf{z}_N) \ge e^{-nC}\right) & \le \frac{\mathbb{E}\left(\psi_N(\mathbf{y}_1, \dots, \mathbf{y}_N, \mathbf{z}_1, \dots, \mathbf{z}_N)\right)}{e^{-NC}}\\
    \nonumber
    & \le (2r)e^{-N(\epsilon^2/2 - C)}
\end{align}
Thus letting $C < \epsilon^2 / 2$, we have that $\sum_{N=1}^\infty P\left(\psi_N(\mathbf{y}_1, \dots, \mathbf{y}_N, \mathbf{z}_1, \dots, \mathbf{z}_N) \ge e^{-NC}\right) < \infty$. Thus by the Borel-Cantelli lemma, we know that 
$$P\left(\limsup_{N\rightarrow \infty}P\left(\psi_N(\mathbf{y}_1, \dots, \mathbf{y}_N, \mathbf{z}_1, \dots, \mathbf{z}_N) \ge e^{-NC} \right)\right) = 0$$
Thus we have that $\psi_N(\mathbf{y}_1, \dots, \mathbf{y}_N,\mathbf{z}_1,\dots, \mathbf{z}_N) \rightarrow 0$ a.s. $[P_{\boldsymbol{\omega}_0}]$ (Condition (a)). To prove condition (b), we will first start by taking the expectation with respect to $P_{\boldsymbol{\omega}_0}$:
\begin{align}
    \nonumber &
    \mathbb{E}_{P_{\boldsymbol{\omega}_0}^N}\left(e^{\beta N}\left( 1- \psi_N(\mathbf{y}_1, \dots, \mathbf{y}_N,\mathbf{z}_1,\dots, \mathbf{z}_N)\right)\int_{\mathcal{U}^c}\prod_{i=1}^N \frac{f_i(\mathbf{y}_i; \boldsymbol{\omega})}{f_i(\mathbf{y}_i; \boldsymbol{\omega}_0)}\text{d}\boldsymbol{\Pi}(\boldsymbol{\omega})\right)\\
    \nonumber
     = & \int_{\boldsymbol{\mathcal{S}}^N} \left(e^{\beta N}\left( 1- \psi_N(\mathbf{y}_1, \dots, \mathbf{y}_N,\mathbf{z}_1,\dots, \mathbf{z}_N)\right)\int_{\mathcal{U}^c}\prod_{i=1}^N \frac{f_i(\mathbf{y}_i; \boldsymbol{\omega})}{f_i(\mathbf{y}_i; \boldsymbol{\omega}_0)}\text{d}\boldsymbol{\Pi}(\boldsymbol{\omega})\right) \text{d}P_{\boldsymbol{\omega}_0}^N\\
     \nonumber
     = & \int_{\mathcal{U}^c}\left(\prod_{i=1}^N \int_{\mathcal{S}} e^{\beta N}\left( 1- \psi_N(\mathbf{y}_1, \dots, \mathbf{y}_N,\mathbf{z}_1,\dots, \mathbf{z}_N)\right)f_i(\mathbf{y}_i; \boldsymbol{\omega})\text{d}\mathbf{y}_i\right)\text{d}\boldsymbol{\Pi}(\boldsymbol{\omega})\\
     \nonumber
     = & e^{\beta N} \int_{\mathcal{U}^c}\mathbb{E}_{P_{\boldsymbol{\omega}}^N}\left( 1- \psi_N(\mathbf{y}_1, \dots, \mathbf{y}_N,\mathbf{z}_1,\dots, \mathbf{z}_N) \right)\text{d}\boldsymbol{\Pi}(\boldsymbol{\omega})\\
     \nonumber
     \le & e^{\beta_1N}e^{-N\epsilon^2/2},
\end{align} 
where the last inequality is from equation (\ref{test_bound2}). Thus by Markov's inequality and letting $\beta_1 < \epsilon^2/2$, we have that 
\begin{align}
    \nonumber
    & P\left( e^{\beta N}\left( 1- \psi_N(\mathbf{y}_1, \dots, \mathbf{y}_N,\mathbf{z}_1,\dots, \mathbf{z}_N)\right)\int_{\mathcal{U}^c}\prod_{i=1}^N \frac{f_i(\mathbf{y}_i; \boldsymbol{\omega})}{f_i(\mathbf{y}_i; \boldsymbol{\omega}_0)}\text{d}\boldsymbol{\Pi}(\boldsymbol{\omega}) \ge e^{-N((\epsilon^2/2 - \beta_1)/2)} \right)\\
    \nonumber
    \le & \frac{\mathbb{E}_{P_{\boldsymbol{\omega}_0}^N}\left(e^{\beta N}\left( 1- \psi_N(\mathbf{y}_1, \dots, \mathbf{y}_N,\mathbf{z}_1,\dots, \mathbf{z}_N)\right)\int_{\mathcal{U}^c}\prod_{i=1}^N \frac{f_i(\mathbf{y}_i; \boldsymbol{\omega})}{f_i(\mathbf{y}_i; \boldsymbol{\omega}_0)}\text{d}\boldsymbol{\Pi}(\boldsymbol{\omega})\right)}{e^{-N((\epsilon^2/2 - \beta_1)/2)}} \\
    \nonumber
    \le &  e^{-N((\epsilon^2/2 - \beta_1)/2)}
\end{align}
Letting $E_N$ be the event that $e^{\beta N}$ $\left( 1- \psi_N(\mathbf{y}_1, \dots, \mathbf{y}_N,\mathbf{z}_1,\dots, \mathbf{z}_N)\right)$
$\int_{\mathcal{U}^c}\prod_{i=1}^N \frac{f_i(\mathbf{y}_i; \boldsymbol{\omega})}{f_i(\mathbf{y}_i; \boldsymbol{\omega}_0)}\text{d}\boldsymbol{\Pi}(\boldsymbol{\omega})$
$\ge e^{-N((\epsilon^2/2 - \beta_1)/2)}$, we have that $\sum_{i=1}^\infty P(E_N) < \infty$. Thus by the Borel-Cantelli lemma, we have that 
$$e^{\beta N}\left( 1- \psi_N(\mathbf{y}_1, \dots, \mathbf{y}_N,\mathbf{z}_1,\dots, \mathbf{z}_N)\right)\int_{\mathcal{U}^c}\prod_{i=1}^N \frac{f_i(\mathbf{y}_i; \boldsymbol{\omega})}{f_i(\mathbf{y}_i; \boldsymbol{\omega}_0)}\text{d}\boldsymbol{\Pi}(\boldsymbol{\omega}) \rightarrow 0$$ a.s. $[P_{\boldsymbol{\omega}_0}]$ for $0 <  \beta_1 < \epsilon^2/2$. Therefore, we have proved conditions (a), (b), and (c). Thus by letting $\beta$ in condition (c) be such that $\beta = \beta_1$, where $0 <  \beta_1 < \epsilon^2/2$, we can see that 
$\boldsymbol{\Pi}_N(\mathcal{U}^c| \mathbf{y}_1, \dots, \mathbf{y}_N) \rightarrow 0$ a.s. $[P_{\boldsymbol{\omega}_0}]$ for every weak neighborhood, $\mathcal{U}$ of $\boldsymbol{\omega}_0$.

\section{Computation}
\label{sec: Computation}
\subsection{Posterior Distributions and Computation}
\label{Posterior_appendix}
In this section, we will discuss the computational strategy used to perform Bayesian inference. In cases where the posterior distribution is a known distribution, a Gibbs update will be performed. We will let $\boldsymbol{\Theta}$ be the collection of all parameters, and $\boldsymbol{\Theta}_{-\zeta}$ be the collection of all parameters, excluding the $\zeta$ parameter. We will first start with the $\boldsymbol{\phi}_{km}$ parameters, for $j = 1,\dots, K$ and $m = 1, \dots, M$. Let $\mathbf{D}_{km} = \tilde{\tau}_{mk}^{-1} diag\left(\gamma_{k1m}^{-1}, \dots, \gamma_{kPm}^{-1}\right)$. By letting
$$\begin{aligned}
\mathbf{m}_{jm} = & \frac{1}{\sigma^2} \sum_{i=1}^N \left(\chi_{im} \left(\mathbf{y}_iZ_{ij} -  Z_{ij}^2 \boldsymbol{\nu}_{j} - Z_{ij}^2\sum_{n \ne m}\chi_{in} \boldsymbol{\phi}_{jn} - \sum_{k \ne j} Z_{ij}Z_{ik}\left[\boldsymbol{\nu}_{k} + \sum_{n=1}^M \chi_{in} \boldsymbol{\phi}_{kn} \right] \right) \right)
\end{aligned}$$
and
$$\mathbf{M}_{jm}^{-1} = \frac{1}{\sigma^2}\sum_{i=1}^N \left(Z_{ij}^2\chi_{im}^2\right)\mathbf{I}_P + \mathbf{D}_{km}^{-1},$$
 we have that 
$$\boldsymbol{\phi}_{jm} | \boldsymbol{\Theta}_{-\boldsymbol{\phi}_{jm} }, \mathbf{y}_1, \dots, \mathbf{y}_N \sim \mathcal{N}(\mathbf{M}_{jm}\mathbf{m}_{jm}, \mathbf{M}_{jm}).$$

The posterior distribution of $\delta_1$ is 
$$\begin{aligned}
\delta_{1k} | \boldsymbol{\Theta}_{-\delta_{1k}}, \mathbf{y}_1, \dots, \mathbf{y}_N \sim & \Gamma\left(a_{1k} + (PM/2), 1 + \frac{1}{2} \sum_{r=1}^P \gamma_{k,r,1}\phi_{k,r,1}^2  \right.  \\
& \left. + \frac{1}{2}\sum_{m=2}^M\sum_{r=1}^P \gamma_{k,r,m}\phi_{k,r,m}^2\left( \prod_{j=2}^m \delta_{j} \right)\right).
\end{aligned}$$
The posterior distribution for $\delta_{ik}$, for $i = 2, \dots, M$, is 
$$\begin{aligned}
\delta_{ik} | \boldsymbol{\Theta}_{-\delta_{ik}}, \mathbf{y}_1, \dots, \mathbf{y}_N \sim & \Gamma\Bigg( a_2 + (P(M - i + 1)/2), 1   \\
& \left. +\frac{1}{2}\sum_{m = i}^M \sum_{r=1}^P \gamma_{k,r,m}\phi_{k,r,m}^2\left( \prod_{j=1; j \ne i}^m \delta_{j} \right)\right).
\end{aligned}$$
The posterior distribution for $a_{1k}$ is not a commonly known distribution, however we have that
$$P(a_{1k}|\boldsymbol{\Theta}_{-a_{1k}}, \mathbf{y}_1, \dots, \mathbf{y}_N) \propto \frac{1}{\Gamma(a_{1k})}\delta_{1k}^{a_{1k} -1} a_{1k}^{\alpha_{1} -1} exp \left\{-a_{1k}\beta_{1} \right\}.$$
Since this is not a known kernel of a distribution, we will have to use Metropolis-Hastings algorithm. Consider the proposal distribution $Q(a_{1k}'| a_{1k}) = \mathcal{N}\left(a_{1k}, \epsilon_1\beta_{1}^{-1}, 0, + \infty\right)$ (Truncated Normal) for some small $\epsilon_1 > 0$. Thus the probability of accepting any step is
$$A(a_{1k}',a_{1k}) = \min \left\{1, \frac{P\left(a_{1k}'| \boldsymbol{\Theta}_{-a_{1k}'}, \mathbf{y}_1, \dots, \mathbf{y}_N\right)}{P\left(a_{1k}| \boldsymbol{\Theta}_{-a_{1k}}, \mathbf{y}_1, \dots, \mathbf{y}_N\right)} \frac{Q\left(a_{1k}|a_{1k}'\right)}{Q\left(a_{1k}'|a_{1k}\right)}\right\}.$$

Similarly for $a_{2k}$, we have
$$P(a_{2k} | \boldsymbol{\Theta}_{-a_{2k}}, \mathbf{y}_1, \dots, \mathbf{y}_N) \propto \frac{1}{\Gamma(a_{2k})^{M-1}}\left(\prod_{i=2}^M\delta_{ik}^{a_{2k} -1}\right) a_{2k}^{\alpha_{2} -1} exp \left\{-a_{2k}\beta_{2} \right\}.$$
We will use a similar proposal distribution, such that $Q(a_{2k}'| a_{2k}) = \mathcal{N}\left(a_{2k}, \epsilon_2\beta_{2}^{-1}, 0, + \infty\right)$ for some small $\epsilon_2 > 0$. Thus the probability of accepting any step is
$$A(a_{2k}',a_{2k}) = \min \left\{1, \frac{P\left(a_{2k}'| \boldsymbol{\Theta}_{-a_{2k}'}, \mathbf{y}_1, \dots, \mathbf{y}_N\right)}{P\left(a_{2k}| \boldsymbol{\Theta}_{-a_{2k}}, \mathbf{y}_1, \dots, \mathbf{y}_N\right)} \frac{Q\left(a_{2k}|a_{2k}'\right)}{Q\left(a_{2k}'|a_{2k}\right)}\right\}.$$
For the $\gamma_{j,r,m}$ parameters, for $j = 1, \dots K$, $r = 1, \dots, P$, and $m = 1, \dots, M$, we have 
$$\gamma_{j,r,m}| \boldsymbol{\Theta}_{-\gamma_{j,r,m}}, \mathbf{y}_1, \dots, \mathbf{y}_N \sim \Gamma\left(\frac{\nu_\gamma + 1}{2},\frac{\phi_{j,r,m}^2\tilde{\tau}_{mj} + \nu_\gamma}{2} \right).$$
The posterior distribution for the $\mathbf{z}_i$ parameters are not a commonly known distribution, so we will have to use the Metropolis-Hastings algorithm. We know that
$$\begin{aligned}
p(\mathbf{z}_i| \boldsymbol{\Theta}_{-\mathbf{z}_i}, \mathbf{y}_1, \dots, \mathbf{y}_N) & \propto \prod_{k=1}^K Z_{ik}^{\alpha_3\pi_k - 1}\\
& \times exp\left\{-\frac{1}{2\sigma^2}\left(\mathbf{y}_i -  \sum_{k=1}^K Z_{ik}\left(\boldsymbol{\nu}_k + \sum_{n=1}^M\chi_{in}\boldsymbol{\phi}_{kn}\right)\right)'\right.\\
& \left.\left(\mathbf{y}_i -  \sum_{k=1}^K Z_{ik}\left(\boldsymbol{\nu}_k + \sum_{n=1}^M\chi_{in}\boldsymbol{\phi}_{kn}\right)\right)\right\}.
\end{aligned}$$
We will use $Q(\mathbf{z}_i'| \mathbf{z}_i) = Dir(a_{\mathbf{z}} \mathbf{z}_i)$ for some large $a_{\mathbf{z}} \in \mathbb{R}^+$ as the proposal distribution. Thus the probability of accepting a proposed step is 
$$A(\mathbf{z}_i', \mathbf{z}_i) = \min \left\{1, \frac{P\left(\mathbf{z}_i'| \boldsymbol{\Theta}_{-\mathbf{z}_i'}, \mathbf{y}_1, \dots, \mathbf{y}_N\right)}{P\left(\mathbf{z}_i| \boldsymbol{\Theta}_{-\mathbf{z}_i}, \mathbf{y}_1, \dots, \mathbf{y}_N\right)} \frac{Q\left(\mathbf{z}_i|\mathbf{z}_i'\right)}{Q\left(\mathbf{z}_i'|\mathbf{z}_i\right)}\right\}.$$
Similarly, a Gibbs update is not available for an update of the $\boldsymbol{\pi}$ parameters. We have that 
$$\begin{aligned}
p(\boldsymbol{\pi}|\boldsymbol{\Theta}_{-\boldsymbol{\pi}}, \mathbf{y}_1,\dots, \mathbf{y}_N) & \propto \prod_{k=1}^K \pi_k^{c_k - 1} \\
& \times \prod_{i=1}^N\frac{1}{B(\alpha_3\boldsymbol{\pi})}\prod_{k=1}^K Z_{ik}^{\alpha_3\pi_k - 1}.
\end{aligned}$$
Letting out proposal distribution be such that $Q(\boldsymbol{\pi}'| \boldsymbol{\pi}) = Dir(a_{\boldsymbol{\pi}} \boldsymbol{\pi})$, for some large $a_{\boldsymbol{\pi}} \in \mathbb{R}^+$, we have that our probability of accepting any proposal is
$$A(\boldsymbol{\pi}', \boldsymbol{\pi}) = \min \left\{1, \frac{P\left(\boldsymbol{\pi}'| \boldsymbol{\Theta}_{-\boldsymbol{\pi}'}, \mathbf{y}_1, \dots, \mathbf{y}_N\right)}{P\left(\boldsymbol{\pi}| \boldsymbol{\Theta}_{-\boldsymbol{\pi}}, \mathbf{y}_1, \dots, \mathbf{y}_N\right)} \frac{Q\left(\boldsymbol{\pi}|\boldsymbol{\pi}'\right)}{Q\left(\boldsymbol{\pi}'|\boldsymbol{\pi}\right)}\right\}.$$
The posterior distribution of $\alpha_3$ is also not a commonly known distribution, so we will use the Metropolis-Hastings algorithm to sample from the posterior distribution. We have that 
$$\begin{aligned}
p(\alpha_3|\boldsymbol{\Theta}_{-\alpha_3}, \mathbf{y}_1, \dots, \mathbf{y}_N) & \propto e^{-b\alpha_3} \\
& \times \prod_{i=1}^N\frac{1}{B(\alpha_3\boldsymbol{\pi})}\prod_{k=1}^K Z_{ik}^{\alpha_3\pi_k - 1}.
\end{aligned}$$
Using a proposal distribution such that $Q(\alpha_3'|\alpha_3) = \mathcal{N}(\alpha_3, \sigma^2_{\alpha_3}, 0, +\infty)$ (Truncated Normal), we are left with the probability of accepting a proposed state as
$$A(\alpha_3',\alpha_3) = \min \left\{1, \frac{P\left(\alpha_3'| \boldsymbol{\Theta}_{-\alpha_3'}, \mathbf{y}_1, \dots, \mathbf{y}_N\right)}{P\left(\alpha_3| \boldsymbol{\Theta}_{-\alpha_3}, \mathbf{y}_1, \dots, \mathbf{y}_N\right)} \frac{Q\left(\alpha_3|\alpha_3'\right)}{Q\left(\alpha_3'|\alpha_3\right)}\right\}.$$
 
Letting
$$\mathbf{B}_j = \left(\frac{1}{\tau_j}\mathbf{I}_P + \frac{1}{\sigma^2}\mathbf{I}_P \sum_{i =1}^N Z_{ij}^2\right)^{-1}$$
and
$$\mathbf{b}_{j} = \frac{1}{\sigma^2}\sum_{i=1}^NZ_{ij}\left(\mathbf{y}_i - \left(\sum_{k\ne j}Z_{ik}\boldsymbol{\nu}_{k}\right)  - \left(\sum_{k=1}^K \sum_{m=1}^MZ_{ik}\chi_{im}\boldsymbol{\phi}_{km} \right)\right),$$
we have that 
$$\boldsymbol{\nu}_j| \boldsymbol{\Theta}_{-\boldsymbol{\nu}_j}, \mathbf{y}_1, \dots, \mathbf{y}_N \sim \mathcal{N}(\mathbf{B}_j\mathbf{b}_j, \mathbf{B}_j),$$
for $j = 1, \dots, K$. Thus we can perform a Gibbs update to update our $\boldsymbol{\nu}_.$ parameters. The $\tau_l$ parameters, for $l = 1, \dots K$, can also be updated by using a Gibbs update since the posterior distribution is:
$$\tau_l| \boldsymbol{\Theta}_{-\tau_l}, \mathbf{y}_1, \dots, \mathbf{y}_N \sim IG\left(\alpha + P/2, \beta + \frac{1}{2}\boldsymbol{\nu}'_l\boldsymbol{\nu}_l\right).$$
The parameter $\sigma^2$ can be updated by using a Gibbs update. If we let 
$$\beta_{\sigma} =\frac{1}{2}\sum_{i=1}^N\left(\mathbf{y}_i -  \sum_{k=1}^K Z_{ik}\left(\boldsymbol{\nu}_k + \sum_{m=1}^M\chi_{im}\boldsymbol{\phi}_{km}\right)\right)'\left(\mathbf{y}_i -  \sum_{k=1}^K Z_{ik}\left(\boldsymbol{\nu}_k + \sum_{m=1}^M\chi_{im}\boldsymbol{\phi}_{km}\right)\right)$$
then we have
$$\sigma^2| \boldsymbol{\Theta}_{-\sigma^2}, \mathbf{y}_1, \dots, \mathbf{y}_N  \sim  IG\left(\alpha_0 + \frac{PN}{2} , \beta_0 +\beta_{\sigma}\right),$$
where $n_i$ are the number of time points observed for the $i^{th}$ observed function. Lastly, we can update the $\chi_{im}$ parameters, for $i = 1, \dots, N$ and $m = 1, \dots, M$, using a Gibbs update. If we let 
$$\mathbf{w}_{im} = \frac{1}{\sigma^2}\left(\left(\sum_{k = 1}^K Z_{ik}\boldsymbol{\phi}_{km}\right)'\left(\mathbf{y}_i - \sum_{k = 1}^K Z_{ik}\left(\boldsymbol{\nu}_k  + \sum_{n\ne m}\chi_{in}\boldsymbol{\phi}_{kn}\right)\right)\right)$$
and 
$$\mathbf{W}^{-1}_{im} = 1 +  \frac{1}{\sigma^2}\left(\left(\sum_{k = 1}^K Z_{ik}\boldsymbol{\phi}_{km}\right)'\left(\sum_{k = 1}^K Z_{ik}\boldsymbol{\phi}_{km}\right)\right),$$
then we have that 
$$\chi_{im}| \boldsymbol{\zeta}_{-\chi_{im}}, \mathbf{y}_1, \dots, \mathbf{y}_N \sim \mathcal{N}(\mathbf{W}_{im}\mathbf{w}_{im}, \mathbf{W}_{im}).$$

In our paper, we have relaxed the assumption that the $\boldsymbol{\Phi}$ are mutually orthogonal parameters. We have shown that we can still maintain many of the desirable properties, while not having to sample in a constrained space. This relaxation makes implementation easier, and may actually help with mixing of the Markov chain. However, we realize that users may want to enforce that the $\boldsymbol{\Phi}$ parameters are orthogonal and therefore can be interpreted as scaled eigenvectors. Using the approach described by \citet{kowal2017bayesian}, we will describe how to sample in this constrained space. 

In order to impose the orthogonality constraint, we have that 
$$\boldsymbol{\Phi}_i' \boldsymbol{\Phi}_j  = \sum_{k=1}^{K} \boldsymbol{\phi}_{ik}' \boldsymbol{\phi}_{jk}  = 0,$$
for some $i$ such that $1 \le i \le KP$ and for all $j \ne i$.
Letting 
$$\mathbf{L}_{-ip} = \begin{bmatrix} 
\boldsymbol{\phi}_{i1}\\
\vdots \\
\boldsymbol{\phi}_{i(p-1)} \\
\boldsymbol{\phi}_{i(p+1)} \\
\vdots \\
\boldsymbol{\phi}_{i(KP)} \\
\end{bmatrix} \;\text{ and }\; \mathbf{c}_{-ip} =\begin{bmatrix}
\sum_{k \ne i} \boldsymbol{\phi}_{kp}'\boldsymbol{\phi}_{k1}\\
\vdots \\
\sum_{k \ne i} \boldsymbol{\phi}_{kp}'\boldsymbol{\phi}_{k(p-1)}\\
\sum_{k \ne i}\boldsymbol{\phi}_{kp}'\boldsymbol{\phi}_{k(p+1)}\\
\vdots\\
\sum_{k \ne i} \boldsymbol{\phi}_{kp}'\boldsymbol{\phi}_{k(KP)}\\
\end{bmatrix},$$
we can write the constraint as 
$$\boldsymbol{\phi}_{ip}\mathbf{L}_{-ip}  = - \mathbf{c}_{-ip},$$
for $1 \le i \le KP$ and $1 \le p \le K$. Using the results in \citet{kowal2017bayesian}, we have that 
$\boldsymbol{\phi}_{ip} \sim \mathcal{N}(\tilde{\mathbf{M}}_{ip}\mathbf{m}_{ip}, \tilde{\mathbf{M}}_{ip})$, where $$\tilde{\mathbf{M}}_{ip} = \mathbf{M}_{ip} - \mathbf{M}_{ip}\mathbf{L}_{-ip}\left(\mathbf{L}_{-ip}' \mathbf{M}_{ip} \mathbf{L}_{-ip} \right)^{-1} \left(\mathbf{L}_{-ip}'\mathbf{M}_{ip} + \mathbf{c}_{-ip} \right).$$ Like in \citet{kowal2017bayesian}, $\mathbf{M}_{ip}$ and $\mathbf{m}_{ip}$ are such that when we relax the orthogonal constraints, we have $\boldsymbol{\phi}_{ip} \sim \mathcal{N}(\mathbf{M}_{ip}\mathbf{m}_{ip}, \mathbf{M}_{ip})$. By using this alternate sampling scheme, one can ensure the orthogonality of the $\boldsymbol{\Phi}$ parameters.

\subsection{Multiple Start Algorithm}
\label{MSA_appendix}
Due to the flexible nature of our model, we often end up with multimodal posterior distributions, which makes posterior inference challenging. In addition to tempered transitions (described in Section \ref{Temp_trans_appendix}), we implement an algorithm called the multiple start algorithm (MSA) in order to obtain a good starting position for our Markov chain. The MSA, algorithm \ref{alg:MSA}, starts by first trying to recover the mean and allocation structure. Once a suitable starting point for the mean and allocation parameters are found, we then estimate to covariance structure conditioned on the starting point for the mean and allocation parameters. 

\begin{algorithm}
\caption{Multiple Start Algorithm}\label{alg:MSA}
\begin{algorithmic}
\Require $\text{n\_try1},\text{n\_try2}, Y, K, n\_MCMC1, n\_MCMC2, \dots$
\State $P \gets \texttt{BPMM\_Nu\_Z(Y, K, n\_MCMC1, \dots)}$ \Comment{Returns the likelihood and estimates for $\boldsymbol{\nu}$ and $\mathbf{Z}$}
\State $\text{max\_likelihood} \gets P[\text{``likelihood''}]$
\State $i \gets 1$
\While{$i \le \text{n\_try1}$}
\State $P_i \gets \texttt{BPMM\_Nu\_Z(Y, K, n\_MCMC1, \dots)}$
\If{max\_likelihood $< P_i[\text{``likelihood''}]$}
    \State \text{max\_likelihood} $\gets P[\text{``likelihood''}]$
    \State $P \gets P_i$
\EndIf
\State $i \gets i + 1$
\EndWhile
\State $\theta \gets \texttt{BPMM\_Theta(P, Y, K, n\_MCMC2, \dots)}$ \Comment{Returns estimates for the rest of the parameters}
\State $\text{max\_likelihood} \gets \theta[\text{``likelihood''}]$
\State $i \gets 1$
\While{$i \le \text{n\_try2}$}
\State $\theta_i \gets \texttt{BPMM\_Theta(P, Y, K, n\_MCMC2, \dots)}$
\If{max\_likelihood $< \theta_i[\text{``likelihood''}]$}
    \State \text{max\_likelihood} $\gets \theta_i[\text{``likelihood''}]$
    \State $\theta \gets \theta_i$
\EndIf
\State $i \gets i + 1$
\EndWhile
\State \Return $(\theta, P)$ \Comment{Returns estimates for all model parameters}
\end{algorithmic}
\end{algorithm}

We can see that the MSA primarily calls two functions, $\texttt{BPMM\_Nu\_Z(\dots)}$ and $\texttt{BPMM\_Theta(\dots)}$. The first function, $\texttt{BPMM\_Nu\_Z(\dots)}$, finds initial starting points for the $\boldsymbol{z}_i$ parameters, $\boldsymbol{\nu}_k$ parameters, and related hyperparameters, while setting $\chi{im}$ and $\boldsymbol{\phi}_{km}$ equal to 0 (or $\mathbf{0}$). The second function, $\texttt{BPMM\_Theta(\dots)}$, finds initial starting points for the $\chi_{im}$ parameters, $\boldsymbol{\phi}$ parameters, $\sigma^2$, and related hyperparameters, conditioning on the initial starting point of the $\mathbf{z}_i$ and $\boldsymbol{\nu}_k$ parameters. In order to get the best results, we recommend standardizing the raw data before performing inference. The multiple start algorithm can be easily implemented in R using the accompanying software package to this paper. 

\subsection{Tempered Transitions}
\label{Temp_trans_appendix}
As stated in the previous section, the posterior distribution may often be multimodal, which often causes traditional MCMC methods to get stuck in local modes. In order to be able to move across modes, we implement tempered transitions, which will allow us to traverse areas of low posterior probability.

Following the works of \citet{behrens2012tuning} and \citet{pritchard2000inference}, we will only temper the likelihood of the model, which can often be written as 
\begin{equation}
\label{eq: general_likelihood}
    p(x) \propto \pi(x)exp\left(-\beta_h h(x)\right),
\end{equation}
where $\beta_h$ controls how much the distribution is tempered. We will assume $1 = \beta_0 < \dots < \beta_h < \dots < \beta_{N_t}$ and that the hyperparameters $N_t$ and $\beta_{N_t}$ are user specified. For larger and more complex models, we will often need a larger $N_t$, however our tempered transitions will be more computationally expensive with larger $N_t$. We will assume that the parameters $\beta_h$ follow a geometric scheme. We can rewrite our likelihood to fit equation (\ref{eq: general_likelihood}):
$$\begin{aligned}
p_h(\mathbf{y}_i|\boldsymbol{\Theta})  \propto & exp\left\{- \beta_h\left(\frac{1}{2}log(\sigma^2) + \frac{1}{2\sigma^2} \left(\mathbf{y}_i -  \sum_{k=1}^K Z_{ik}\left(\boldsymbol{\nu}_k + \sum_{m=1}^M\chi_{im}\boldsymbol{\phi}_{km}\right)\right)'\right.\right.\\
& \left. \left. \left(\mathbf{y}_i -  \sum_{k=1}^K Z_{ik}\left(\boldsymbol{\nu}_k + \sum_{m=1}^M\chi_{im}\boldsymbol{\phi}_{km}\right)\right) \right)\right\} \\
& =\left(\sigma^2\right)^{-\beta_h / 2}exp\left\{-\frac{\beta_h}{2\sigma^2}\left(\mathbf{y}_i -  \sum_{k=1}^K Z_{ik}\left(\boldsymbol{\nu}_k + \sum_{m=1}^M\chi_{im}\boldsymbol{\phi}_{km}\right)\right)'\right.\\
 & \left. \left(\mathbf{y}_i -  \sum_{k=1}^K Z_{ik}\left(\boldsymbol{\nu}_k + \sum_{m=1}^M\chi_{im}\boldsymbol{\phi}_{km}\right)\right)\right\}.
\end{aligned}$$

Let $\boldsymbol{\Theta}_h$ be the set of parameters generated from the model using the tempered likelihood associated with $\beta_h$. The tempered transition algorithm can be summarized by the following steps:
\begin{enumerate}
    \item Start with initial state $\boldsymbol{\Theta}_0$.
    \item Transition from $\boldsymbol{\Theta}_0$ to $\boldsymbol{\Theta}_1$ using the tempered likelihood associated with $\beta_1$.
    \item Continue in this manner until we transition from $\boldsymbol{\Theta}_{N_t - 1}$ to $\boldsymbol{\Theta}_{N_t}$ using the tempered likelihood associated with $\beta_{N_t}$.
    \item Transition from $\boldsymbol{\Theta}_{N_t}$ to $\boldsymbol{\Theta}_{N_t +1}$ using the tempered likelihood associated with $\beta_{N_t}$.
    \item Continue in this manner until we transition from $\boldsymbol{\Theta}_{2N_t -1}$ to $\boldsymbol{\Theta}_{2N_t}$ using $\beta_1$.
    \item Accept transition from $\boldsymbol{\Theta}_0$ to $\boldsymbol{\Theta}_{2N_t}$ with probability 
    $$\min \left\{1, \prod_{h=0}^{N_t - 1} \frac{\prod_{i=1}^N p_{h+1}(\mathbf{y}_i)|\boldsymbol{\Theta}_h)}{\prod_{i=1}^N p_{h}(\mathbf{y}_i|\boldsymbol{\Theta}_h)} \prod_{h=N_t + 1}^{2N_t} \frac{\prod_{i=1}^N p_{h}(\mathbf{y}_i)|\boldsymbol{\Theta}_h)}{\prod_{i=1}^N p_{h+1}(\mathbf{y}_i)|\boldsymbol{\Theta}_h)}\right\}.$$
\end{enumerate}
Since we only temper the likelihood, many of the posterior distributions from Section \ref{Posterior_appendix} can be used. Thus we will only have to modify the posterior distributions for the $\boldsymbol{\nu}$, $\sigma^2$, $\chi$, $\boldsymbol{\phi}$, and $\mathbf{Z}$ parameters. We will start with the $(\boldsymbol{\phi})_h$ parameters. Letting 
$$\begin{aligned}
(\mathbf{m}_{km})_h = & \frac{\beta_h}{(\sigma^2)_h} \sum_{i=1}^N \left((\chi_{im})_h \left(\mathbf{y}_i(Z_{ij})_h -  (Z_{ij})h^2 (\boldsymbol{\nu}_{j})_h - (Z_{ij})_h^2\sum_{n \ne m}(\chi_{in})_h (\boldsymbol{\phi}_{jn})_h \right. \right.\\ 
& \left. \left.- \sum_{k \ne j} (Z_{ij})_h(Z_{ik})_h\left[(\boldsymbol{\nu}_{k})_h + \sum_{n=1}^M (\chi_{in})_h (\boldsymbol{\phi}_{kn})_h \right] \right) \right),
\end{aligned}$$
and
$$(\mathbf{M}_{km})_h^{-1} = \frac{\beta_h}{(\sigma^2)_h}\sum_{i=1}^N \left((Z_{ij})_h^2(\chi_{im})_h^2\right)\mathbf{I}_P + \left(\mathbf{D}_{km}\right)_h^{-1},$$
 we have that 
$$(\boldsymbol{\phi}_{km})_h  | \boldsymbol{\Theta}_{-(\boldsymbol{\phi}_{km})_{h}}, \mathbf{y}_1, \dots, \mathbf{y}_N\sim \mathcal{N}((\mathbf{M}_{km})_h(\mathbf{m}_{km})_h, (\mathbf{M}_{km})_h).$$
The posterior distribution of the $(\mathbf{Z})_h$ parameters are still not commonly known distributions, so we have to use the Metropolis-Hastings algorithm. Thus, we have that
$$\begin{aligned}
p((\mathbf{z}_i)_h| (\boldsymbol{\Theta}_{-(\mathbf{z}_i)_h})_h, \mathbf{y}_1, \dots, \mathbf{y}_N) & \propto \prod_{k=1}^K (Z_{ik})_h^{(\alpha_3)_h(\pi_k)_h - 1}\\
&  \times exp\left\{-\frac{\beta_h}{2(\sigma^2)_h}\left(\mathbf{y}_i -  \sum_{k=1}^K (Z_{ik})_h\left((\boldsymbol{\nu}_k)_h + \sum_{n=1}^M(\chi_{in})_h(\boldsymbol{\phi}_{kn})_h\right)\right)'\right.\\
& \left.\left(\mathbf{y}_i -  \sum_{k=1}^K (Z_{ik})_h\left((\boldsymbol{\nu}_k)_h + \sum_{n=1}^M(\chi_{in})_h(\boldsymbol{\phi}_{kn})_h\right)\right)\right\}.
\end{aligned}$$
We will use $Q((\mathbf{z}_i)_h'| (\mathbf{z}_i)_{h-1}) = Dir(a_{\mathbf{z}} (\mathbf{z}_i)_{h-1})$ for some large $a_{\mathbf{z}} \in \mathbb{R}^+$ as the proposal distribution. Thus the probability of accepting a proposed step is 
$$A((\mathbf{z}_i)_{h}', (\mathbf{z}_i)_{h-1}) = \min \left\{1, \frac{P\left((\mathbf{z}_i)_{h}'| (\boldsymbol{\Theta}_{-(\mathbf{z}_i)_{h}'})_h, \mathbf{y}_1, \dots, \mathbf{y}_N\right)}{P\left((\mathbf{z}_i)_{h-1}| \boldsymbol{\Theta}_{-(\mathbf{z}_i)_{h-1}}, \mathbf{y}_1, \dots, \mathbf{y}_N\right)} \frac{Q\left((\mathbf{z}_i)_{h-1}|(\mathbf{z}_i)_{h}'\right)}{Q\left((\mathbf{z}_i)_{h}'|(\mathbf{z}_i)_{h-1}\right)}\right\}.$$
Next, letting 
$$(\mathbf{B}_j)_h =  \left(\frac{\beta_h}{(\tau_j)_h}\mathbf{I}_P + \frac{1}{(\sigma^2)_h}\mathbf{I}_P \sum_{i =1}^N (Z_{ij})_h^2\right)^{-1}$$
and
$$\begin{aligned}
\mathbf{b}_{j} = \frac{\beta_h}{(\sigma^2)_h}\sum_{i=1}^N(Z_{ij})_h\left(\mathbf{y}_i - \left(\sum_{k\ne j}(Z_{ik})_h(\boldsymbol{\nu}_{k})_h\right)  - \left(\sum_{k=1}^K \sum_{m=1}^M(Z_{ik})_h(\chi_{im})_h(\boldsymbol{\phi}_{km})_h \right)\right)
\end{aligned}$$
we have that 
$$(\boldsymbol{\nu}_j)_h| \boldsymbol{\Theta}_{-(\boldsymbol{\nu}_j)_h}, \mathbf{y}_1, \dots, \mathbf{y}_N \sim \mathcal{N}((\mathbf{B}_j)_h(\mathbf{b}_j)_h, (\mathbf{B}_j)_h).$$
The posterior distribution for $(\sigma^2)_h$ is a distribution that can be easily sampled from, so  we can use Gibbs sampling to get posterior draws. Letting 
\begin{align}
    \nonumber (\beta_{\sigma})_h &= \frac{\beta_h}{2}\sum_{i=1}^N\left[\left(\mathbf{y}_i -  \sum_{k=1}^K (Z_{ik})_h\left((\boldsymbol{\nu}_k)_h + \sum_{m=1}^M(\chi_{im})_h(\boldsymbol{\phi}_{km})_h\right)\right)'\right. \\
    \nonumber & \left.\left(\mathbf{y}_i -  \sum_{k=1}^K (Z_{ik})_h\left((\boldsymbol{\nu}_k)_h + \sum_{m=1}^M(\chi_{im})_h(\boldsymbol{\phi}_{km})_h\right)\right)\right]
\end{align}
 we have
$$(\sigma^2)_h| \boldsymbol{\Theta}_{-(\sigma^2)_h}, \mathbf{y}_1, \dots, \mathbf{y}_N  \sim  IG\left(\alpha_0 + \frac{\beta_h PN}{2} , \beta_0 +(\beta_{\sigma})_h\right).$$
Lastly, we can sample from $(\chi)_h$ using a Gibbs sampler to get posterior draws. Letting 
$$(\mathbf{w}_{im})_h = \frac{\beta_h}{(\sigma^2)_h}\left(\left(\sum_{k = 1}^K (Z_{ik})_h(\boldsymbol{\phi}_{km})_h\right)'\left(\mathbf{y}_i - \sum_{k = 1}^K (Z_{ik})_h\left((\boldsymbol{\nu}_k)_h  + \sum_{n\ne m}(\chi_{in})_h(\boldsymbol{\phi}_{kn})_h\right)\right)\right)$$
and 
$$(\mathbf{W}^{-1}_{im})_h = 1 +  \frac{\beta_h}{(\sigma^2)_h}\left(\left(\sum_{k = 1}^K (Z_{ik})_h(\boldsymbol{\phi}_{km})_h\right)'\left(\sum_{k = 1}^K (Z_{ik})_h(\boldsymbol{\phi}_{km})_h\right)\right),$$
then we have that 
$$(\chi_{im})_h| \boldsymbol{\zeta}_{-(\chi_{im})_h}, \mathbf{y}_1, \dots, \mathbf{y}_N \sim \mathcal{N}((\mathbf{W}_{im})_h(\mathbf{w}_{im})_h, (\mathbf{W}_{im})_h).$$
As stated before, complex models will often require large $N_t$ to accept the tempered transition proposed states. Unfortunately, this can be very computationally expensive, which is why we recommend using a mixture of tempered transitions and standard sampling techniques as described in Section \ref{Posterior_appendix}. From proposition 1 of \citet{roberts2007coupling}, we know that an independent mixture of tempered transitions and untempered transitions will preserve the stationary distribution of the Markov chain.

\subsection{Membership Rescale Algorithm}
\label{appendix_MRA}
As discussed in Section 2.4 of the main text, our model can be unidentifiable. To make the clusters more interpretable, we will rescale the allocation parameters $Z_{ik}$ such that in the two-feature model, at least one observation completely belongs to each feature. This specific assumption that one observation belongs entirely in each feature is known as the \textit{seperability} condition \citep{papadimitriou1998latent, mcsherry2001spectral, azar2001spectral, chen2022learning}. Thus in order to ensure identifiability, algorithm \ref{alg:MTA} can be used when we only have two features. In the case where there are more than two features, the assumption of seperability can be relatively strong, and weaker geometric assumptions, such as the \textit{sufficiently scattered} condition \citep{huang2016anchor, jang2019minimum, chen2022learning} can be used to ensure identifiability. From \citet{chen2022learning}, we have that an allocation matrix $\mathbf{Z}$ is sufficiently scattered if:
\begin{enumerate}
    \item cone$(\mathbf{Z}')^* \subseteq \mathcal{K}$
    \item cone$(\mathbf{Z}')^* \cap bd\mathcal{K} \subseteq \{\lambda \mathbf{e}_f, f = 1, \dots, k, \lambda \ge 0 \}$
\end{enumerate}
where $\mathcal{K} := \left\{\mathbf{x}\in \mathbb{R}^K | \|\mathbf{x}\|_2 \le \mathbf{x}'\mathbf{1}_K\right\}$, $bd\mathcal{K}  := \left\{\mathbf{x}\in \mathbb{R}^K | \|\mathbf{x}\|_2 = \mathbf{x}'\mathbf{1}_K\right\}$, cone$(\mathbf{Z}')^* := \left\{\mathbf{x} \in \mathbb{R}^K | \mathbf{x}\mathbf{Z}' \ge 0 \right\}$, and $\mathbf{e}_f$ is a vector with the $i^{th}$ element equal to 1 and zero elsewhere. The first condition can be interpreted as the allocation parameters should form a convex polytope that contains the dual cone $\mathcal{K}^*$. Thus we have that 
$$ \text{Conv}(\mathbf{Z}')\subseteq \mathcal{K}^*,$$
where $\mathcal{K}^* := \left\{\mathbf{x}\in \mathbb{R}^K |  \mathbf{x}'\mathbf{1}_K \ge \sqrt{k -1}\|\mathbf{x}\|_2\right\}$ and $\text{Conv}(\mathbf{Z}'):= \{\mathbf{x} \in \mathbb{R}^K | \mathbf{x} = \mathbf{Z}'\lambda, \lambda \in \Delta^N \}$, where $\Delta^k$ denotes the N-dimensional simplex. Ensuring that these conditions are met in our proposed model is non-trivial. The major non-identifiabilty problem we wish to solve is the \textit{rescaling problem} discussed in Section 2.4 of the main text. 
Therefore, we will focus on trying to promote allocation structures so that the first condition is satisfied. Similarly to the case of two functional features, we aim to find a linear transformation such that the convex polytope of our transformed allocation parameters covers the most area. Thus, letting $\mathbf{T} \in \mathbb{R}^K \times \mathbb{R}^K$ be our transformation matrix, we aim to solve the following optimization problem:
\begin{align}
    \nonumber \max_{\mathbf{T}} & \;\; |\text{Conv}(\mathbf{T}\mathbf{Z}')|\\
    \nonumber s.t. &  \;\; \mathbf{z}_i\mathbf{T} \in \mathcal{C}\;\; \forall \;i,
\end{align}
where $|\text{Conv}(\mathbf{T}\mathbf{Z}')|$ denotes the volume of the convex polytope constructed by the allocation parameters. Since the second condition is likely not met, we cannot ensure that our model is identifiable. However, the model is more interpretable, making inference easier for the end user. Once the transformation matrix ($\mathbf{T}$) is found, we can rescale the allocation parameters ($\mathbf{z}_i$) and the corresponding mean and covariance parameters ($\boldsymbol{\nu}_k$ and $\boldsymbol{\phi}_{km}$). In practice, maximizing the area of a convex polytope can be difficult as formulated. Thus, in the case when we have a three-feature model, we will first project the data onto a two-dimensional space (since we have the condition that $Z_{i1} + Z_{i2} + Z_{i3} = 1$, we do not lose any information). From there, we find the minimum minimum enclosing triangle that encloses the convex polytope created by the allocation parameters in the two-dimensional space \citep{parvu2016implementation}. We then use the vertices of the triangle to transform the data to maximize the area of the convex polytope.

\begin{algorithm}
\caption{Membership Rescale Algorithm}\label{alg:MTA}
\begin{algorithmic}
\Require $\mathbf{Z}, \boldsymbol{\nu}, \boldsymbol{\Phi}, M$
\State $T \gets \texttt{matrix}(0, 2, 2)$ \Comment{Initialize inverse transformation matrix (2 x 2)}
\State $i \gets 1$
\While{$i \le 2$}
\State $\text{max\_ind} \gets \texttt{max\_ind}(\mathbf{Z}[,i])$ \Comment{Find index of max entry in $i^{th}$ column}
\State $T[i,] \gets (\mathbf{Z}[\text{max\_ind},])$
\State $i \gets i + 1$
\EndWhile
\State $\mathbf{Z}\_t \gets \mathbf{Z} \times \texttt{inv}(T)$ \Comment{Transform the $\mathbf{Z}$ parameters}
\State $\boldsymbol{\nu}\_t \gets T \times \boldsymbol{\nu}$ \Comment{Transform the $\boldsymbol{\nu}$ parameters}
\State $i \gets 1$
\While{$i \le M$}
\State $\boldsymbol{\Phi}\_t[,,i] \gets T \times \boldsymbol{\Phi}[,,i]$ \Comment{Transform the $\boldsymbol{\Phi}$ parameters}
\State $i \gets i + 1$
\EndWhile
\State \Return $(\mathbf{Z}\_t, \boldsymbol{\nu}\_t, \boldsymbol{\Phi}\_t)$
\end{algorithmic}
\end{algorithm}

\section{Simulations and Case Studies}
\label{sec: sim_case_studies}
\subsection{Simulation Study 1}
\label{SS1_appendix}
In this simulation study, we empirically study the convergence properties of our model on simulated data. In this simulation study, we considered a two-feature mixed membership model, where the observed data are 10-dimensional vectors ($y \in \mathbb{R}^{10}$). To generate the data set, we first generate the model parameters in the following way:
$$\boldsymbol{\nu}_k \sim \mathcal{N}(\mathbf{0}_{10}, 9\mathbf{I}_{10}),$$
$$\chi_{im} \sim \mathcal{N}(0,1),$$
for $k = 1,2$ and $m = 1, \dots, 4$.
Next, we generated the $\boldsymbol{\phi}$ parameters in the following way:
$$\boldsymbol{\phi}_{km} = \mathbf{q}_{km},$$
where $\mathbf{q}_{k1} \sim \mathcal{N}(\mathbf{0}_10, \mathbf{I}_10)$, $\mathbf{q}_{k2} \sim \mathcal{N}(\mathbf{0}_10, 0.49\mathbf{I}_10)$, $\mathbf{q}_{k3} \sim \mathcal{N}(\mathbf{0}_10, 0.25\mathbf{I}_10)$, and $\mathbf{q}_{k4} \sim \mathcal{N}(\mathbf{0}_10, 0.09\mathbf{I}_10)$. The allocation parameters $\mathbf{z}_i$ were drawn from a mixture of Dirichlet distributions. Approximately 30\% of the allocation parameters were dawn from a Dirichlet distribution with $\alpha_1 = 10$ and $\alpha_2 = 1$. Another roughly 30\% were drawn from a Dirichlet distribution with $\alpha_1 = 1$ and $\alpha_2 = 10$. The final roughly 40\% of the allocation parameters were drawn from a Dirichlet distribution with $\alpha_1 = \alpha_2 = 1$. Lastly, $\sigma^2$ was set to 0.01 for this simulation study. Once all the parameters were drawn, we generated a dataset and repeated this process 50 times for each of the three different sample sizes ($i = 50,250,1000$). A mixed membership model was then fit for each of the datasets using 200,000 MCMC iterations saving only every 100 iterations (n\_try1 = 150, n\_try2 = 10, n\_MCMC1 = 4000, n\_MCMC2 = 4000, M = 4). Lastly, convergence metrics were calculated and displayed in Figure 2 of the main text.

\subsection{Simulation Study 2}
\label{SS2_appendix}
In this simulation study, we evaluate the performance of various information criteria in choosing the number of features in our proposed mixed membership model. To evaluate the information criteria (BIC, AIC, and DIC), we fit multiple mixed membership models with as little as 2 features to as many as 5 features ($K = 2, \dots, 5$) on 50 different datasets. The datasets were generated from our proposed mixed membership model with 3 features. In order to generate the datasets, we first randomly generated the parameters of our model. For each of the 50 datasets, the parameters were drawn in the following way:
$$\boldsymbol{\nu}_k \sim \mathcal{N}(\mathbf{0}_{20}, 10\mathbf{I}_{20}),$$
$$\boldsymbol{\phi}_{i1} \sim \mathcal{N}(\mathbf{0}_{20}, \mathbf{I}_{20}),$$
$$\boldsymbol{\phi}_{i2} \sim \mathcal{N}(\mathbf{0}_{20}, 0.5\mathbf{I}_{20}),$$
$$\boldsymbol{\phi}_{i3} \sim \mathcal{N}(\mathbf{0}_{20}, 0.2\mathbf{I}_{20}),$$
$$\chi_{im} \sim \mathcal{N}(0,1),$$
where $i = 1, \dots, 200$ and $m = 1, \dots, 3$. Similarly to simulation study 1, the allocation parameters were drawn from a mixture of Dirichlet distributions. Roughly 20\% of the allocation parameters were dawn from a Dirichlet distribution with $\alpha_1 = 10$ and $\alpha_2 = 1$. Another roughly 20\% were drawn from a Dirichlet distribution with $\alpha_1 = 1$ and $\alpha_2 = 10$. The final roughly 60\% of the allocation parameters were drawn from a Dirichlet distribution with $\alpha_1 = \alpha_2 = 1$. Similarly, we set $\sigma^2$ equal to 0.01 for all 50 datasets. Once all of the parameters were drawn, the 200 observation ($\mathbf{y}_i \in \mathbb{R}^{20}$) datasets were drawn. Once the data sets were created, we fit 4 models ($K = 2, \dots, 5)$ using a MCMC with 100,000 iterations, saving only every 10 iterations, with the following hyperparameters: n\_try1 = 50, n\_try2 = 5, n\_MCMC1 = 4000, n\_MCMC2 = 10000, M = 4.

The first IC we considered for this simulation study is the Bayesian information criterion (BIC).
The BIC, proposed by \citet{schwarz1978estimating}, is defined as:
$$\text{BIC} = 2\log P\left(\mathbf{Y}|\hat{\boldsymbol{\Theta}}\right) - d\log (N)$$
where $d$ is the number of parameters, $\hat{\boldsymbol{\Theta}}$ is the collection of maximum likelihood estimators (MLE) of our parameters, and $\mathbf{Y} = \{\mathbf{y}_i\}_{i=1}^N$ is the collection of our observed vectors. In the case of our mixed membership model, we have that
\begin{equation}
\text{BIC} = 2\log P\left(\mathbf{Y}|\hat{\boldsymbol{\nu}}, \hat{\boldsymbol{\Phi}}, \hat{\sigma}^2, \hat{\mathbf{Z}}, \hat{\boldsymbol{\chi}}\right) - d\log (N)
    \label{BIC}
\end{equation}
where $d = (N + P)K + 2MKP + 4K + (N + K)M + 2 $.

Similarly, the Akaike IC (AIC), proposed by \citet{akaike1974new}, can be written as
\begin{equation}
    \text{AIC} = -2\log P\left(\mathbf{Y}|\hat{\boldsymbol{\nu}}, \hat{\boldsymbol{\Phi}}, \hat{\sigma}^2, \hat{\mathbf{Z}}, \hat{\boldsymbol{\chi}}\right) + 2d.
    \label{AIC}
\end{equation}

Following the work of \citet{roeder1997practical}, we will use the posterior mean instead of the MLE for our estimates of BIC and AIC. Due to identifiability problems, the posterior mean of the mean component in equation (\ref{eq: likelihood}) for each observation, as well as the posterior mean of $\sigma^2$, will be used to estimate the BIC and AIC instead of estimates of the posterior mean for each individual parameter.

The modified Deviance IC (DIC), proposed by \citet{celeux2006deviance}, is advantageous to the original DIC proposed by \citet{spiegelhalter2002bayesian} when we have a posterior distribution with multiple modes, and when identifiability may be a problem. The modified DIC (referred to as $\text{DIC}_3$ in \citet{celeux2006deviance}) is specified as:
\begin{equation}
    \text{DIC} = -4 \mathbb{E}_{\boldsymbol{\Theta}}[\log f(\mathbf{Y}|\boldsymbol{\Theta})|\mathbf{Y}] + 2 log \hat{f}(\mathbf{Y})
    \label{DIC}
\end{equation}
where $\hat{f}(\mathbf{y}_i) = \frac{1}{N_{MC}}\sum_{l=1}^{N_{MC}}P\left(\mathbf{y}_i|\boldsymbol{\nu}^{(l)}, \boldsymbol{\Phi}^{(l)}, \left(\sigma^2\right)^{(l)}, \mathbf{Z}^{(l)}\right)$, $\hat{f}(\mathbf{Y}) = \prod_{i=1}^{N}\hat{f}(\mathbf{y}_i)$, and $N_{MC}$ is the number of MCMC samples used for estimating $\hat{f}(\mathbf{y}_i)$. We can approximate $\mathbb{E}_{\boldsymbol{\Theta}}[\log  f(\mathbf{Y}|\boldsymbol{\Theta})|\mathbf{Y}]$ by using the MCMC samples, such that
$$\mathbb{E}_{\boldsymbol{\Theta}}[\log  f(\mathbf{Y}|\boldsymbol{\Theta})|\mathbf{Y}] \approx \frac{1}{N_{MC}} \sum_{l=1}^{N_{MC}}\sum_{i=1}^{N}\log \left[P\left(\mathbf{y}_i|\boldsymbol{\nu}^{(l)},\boldsymbol{\Phi}^{(l)}, \left(\sigma^2\right)^{(l)}, \mathbf{Z}^{(l)}\right)\right].$$

\subsection{EEG Case Study}
\label{EEG_appendix}
\begin{figure}[H]
    \centering
    \includegraphics[width=.49\textwidth]{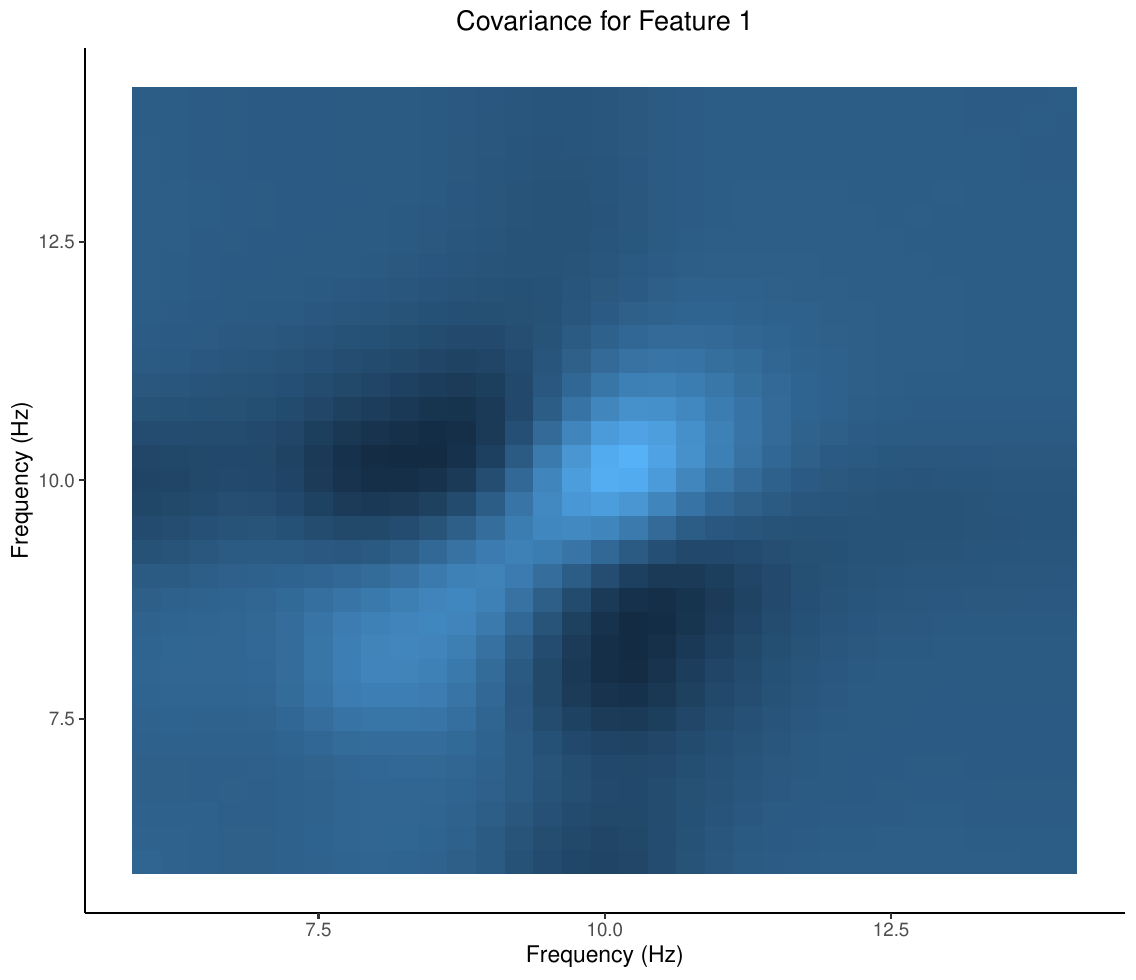}
    \includegraphics[width=.49\textwidth]{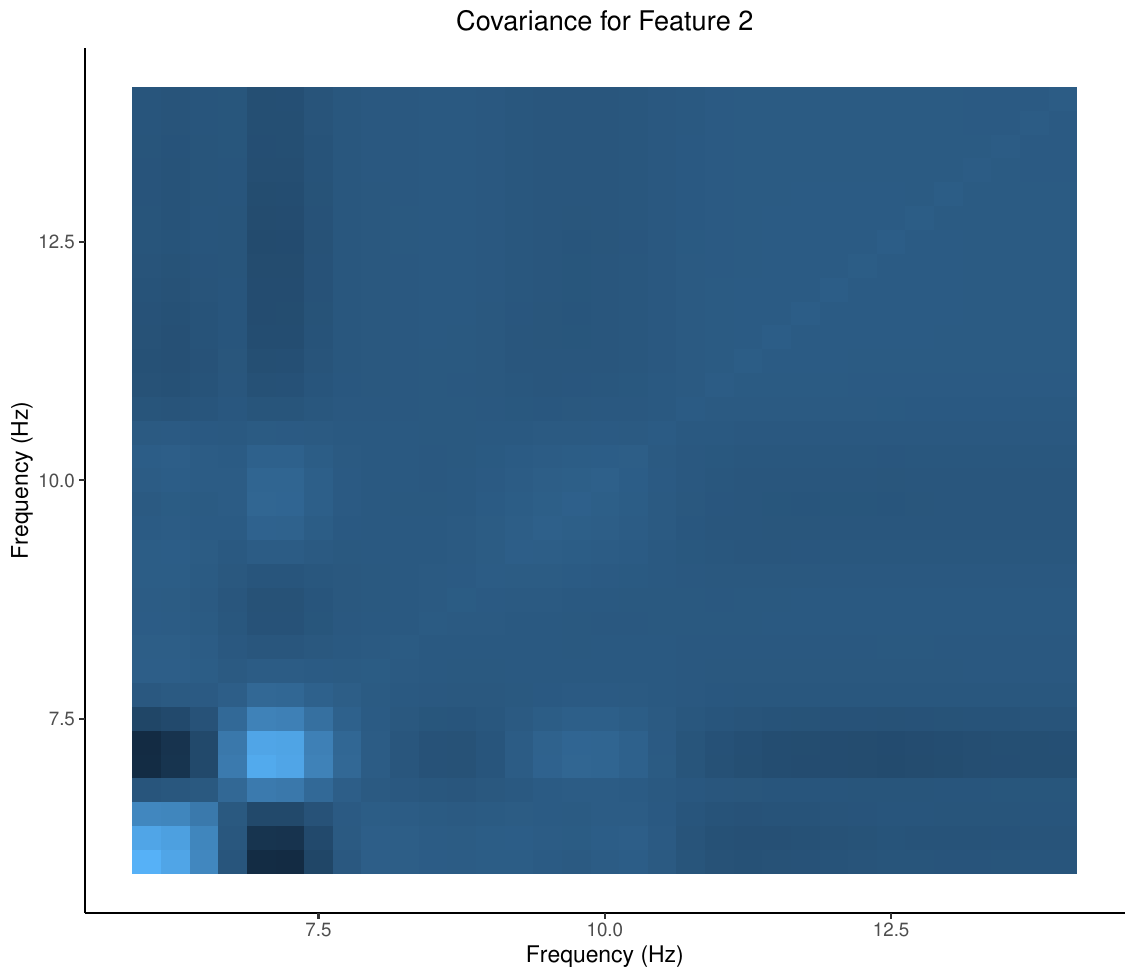}
    \includegraphics[width=.49\textwidth]{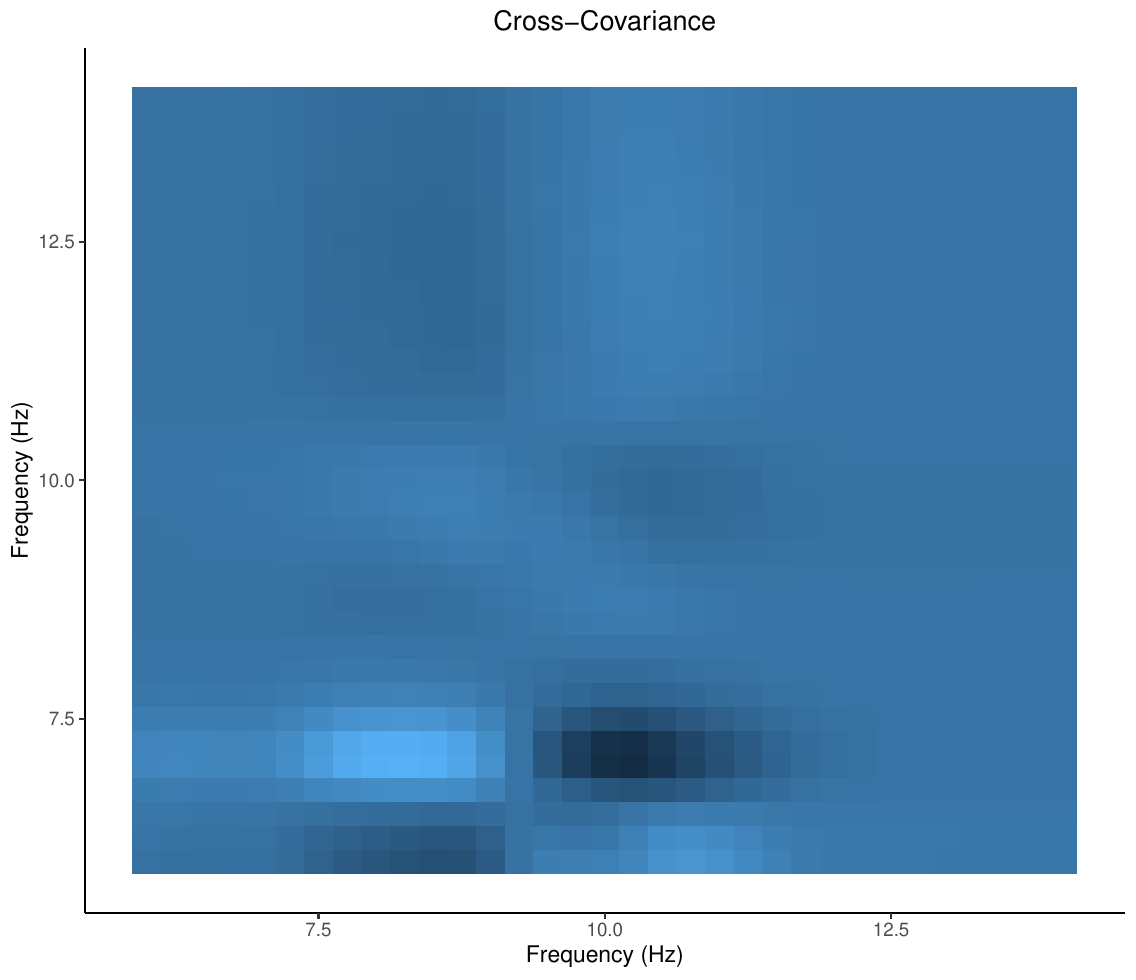}
    \caption{Visualization of the covariance structure for the two feature mixed membership model. Light blue represents positive covariance, while dark blue represents negative covariance.}
    \label{fig:EEG_Cov}
\end{figure}

In this case study, we analyze resting-state EEG data from typically developing (TD) children and children with Autism spectrum disorder (ASD) \citep{dickinson2018peak}. For this case study, we fit a 2 feature mixed membership model and a 3 feature mixed membership model with 5 eigenvectors ($M = 5$) for each model. Using AIC and BIC to help inform our choice on the number of features, we find that the 2 feature model seems to be a better model for the data ($AIC_2 = -12905.6$, $AIC_3 = -12204.5$, $BIC_2 = 9236.7$, $BIC_3 = 7328.0$, $DIC_2 = -14010.5$, $DIC_3 = -14197.9$). To get a good starting position, we used the multiple start algorithm (algorithm \ref{alg:MSA}) with n\_try1 = 50, n\_try2= 50, n\_MCMC1 = 8000, and n\_MCMC2 = 8000. Once we had our initial starting position, we ran a Markov chain for 500,000 iterations, saving only every 10 iterations. Figure \ref{fig:EEG_Cov} shows the recovered covariance structure from our mixed membership model. We can see that the covariance structure for feature 1 accounts for the shift in the alpha peak that was found in \citet{scheffler2019covariate}. On the other hand, we can see that most of the variation in feature 2 is in the low frequency range, which is where we expect the most pink noise.

\subsection{Molecular Subtypes of Breast Cancer}
\label{BC_appendix}

\begin{figure}[H]
    \centering
    \includegraphics[width=.49\textwidth]{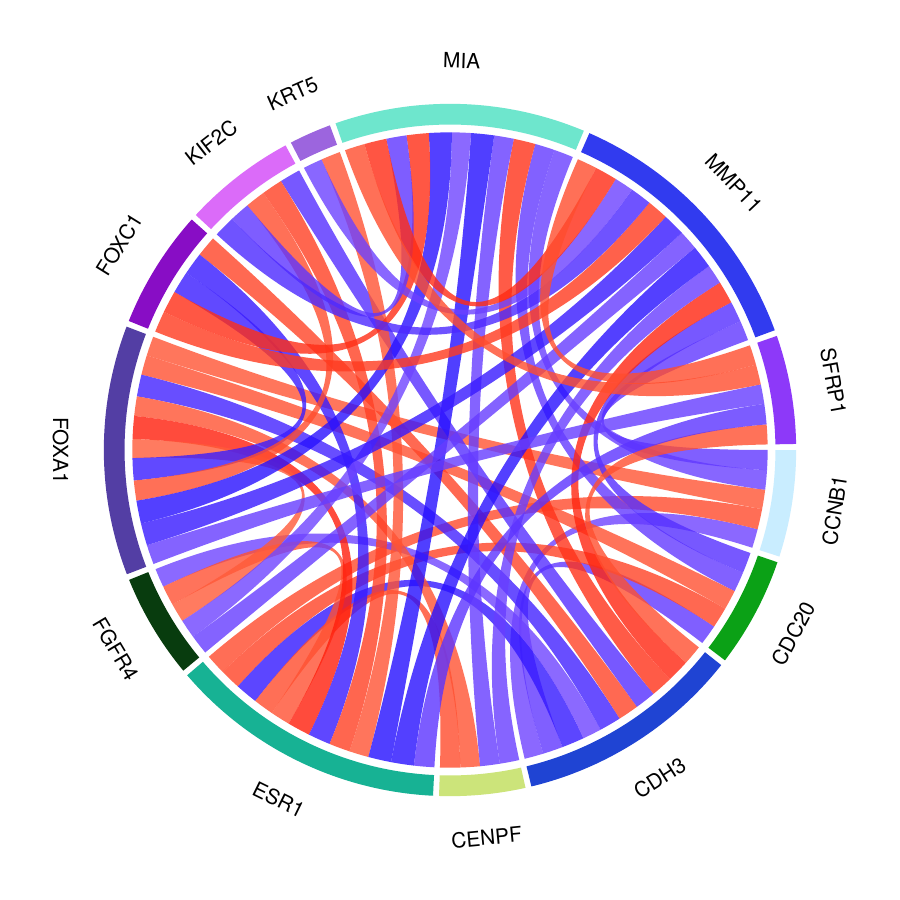}
    \includegraphics[width=.49\textwidth]{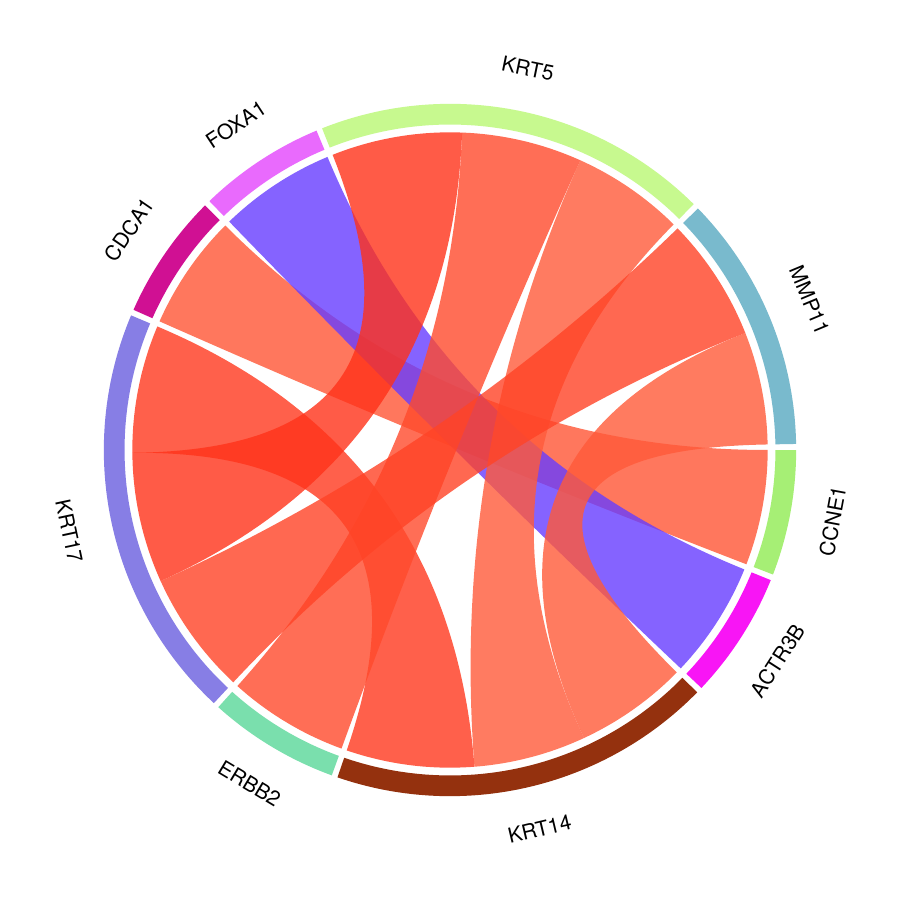}
    \includegraphics[width=.49\textwidth]{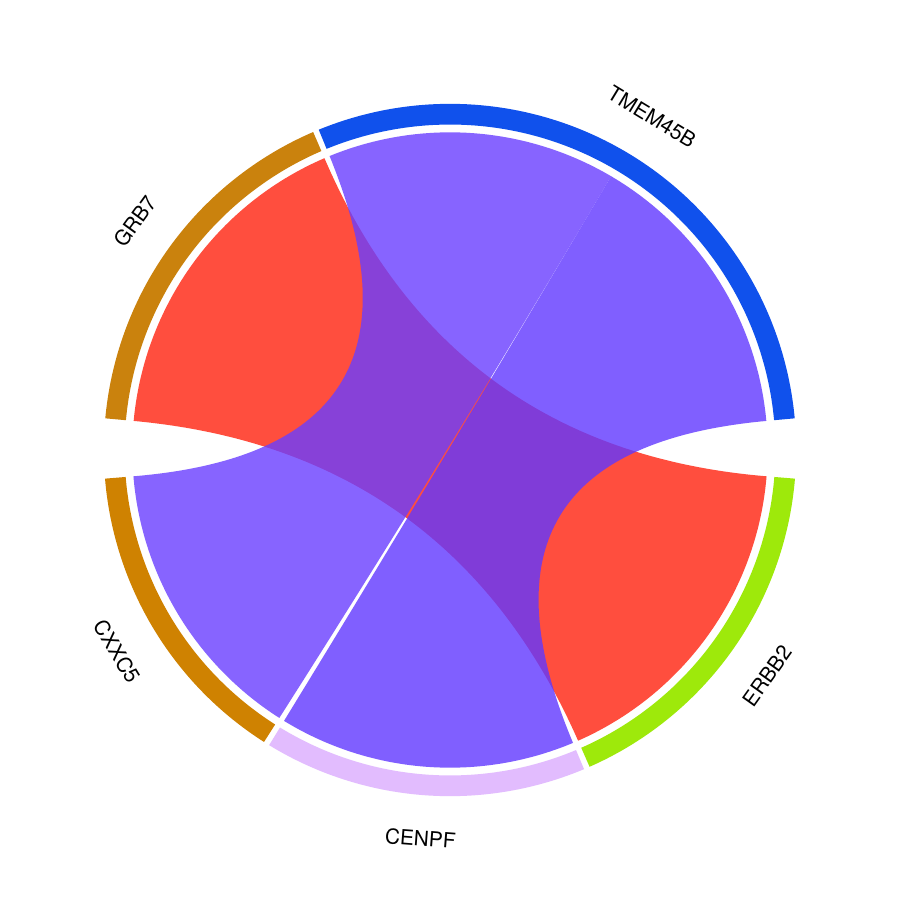}
    \caption{Visualization of the correlation structure of the each feature (Feature 1: Top Left, Feature 2: Top Right, Feature 3: Bottom Middle). Positive correlation is depicted by a red chord, while negative correlation is depicted by a blue chord. Pairwise correlations of less than 0.8 were omitted from the diagrams above.}
    \label{fig:BC_corr}
\end{figure}

For this case study, we used the data provided in \citet{parker2009supervised}, and only used the observations labeled as LumA, Her2, or Basal  ($N =115$). The data set contained some missing values, so we used MICE \citep{van2011mice} to impute the missing data. To get a good starting position, we use the multiple start algorithm (algorithm \ref{alg:MSA}) with n\_try1 = 50, n\_try2 = 6, n\_MCMC1 = n\_MCMC2 = 10000, and K = 3. Using the informed starting position, we then ran our Markov chain for 500,000 iterations, saving every $10^{th}$ iteration. The parameters were then rescaled for ease of interpretation by using the membership rescale algorithm (algorithm \ref{alg:MTA}). From Figure \ref{fig:BC_corr}, we can see the correlation structure in each of the 3 features. We can see that there is relatively high correlation between many of the genes in feature 1 (corresponding to the LumA cancer subtype). 

\section{Factor Models and Mixed Membership Models}
\label{sec: factor}
\begin{figure}
    \centering
    \includegraphics[width = 0.9\textwidth]{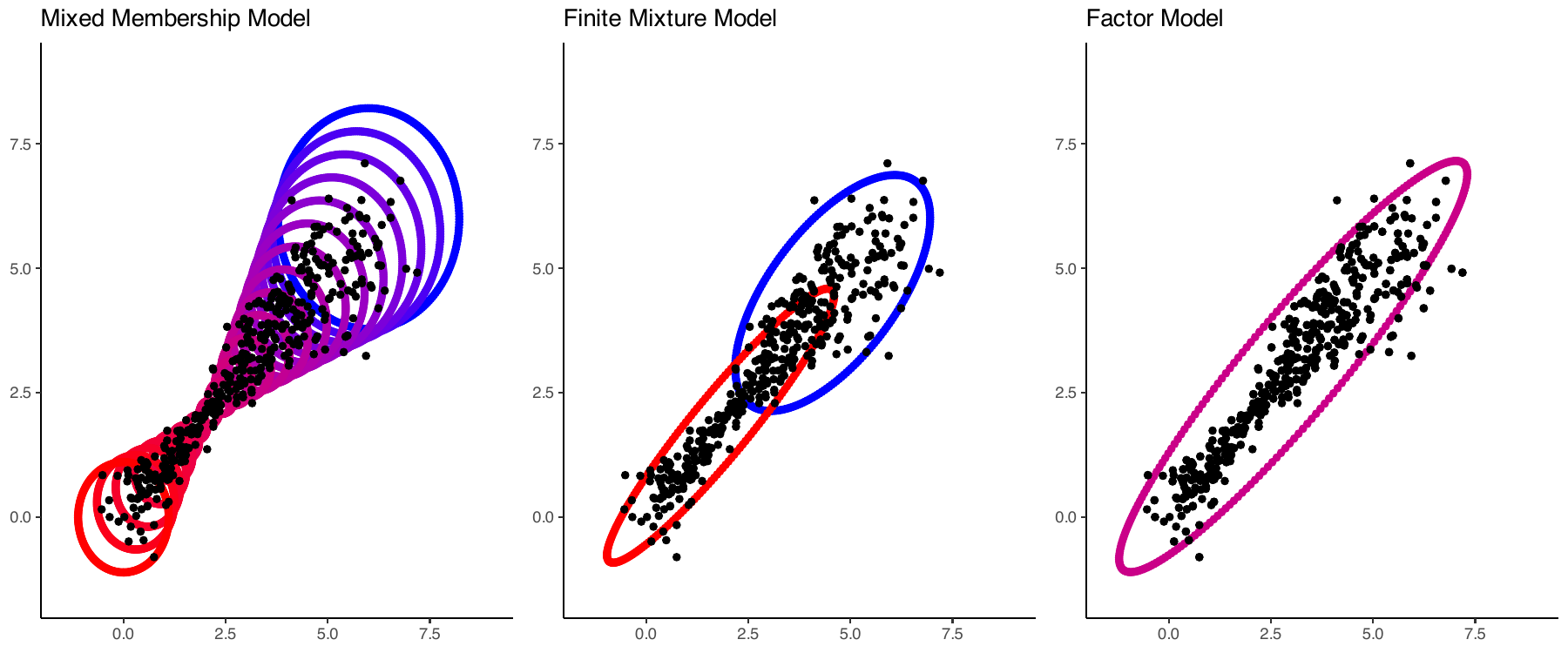}
    \caption{Comparative visualization of the differences between mixed membership models, finite mixture models, and factor models. Each of the models was fit on the same set of data, illustrated by the black dots.}
    \label{fig: MM_FMM_FM}
\end{figure}

Mixed membership models for continuous data, as encoded in our representation in (\ref{eq: fmm_sum}), are related to latent factor models, as they rely on similar additive structures. Nevertheless, mixed membership models obtain an alternative decomposition of data variance, leading to a different interpretation of the model parameters. An illustration of the differences between Gaussian finite mixture models, factor models, and our proposed mixed membership model can be seen in Figure \ref{fig: MM_FMM_FM}.

Factor models are a common tool used in multivariate analysis to model  dependence in high-dimensions through a lower-dimensional linear combination  of latent \textit{factors} \citep{bernardo2003bayesian, carvalho2008high, bhattacharya2011sparse}. The general form of a factor model can be written as
$$\mathbf{y}_i - \boldsymbol{\mu}= \mathbf{B}\boldsymbol{\lambda}_i + \boldsymbol{\nu}_i,$$
where $\mathbf{B} \in \mathbb{R}^{P \times K}$ is known as a matrix of \textit{factor loadings} and $\boldsymbol{\lambda}_i \sim \mathcal{N}_P(\mathbf{0}, \mathbf{I}_P)$ are known as \textit{latent factors}. The parameters $\boldsymbol{\nu}_i \sim \mathcal{N}_P(\mathbf{0}, \boldsymbol{\Sigma})$ are parameters accounting for random error, where $\boldsymbol{\Sigma}$ is a $P \times P$ diagonal matrix. Integrating out the latent factors, factor models generally assume the following distribution on our data:
$$\mathbf{y}_i \sim \mathcal{N}(\boldsymbol{\mu}, \mathbf{B}\mathbf{B}' + \boldsymbol{\Sigma}).$$
Using this parameterization, we can see how factor models are useful in estimating high-dimensional covariance matrices using a low-dimensional representation using factors.

Factor models can be written in an alternative representation that look similar to the mixed membership model in equation (13). Treating the latent factors in a similar fashion as the allocation parameters in equation (13), we arrive at
$$\mathbf{y}_i|\boldsymbol{\lambda}_i \sim \mathcal{N}\left(\boldsymbol{\mu} + \sum_{k=1}^K \lambda_{ik}\mathbf{b}_k, \boldsymbol{\Sigma}\right),$$
where $\lambda_{ik}$ is the $k^{th}$ element of $\boldsymbol{\lambda}_i$ and $\mathbf{b}_k$ is the $k^{th}$ column of $\mathbf{B}$. If we try to interpret $\lambda_{ik}$ in a similar way as the allocation parameters in our proposed mixed membership model, we have that the mean of the $k^{th}$ feature becomes $\boldsymbol{\mu} + \mathbf{b}_k$. While the form of factor models may seem similar to our proposed mixed membership model, there are two key differences between the models. The first, and most important difference, is that $\lambda_i$ do not lie on the unit simplex. This constraint greatly affects the estimation of the feature specific means, more than just a simple rescaling of the means. Constraining the allocation parameters also helps extremely with interpretability. Since $\mathbf{z}_i$ lie on the unit simplex, we can interpret the elements $Z_{ik}$ as the $i^{th}$ observation's proportion of membership to the $k^{th}$ feature. On the other hand, $\sum_{k=1}^K\lambda_{ik}$ is not necessarily equal to 1, meaning we cannot interpret the $\lambda_{ik}$ parameters in a similar fashion. Moreover, the $\lambda_{ik}$ parameters can be negative, making interpretability of the $\lambda_{ik}$ parameters challenging. The second key difference is that the factor model conditional on the latent factors has the same covariance, $\boldsymbol{\Sigma}$. Thus using a factor model, we cannot estimate the correlation structure stratified by feature (i.e. Figures 1 and 2 in the Supplementary Materials).

\begin{figure}
    \centering
    \includegraphics[width = 0.95\textwidth]{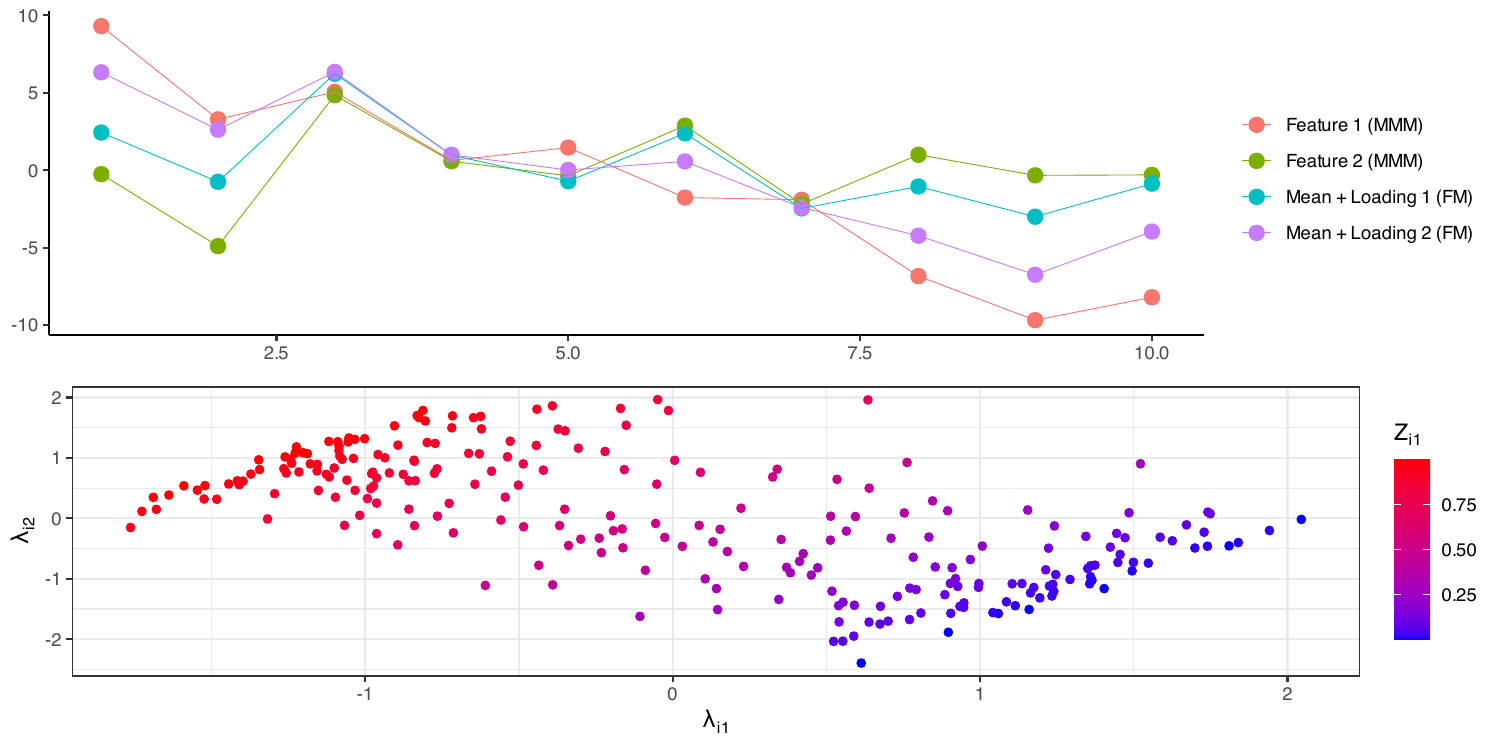}
    \caption{Comparison between a factor model and our mixed membership model, fit on simulated data. The top subfigure illustrates the difference in the mean components between the two models, while the bottom subfigure illustrates the difference between the latent factors of a factor model and allocation parameters of a mixed membership model.}
    \label{fig: FA_vs_MMM}
\end{figure}

An illustration of the differences between factor models and mixed membership models can be seen in Figure \ref{fig: FA_vs_MMM}. To compare the differences between factor models and mixed membership models, we simulated 250 data points ($\mathbf{y}_i \in \mathbb{R}^{10}$) and fit a factor analysis model with 2 factors, as well as a mixed membership model with 2 features. Even though factor models and mixed membership models have a similar additive mean structure, we can see that the estimated means significantly differ due to the added constraint on the allocation parameters in a mixed membership model. Figure \ref{fig: FA_vs_MMM} also illustrates that the allocation parameters ($\mathbf{z}_i$) are closely related to both factors in a factor model. However, trying to interpret the factors as membership to a cluster or feature is challenging because the factors lie $\mathbb{R}^2$, which is an unconstrained space. On the other hand, the allocation parameters can simply be represented on the unit interval, allowing for easy interpretation. Therefore, while there are similarities between factor models and mixed membership models, we can see that there are substantial differences between the two models. We maintain that while factor models are a useful tool to estimated the covariance structure of high dimensional data, they are not well suited for the clustering-type problems discussed in this manuscript.

\bibliographystyle{elsarticle-harv}
\bibliography{BPMM}